\documentclass[11pt]{article}
\usepackage[utf8]{inputenc}
\usepackage[letterpaper]{geometry}
\usepackage{xcolor}
\usepackage{mathrsfs}
\usepackage{amsmath}
\usepackage{amssymb}
\usepackage{setspace}
\usepackage{wasysym}
\usepackage{esint}
\usepackage[authoryear,longnamesfirst]{natbib}
\usepackage[bookmarks=true,bookmarksnumbered=false,bookmarksopen=true,bookmarksopenlevel=1,
 breaklinks=true,pdfborder={0 0 1},backref=false,colorlinks=true, urlcolor=blue, citecolor=blue]
 {hyperref}
\hypersetup{citecolor=blue}

\usepackage{booktabs}   
\usepackage{multirow}
\usepackage{threeparttable}
\usepackage{array}  
\usepackage{siunitx}  
\usepackage{dcolumn}
\newcolumntype{d}[1]{D{.}{.}{#1}}

\usepackage{tikz}
\usetikzlibrary{decorations.pathreplacing}

\makeatletter

\usepackage{amsfonts}\usepackage{mathrsfs}\usepackage{bm}\usepackage{verbatim}\usepackage{dsfont}\usepackage{tabularx}
\usepackage{lipsum}\usepackage{ltablex}\usepackage{array}\usepackage{makecell}\usepackage{caption}\usepackage{multirow}\usepackage{float}\usepackage{wasysym}\usepackage{graphicx}\usepackage{soul}\usepackage{rotating}
\usepackage{graphicx}  
\usepackage[hang,flushmargin]{footmisc}
\usepackage[longnamesfirst,authoryear]{natbib}
\normalbaselines 

\providecommand{\U}[1]{\protect\rule{.1in}{.1in}}
\newtheorem{theorem}{Theorem}[section]

\newtheorem{assumption}{Assumption}

\newtheorem{corollary}[theorem]{Corollary}

\newtheorem{lemma}[theorem]{Lemma}
\newtheorem{proposition}[theorem]{Proposition}\newtheorem{remark}{Remark}[section]

\@addtoreset{lemma}{Appendix B}

\numberwithin{equation}{section}

\newcommand{\Keywords}[1]{\par\noindent {\small{\em Keywords\/}: #1}} 
\newcommand{\JELclass}[1]{\par\noindent {\small{\em JEL classification\/}: #1}} 

\makeatother

\begin{document}
\title{Dimension Reduction for Conditional Density Estimation with Applications
to High-Dimensional Causal Inference}
\author{{Jianhua Mei}\thanks{Research School of Economics, The Australian National University,
Canberra, ACT 0200, Australia. Email: \protect\protect\protect\protect\protect\protect\protect\protect\protect\protect\protect\protect\protect\protect\protect\protect\protect\href{http://jianhua.mei@anu.edu.au}{jianhua.mei@anu.edu.au}.} \\
 {\normalsize{}{}{}{}{}Australian National University} \and
{Fu Ouyang}\thanks{School of Economics, The University of Queensland, St Lucia, QLD 4072,
Australia. Email: \protect\protect\protect\protect\protect\protect\protect\protect\protect\protect\protect\protect\protect\protect\protect\protect\protect\href{http://f.ouyang@uq.edu.au}{f.ouyang@uq.edu.au}.} \\
 {\normalsize{}{}{}{}{}University of Queensland} \and {Thomas
T. Yang}\thanks{Corresponding author. Research School of Economics, The Australian
National University, Canberra, ACT 0200, Australia. Email: \protect\protect\protect\protect\protect\protect\protect\protect\protect\protect\protect\protect\protect\protect\protect\protect\protect\href{http://tao.yang@anu.edu.au}{tao.yang@anu.edu.au}.} \\
 {\normalsize{}{}{}{}{}Australian National University}}
\date{\today}

\maketitle
\onehalfspacing 



\begin{abstract}
We propose a novel and computationally efficient approach for nonparametric conditional density estimation in high-dimensional settings that achieves dimension reduction without imposing restrictive distributional or functional form assumptions. To uncover the underlying sparsity structure of the data, we develop an innovative conditional dependence measure and a modified cross-validation procedure that enables data-driven variable selection, thereby circumventing the need for subjective threshold selection. We demonstrate the practical utility of our dimension-reduced conditional density estimation by applying it to doubly robust estimators for average treatment effects. Notably, our proposed procedure is able to select relevant variables for nonparametric propensity score estimation and also inherently reduces the dimensionality of outcome regressions through a refined ignorability condition. We evaluate the finite-sample properties of our approach through comprehensive simulation studies and an empirical study on the effects of 401(k) eligibility on savings using SIPP data.

\vspace{4mm} 

\JELclass{C12, C14, C21, C52}

\medskip{}

\vspace{2cm}

\end{abstract}

\newpage{}

\onehalfspacing

\section{Introduction}
Understanding the relationship between a response variable and a set of covariates is of central interest in statistics and econometrics. The conditional probability density function fully characterizes this relationship. Beyond being a fundamental object of interest, the conditional density plays a pivotal role in a wide range of estimation and inference methods. For instance, in causal inference, the propensity score--the conditional density of a binary treatment on covariates--is central to inverse propensity weighting (IPW; e.g., \cite{Hirano2003}), doubly robust, and propensity score matching (e.g., \cite{abadie2016matching}) estimators that are designed to adjust for confounding and mitigate selection bias in observational studies. The literature on this topic is vast, and we refer readers to recent comprehensive treatments by \citet{ImbensRubin}, \citet{ding2024first}, and \citet{chernozhukov2024applied}. The importance of conditional densities is also pronounced in modern econometrics, with applications including the estimation of structural models (e.g.,
\citet{GuerreEtal}, \citet{Matzkin2013}, \citet{PerrigneVuong}), nonparametric estimation of nonseparable models (e.g., \citet{AltonjiMatzkin}, \citet{BlundellEtal2017}), and semiparametric estimation of discrete choice models with endogeneity (e.g., \citet{Lewbel2000}), among others.

We investigate the estimation of conditional densities in high-dimensional settings, a task often challenged by the ``curse of dimensionality''. To address this issue, existing approaches commonly impose strong distributional or functional form assumptions (e.g., \citet{MaZhu2013}). However, in real-world data, these assumptions can be rather restrictive. In this paper, we propose an alternative approach that does not depend on these constraints. Instead, we introduce a ``sparsity'' condition, where only a small subset of covariates meaningfully affects the distribution of the response variable. While this sparsity condition limits the applicability of our method in cases with many weak signals, it is inspired by the observation in the well-known work of \citet{HallRacineLi} that irrelevant covariates are surprisingly common in practice. Nonetheless, the primary advantage of our method is that it does not rely on distributional or functional form assumptions.

This paper is further motivated by a practical challenge that arises in the doubly robust estimation of average treatment effects (ATE) in high-dimensional settings. While practitioners may tend to include many covariates to strengthen the ignorability assumption, doing so undermines the overlap condition (because the treatment status becomes more predictable, and the propensity score is therefore more likely to approach 0 or 1) and exacerbates the curse of dimensionality for nonparametric estimators. This, in turn, can hinder the application of modern double/debiased machine learning (DML) methods for robust causal inference. Our identification results, summarized in Proposition \ref{Prop:refine}, offer a direct solution to this challenge. Under the above sparsity condition, these results establish a ``double dimension reduction'': recovering the sparsity structure of the treatment variable not only simplifies the propensity score model but also reduces the dimensionality of the outcome regression models through a refined ignorability condition. This insight, combined with our proposed dimension-reduced conditional density estimator, enables the identification and use of a lower-dimensional set of truly relevant covariates. As a result, it enhances the plausibility of the overlap condition and facilitates flexible nonparametric estimation of both the propensity score and outcome regressions in high-dimensional contexts.

To set the stage, let $Y$ denote a response variable, $\boldsymbol{W}$ be a set of pre-selected covariates known to be relevant to $Y$, and $\boldsymbol{X}$ be a high-dimensional vector of candidate covariates. The covariates in $\boldsymbol{W}$ are typically selected based on domain knowledge, theoretical models, or prior selection in earlier analysis stages; $\boldsymbol{W}$ may also be empty.\footnote{If data sparsity or model interpretability is not a key concern, $\boldsymbol{W}$ may also include practitioner-constructed sufficient statistics, such as leading principal components.} Beyond its intrinsic interest, including well-chosen $\boldsymbol{W}$ can simplify the dimension reduction task when many covariates in $\boldsymbol{X}$ influence $Y$ only through  $\boldsymbol{W}$, as illustrated in our empirical application. Our framework allows the dimension of $\boldsymbol{X}$ to grow at a polynomial rate. The goal is to identify the covariates among $\boldsymbol{X}$ that are relevant to $Y$ given $\boldsymbol{W}$. To this end, we propose a new, model-free measure of conditional dependence between $Y$ and each covariate in $\boldsymbol{X}$, conditional on $\boldsymbol{W}$. This measure is non-negative and equals zero if and only if $Y$ is independent of the covariate given $\boldsymbol{W}$. This conditional dependence measure underpins our dimension reduction approach and constitutes one of the core contributions of the paper.

Building on this measure, we develop a novel, two-stage procedure for dimension-reduced conditional density estimation. In the first stage, we adapt the ``sure independence screening'' (SIS) framework of \cite{FanLv2008} to design a screening procedure that efficiently identifies relevant covariates in a high-dimensional setting. Specifically, we compute the conditional dependence measure for each candidate covariate individually, rank them, and retain those that exhibit the strongest conditional dependence with $Y$ given $\boldsymbol{W}$. This screening step is particularly well-suited for nonparametric density estimations, as directly measuring dependence or regressing $Y$ on high-dimensional covariates without parametric or functional form assumptions is generally infeasible.

In the second stage, we refine the variable selection using a modified cross-validation (CV) procedure, building on \cite{HallRacineLi}, to determine the final set of covariates in an objective and data-driven manner. The key insight of the CV-based refinement is that, in kernel conditional density estimation, the optimal bandwidths for irrelevant covariates, chosen to minimize the integrated squared error (ISE), tend to diverge to their upper extremes, effectively smoothing them out of the final density estimate. Our modification significantly reduces the computational burden of the original algorithm by \cite{HallRacineLi} via restricting the search space, thereby improving numerical stability and computational efficiency, as demonstrated in our simulation studies. The final conditional density estimate is obtained using the selected covariates and the optimal bandwidths from this refinement step.

We demonstrate the practical utility of our dimension reduction method by applying it to the doubly robust estimation of ATE. Our primary contribution in this section is the identification results summarized in Proposition \ref{Prop:refine}. Simulation results show that the doubly robust ATE estimators that incorporate the insights of Proposition \ref{Prop:refine} and leverage our dimension-reduced propensity score estimator exhibit satisfactory finite-sample performance, in sharp contrast to estimators that use the full high-dimensional covariate set. We further illustrate this approach through revisiting the empirical analysis of the effect of 401(k) eligibility on asset accumulation, using rich longitudinal data from the 1996 Survey of Income and Program Participation (SIPP).

Our measure of conditional dependence is a novel generalization of the Kolmogorov-Smirnov (KS) distance-based measure proposed by \citet{MaiZou2015}, and it differs from theirs in several important aspects. First, while their measure is designed exclusively for assessing unconditional (in)dependence (i.e., when $\boldsymbol{W}$ is empty), our measure naturally extends to conditional settings. Second, by introducing fixed reference values when computing the KS distance, our measure substantially reduces computational burden; see Remark \ref{Remark:comparenew} for further discussion.

Two other widely recognized dependence measures in the literature are the (conditional) distance correlation introduced by \citet{Szekely_et_al} and \citet{WangEtal2015}, and the (conditional) rank-based correlation proposed by \citet{Chatterjee2021} and \citet{azadkia2021simple}. In terms of computation, our method is significantly faster than distance correlation but slower than the rank-based approach, as detailed in Remark~\ref{Remark:compare}. As expected, our simulations indicate that the statistical power of our method in small samples lies between the two: lower than that of distance correlation, but higher than that of the rank-based approach. We therefore view our method as particularly useful for moderate sample sizes (e.g., in the thousands to tens of thousands), striking a balance between computational efficiency and small sample statistical power.

This paper also contributes to the recent advances in estimating conditional density functions using a variety of machine learning (ML) methods; see, for example, \cite{izbicki_converting_2017}, \cite{rothfuss2019conditional}, \cite{gao2022lincde}, and the references therein. However, all these studies essentially assume that the researcher has prior knowledge of the finite-dimensional set of relevant covariates. In contrast, our approach provides a data-driven solution to identifying relevant covariates in high-dimensional settings. Additionally, our method complements the rapidly growing literature on causal inference using DML methods, where the propensity score or conditional density is often a key input for constructing robust estimators. Examples include \citet{farrell2015robust}, \citet{ChernozhukovEtal2018}, \cite{chang2020double}, \cite{FarrellEtal2021}, \cite{zhang2024continuous}, and \cite{haddad2024difference}, among many others reviewed in \citet{chernozhukov2024applied}.

The remainder of this paper is organized as follows. Section \ref{SEC:f(y,x)reduce} introduces our new conditional dependence measure, the variable screening and CV refinement procedures, and the dimension-reduced conditional density estimator. Section \ref{sec:theoryapply} demonstrates how the proposed methods can be applied to estimate ATE using doubly robust estimators, with a focus on dimension-reduced propensity score estimation. Section \ref{sec:simulation} evaluates the finite-sample performance of our approach through comprehensive simulation studies, where we also provide practical implementation guidance. Section \ref{sec:application} demonstrates the application of our approach to the analysis of the effect of 401(k) eligibility on savings. Section \ref{SEC:conclusion} concludes the paper.

Supporting materials are provided in the appendices. Appendix \ref{APP:Additional-Results} contains additional results and further discussion of related methods in the literature. Appendix \ref{APP:compute_Rho} presents a computationally efficient algorithm to calculate our proposed dependence measure. Proofs of the main theoretical results are provided in Appendix \ref{APP:mainproof}, while technical lemmas and their proofs are deferred to Appendix \ref{APP:lemmasProofs}. Finally, Appendix \ref{appendix:tables} collects tables summarizing the results of our Monte Carlo simulations and empirical illustration.

\par\bigskip
\noindent\textbf{Notation.} All vectors are column vectors. Capital letters denote random elements, and the corresponding lowercase letters denote their realizations. Boldface letters represent vectors, while regular (non-bold) letters represent scalars. The notation $\textrm{dim}(\boldsymbol{z})$ denotes the dimension of a vector $\boldsymbol{z}$. Superscripts ``$\mathsf{c}$'' and ``$\mathsf{d}$'' are used to indicate continuous and discrete random variables, respectively. We use $\Pr(\cdot)$ and $\mathbb{E}[\cdot]$ to denote probability and expectation, respectively. The function $\mathbf{1}(\cdot)$ is the indicator function, which equals one if the event in the parentheses occurs, and zero otherwise. For two random vectors $\boldsymbol{U}$ and $\boldsymbol{V}$, the notation $\boldsymbol{U}|\cdot \overset{d}{\sim} \boldsymbol{V}|\cdot$ indicates that $\boldsymbol{U}$ and $\boldsymbol{V}$ have identical distributions conditional on $\cdot$. The notation $\boldsymbol{U} \perp \boldsymbol{V}$ ($\boldsymbol{U} \perp \boldsymbol{V}|\cdot$) denotes stochastic independence (conditional on $\cdot$). We use $F(\boldsymbol{u}|\boldsymbol{v})$ and $f(\boldsymbol{u}|\boldsymbol{v})$ to denote the conditional cumulative distribution function (CDF) and probability density function (PDF) of $\boldsymbol{U}$ given $\boldsymbol{V}$, respectively. As $n \to \infty$, the notations $\overset{P}{\rightarrow}$ and $\overset{d}{\rightarrow}$ denote convergence in probability and convergence, respectively. The symbol $C$ represents a generic positive constant whose value may vary from line to line. For deterministic sequences $\left\{ a_{n}\right\} _{n=1}^{\infty}$ and $\left\{ b_{n}\right\} _{n=1}^{\infty}$, we write $a_{n}\propto b_{n}$ (or $a_n=O(b_n)$) to mean that $0<C_{1}\leq\lim\inf_{n\rightarrow\infty}\left\vert a_{n}/b_{n}\right\vert \leq\lim\sup_{n\rightarrow\infty}\left\vert a_{n}/b_{n}\right\vert \leq C_{2}<\infty$ for some $C_1, C_2 > 0$. We write $a_n \lesssim b_n$ if $\limsup_{n \to \infty} |a_n / b_n| \leq C$ for some $C>0$, and $a_n \gtrsim b_n$ if $b_n \lesssim a_n$. We write $a_n \ll b_n$ (or $a_n = o(b_n)$) if $a_n / b_n \to 0$, and $a_n \gg b_n$ if $b_n \ll a_n$. The symbol $\setminus$ denotes set difference.

\section{Dimension Reduction for Conditional Densities\protect\protect\protect\protect\protect} \label{SEC:f(y,x)reduce}
This section develops our dimension reduction methodology for conditional density estimation. We consider a setting with a response variable $Y$, a set of pre-selected covariates $\boldsymbol{W}$ known to be relevant, and a high-dimensional vector of candidate covariates $\boldsymbol{X} = (X_1, \dots, X_p)'$. The goal is to estimate the conditional density $f(Y|\boldsymbol{X}, \boldsymbol{W})$ under a sparsity assumption: only a small, unknown subset of $\boldsymbol{X}$ is truly relevant for this density.

To achieve this, we first develop a new measure of conditional dependence (Section \ref{SEC:newMeasure}) and apply it within an SIS procedure to screen for relevant variables (Section \ref{Sec:Screening}). We then propose a CV refinement step and construct the post-selection conditional density estimator (Section \ref{SEC:post-estimation}). In Section \ref{sec:Propensity-Score}, we discuss how this two-stage procedure can be specifically adapted for propensity score estimation. In what follows, we decompose $\boldsymbol{W}$ into its continuous and discrete components as $\boldsymbol{W} = \left( \boldsymbol{W}^{\mathsf{c}\prime}, \boldsymbol{W}^{\mathsf{d}\prime} \right)'$, where $\boldsymbol{W}^{\mathsf{c}}$ and $\boldsymbol{W}^{\mathsf{d}}$ are $q^{\mathsf{c}} \times 1$ and $q^{\mathsf{d}} \times 1$ subvectors, respectively.

\subsection{New Measure and Statistic \protect\protect\protect\protect} \label{SEC:newMeasure}
This section introduces our measure of conditional dependence. We focus on the dependence between a scalar response variable $Y$ and a single candidate covariate $X$, conditional on $\boldsymbol{W}$. 

Note that $X\perp Y|\boldsymbol{W}=\boldsymbol{w}$ holds if and only if the conditional distribution of $X$ given $(Y, \boldsymbol{W})$ is invariant with respect to the value of $Y$. That is,
\begin{equation}
\left.X\right|Y=y,\boldsymbol{W}=\boldsymbol{w}\stackrel{d}{\sim}\left.X\right|Y=y',\boldsymbol{W}=\boldsymbol{w},\label{eq:compareDist}
\end{equation}
for any values $y$ and $y'$. To quantify departures from this condition, we use the KS distance. Let $\Lambda_X(Y=y,\boldsymbol{W}=\boldsymbol{w})$ be the KS distance between the conditional CDF of $X$ and the same CDF evaluated at a user-specified reference value $y^*$:
\[
\Lambda_{X}\left(Y=y,\boldsymbol{W}=\boldsymbol{w}\right)=\sup_{x}\left|F\left(x|y,\boldsymbol{w}\right)-F\left(x|y^{*},\boldsymbol{w}\right)\right|.
\]
Here, $y^*$ can be set to a fixed value, such as the sample median of $Y$. Clearly, $X\perp Y\mid \boldsymbol{W}=\boldsymbol{w}$ if and only if $\sup_{y}\Lambda_{X}(Y=y,\boldsymbol{W}=\boldsymbol{w})=0$.

It is worth noting that we do not need to use a statistic of the following form:
\begin{equation}
\sup_{(y,y')}\sup_{x}\left|F\left(x|y,\boldsymbol{w}\right)-F\left(x|y',\boldsymbol{w}\right)\right|. \label{eq:tri1}
\end{equation}
In fact, (\ref{eq:tri1}) is strictly positive if and only if $\sup_{y}\sup_{x}\left|F\left(x|y,\boldsymbol{w}\right)-F\left(x|y^{*},\boldsymbol{w}\right)\right|>0$. To see this, suppose there exists a pair $(y,y')$ such that 
\[
\sup_{x}\left|F(x\mid y,\boldsymbol{w})-F(x\mid y',\boldsymbol{w})\right|\geq\delta>0.
\]
Then, by the triangle inequality, we must have 
\[
\max\left\{ \sup_{x}\left|F(x\mid y,\boldsymbol{w})-F(x\mid y^{*},\boldsymbol{w})\right|,\sup_{x}\left|F(x\mid y',\boldsymbol{w})-F(x\mid y^{*},\boldsymbol{w})\right|\right\} \geq\frac{\delta}{2}>0.
\]
The reverse direction is obvious. Using $\Lambda_{X}(Y=y,\boldsymbol{W}=\boldsymbol{w})=0$ (i.e., fixing $y'$ to $y^{*}$ in (\ref{eq:tri1})) can substantially reduce the computational burden, as further discussed in Remark \ref{Remark:comparenew}.

Since $X\perp Y|\boldsymbol{W}$ requires the distributional equivalence in (\ref{eq:compareDist}) to hold for all values of $\left(y,\boldsymbol{w}\right)$, we define our overall (single summary) measure of dependence by taking the expectation of $\Lambda_X$ over the joint distribution of $(Y, \boldsymbol{W})$:
\begin{equation}
\rho=\mathbb{E}\left[\Lambda_{X}\left(Y,\boldsymbol{W}\right)\right].\label{eq:rho_continueX}
\end{equation}
By excluding events with zero probability under the distribution of \((Y, \boldsymbol{W})\), we have that \(X \perp Y \mid \boldsymbol{W}\) if and only if \(\rho = 0\). To see this, note that  
\begin{equation}
\rho = \int \int \Lambda_X(y, \boldsymbol{w}) f(y \mid \boldsymbol{w}) \, dy \, f(\boldsymbol{w}) \, d\boldsymbol{w}. 
\end{equation}
Thus, \(\rho > 0\) if and only if there exists a set of \(\boldsymbol{w}\) with nonzero measure such that, conditional on these \(\boldsymbol{w}\), there exists a set of \(y\) (also with nonzero measure) for which \(\Lambda_X(y, \boldsymbol{w}) > 0\). 

The sample counterpart of $\Lambda_{X}(Y=y,\boldsymbol{W}=\boldsymbol{w})$ can be estimated as 
\begin{equation}
\hat{\Lambda}_{X}\left(Y=y,\boldsymbol{W}=\boldsymbol{w}\right)=\max_{i=1,...,n}\left|\hat{F}_{n}\left(x_{i}|y,\boldsymbol{w}\right)-\hat{F}_{n}\left(x_{i}|y^{*},\boldsymbol{w}\right)\right|,\label{eq:newLambda}
\end{equation}
where the estimated CDFs are given by
\begin{equation}
\hat{F}_{n}\left(x|y,\boldsymbol{w}\right)=\frac{\sum_{i=1}^{n}\mathbf{1}\left(x_{i}\leq x\right)K_{h}\left(y_{i}-y\right)\left[\Pi_{l=1}^{q^{\mathsf{c}}}K_{h}\left(w_{li}^{\mathsf{c}}-w_{l}^{\mathsf{c}}\right)\right]\left[\Pi_{l=1}^{q^{\mathsf{d}}}K_{\lambda}^{\mathsf{d}}\left(w_{li}^{\mathsf{d}},w_{l}^{\mathsf{d}}\right)\right]}{\sum_{i=1}^{n}K_{h}\left(y_{i}-y\right)\left[\Pi_{l=1}^{q^{\mathsf{c}}}K_{h}\left(w_{li}^{\mathsf{c}}-w_{l}^{\mathsf{c}}\right)\right]\left[\Pi_{l=1}^{q^{\mathsf{d}}}K_{\lambda}^{\mathsf{d}}\left(w_{li}^{\mathsf{d}},w_{l}^{\mathsf{d}}\right)\right]},\label{eq:CDFhat}
\end{equation}
Here, $w_{l}^{\mathsf{c}}$ and $w_{l}^{\mathsf{d}}$ denote the $l$-th elements of the continuous and discrete parts of $\boldsymbol{w}$, respectively. $K_{h} ( \cdot ) = h^{-1} K ( \cdot )$, where $K ( \cdot )$ is a standard kernel function for continuous covariates. The kernel for discrete covariates is the Aitchison-Aitken kernel, defined as
\[
K_{\lambda}^{\mathsf{d}} \left( w_{l}^{\mathsf{d}}, w^{\mathsf{d}} \right) = \left( \frac{\lambda}{r_{l} - 1} \right)^{ \mathbf{1} \left( w_{l}^{\mathsf{d}} \neq w^{\mathsf{d}} \right) } \left( 1 - \lambda \right)^{ \mathbf{1} \left( w_{l}^{\mathsf{d}} = w^{\mathsf{d}} \right) },
\]
where $r_{l}$ is the number of distinct atoms in the support of $W_{l}^{\mathsf{d}}$ and $0 \leq \lambda \leq (r_{l} - 1)/r_{l}$. For simplicity, we assume the same bandwidth $h$ and tuning parameter $\lambda$ across all covariates. The use of the Aitchison-Aitken kernel is standard for handling discrete data; see, e.g., Chapter 4 in \cite{LiRacine}.

Finally, our estimator of $\rho$ is defined as
\begin{equation}
\hat{\rho}=\frac{1}{n}\sum_{i=1}^{n}\hat{\Lambda}_{X}\left(Y=y_{i},\boldsymbol{W}=\boldsymbol{w}_{i}\right).\label{eq:rho_continuousX}
\end{equation}
In Appendix \ref{APP:compute_Rho}, we develop a highly efficient algorithm to compute $\hat{\rho}$ at a computational cost proportional to $n^2$. We conclude this section with two remarks: one highlighting the computational improvement of our approach over a direct generalization of \cite{MaiZou2015}, and another comparing our measure with other popular conditional dependence measures in the literature.

\begin{remark} \label{Remark:comparenew} 
\emph{If the measure in \cite{MaiZou2015} were directly generalized to handle conditional dependence, the resulting empirical measure would correspond to a statistic based on (\ref{eq:tri1}), which requires comparing all pairs of observations and results in a computational cost proportional to $n^3$. By establishing the equivalence between (\ref{eq:tri1}) and our formulation $\sup_y\Lambda_{X}\left(Y=y,\boldsymbol{W}=\boldsymbol{w}\right)$, which uses a single (or a finite number of; see, e.g., simulations in Section \ref{sec:simu_1}) fixed reference point(s), our approach reduces the computational cost to $O(n^2)$, yielding a substantial computational improvement.} 
\end{remark} 

\begin{remark} \label{Remark:compare} 
\emph{Our approach can be benchmarked against two prominent classes of conditional dependence measures in the literature. First, compared to the conditional distance correlation (CDC) proposed by \citet{WangEtal2015}, our method is computationally more efficient, achieving a lower computational cost of $O(n^2)$, whereas the CDC has a cost of $O(n^3)$.\footnote{The computational cost of $\mathcal{V}_{n}(\mathbf{W}_{n} \mid Z)$ in equation (2) of their appendix is $O(n^2)$, as can be established using identities in \cite{Szekely_et_al}, specifically their equations (2.8), (2.18), and the proof in their appendix. Since the CDC requires evaluating $\mathcal{V}_{n}(\mathbf{W}_{n} \mid Z)$ at each observation, the total cost scales as $O(n^3)$.} This efficiency gain involves a trade-off, as our simulation results show that in small samples, the statistical power of our measure is generally lower than that of the CDC. Second, compared to the innovative rank-based (conditional) dependence measure from \citet{Chatterjee2021} and \citet{azadkia2021simple}, the trade-off is reversed. Their measure is faster, with computational cost proportional to $n \log n$, but our simulations demonstrate that their method is less powerful. Thus, our approach strikes a practical balance between computational cost and statistical power, making it particularly well-suited for moderately sized datasets.}
\end{remark}

\subsection{Screening and Its Error Bound \protect\protect} \label{Sec:Screening}
We adopt the conditional dependence measure developed in the previous section for our proposed screening procedure. For each candidate covariate $X_j$, where $j = 1, \dots, p$, we define
\[
\rho_j = \mathbb{E} \left[ \Lambda_{X_j} \left( Y, \boldsymbol{W} \right) \right],
\]
as in (\ref{eq:rho_continueX}), with its sample counterpart given by (\ref{eq:rho_continuousX}). We impose the following technical conditions.

\begin{assumption} \label{A:signal_strength} 
\begin{enumerate}
\item[(1)] The elements of $\boldsymbol{X}$ satisfy: 
\[
X_{j}\not\perp Y|\boldsymbol{W},\text{ }j=1,...,s^{*},\text{ and }X_{j}\perp Y|\boldsymbol{W},\text{ }j=s^{*}+1,...,p.
\]
Furthermore, there exists a small positive $c$ such that 
\[
\min_{j=1,...,s^{*}}\rho_{j}=\mathbb{E}\left[\Lambda_{X_{j}}\left(Y,\boldsymbol{W}\right)\right]\gtrsim\left(\log n\right)^{1+c}n^{-r/\left(2r+q^{\mathsf{c}}+1\right)},
\]
where $r$ is the order of the kernel function $K\left(\cdot\right)$, as characterized in (6) of this assumption. 
\item[(2)] $p\propto n^{B_{p}}$ for some $B_{p}\geq0.$ 
\item[(3)] $\left\{\boldsymbol{x}_{i},y_{i},\boldsymbol{w}_{i}\right\}_{i=1}^{n} $
are i.i.d. across $i$. 
\item[(4)] There exist a small positive constant $B_{1}$ such that $0<B_{1}\le f\left(Y,\boldsymbol{W}\right)<\infty.$ 
\item[(5)] All discrete covariates in $\boldsymbol{W}^{\mathsf{d}}$ and $\boldsymbol{X}$ have a uniformly finite number of atoms, with probabilities at each atom uniformly bounded away from zero. 
\item[(6)] The kernel function $K\left(\cdot\right)$ satisfies: $\int K\left(u\right)du=1,$
$\int K^{2}\left(u\right)du<\infty,$ $\int u^{j}K\left(u\right)du=0,$
for $j=1,...,r-1$, and $\int u^{r}K\left(u\right)du<\infty$ for
a positive integer $r\geq2.$ 
\item[(7)] For $j=1,...,p,$ $F\left(X_{j}|Y,\boldsymbol{W}\right)$ and
$f\left(X_{j}|Y,\boldsymbol{W}\right)$ are $r$-th order continuously
differentiable with respect to $Y$ and $\boldsymbol{W}^{\mathsf{c}}$,
and those derivatives are uniformly bounded, for $j=1,...,p$
and all possible values of $\left(X_{j},Y,\boldsymbol{W}\right)$. 
\end{enumerate}
\end{assumption}

Assumption \ref{A:signal_strength}(1) defines the sets of relevant and irrelevant covariates. The second part of Assumption \ref{A:signal_strength}(1) imposes a minimum signal strength condition, ensuring that the dependence of each relevant covariate on $Y$ is sufficiently large to be detected in finite samples. Assumption \ref{A:signal_strength}(2) allows $p$ to grow at any polynomial rate. This can be relaxed to an exponential rate at the cost of a slight
sacrifice of power as reflected in the signal strength requirement in Assumption \ref{A:signal_strength}(1). Assumption \ref{A:signal_strength}(3) requires a random sample. Assumptions \ref{A:signal_strength}(4) and (5) guarantee that the denominators in $\hat{\rho}_j$ are uniformly bounded away from zero with high probability. Finally, Assumptions \ref{A:signal_strength}(6) and (7) are standard kernel regularity and smoothness conditions used to control the bias in nonparametric estimation.

We define the set containing the indices of all relevant variables as $\mathcal{M}^{*} = \left\{ 1,...,s^{*}\right\}$. Its corresponding estimator, $\hat{\mathcal{M}}$, is constructed by collecting the indices of all covariates whose estimated dependence measure, $\hat{\rho}_j$, exceeds a certain threshold $Cn^{-r/\left(2r+q^{\mathsf{c}}+1\right)}\log n$ for a fixed positive constant $C$, i.e.,  
\begin{equation}
\hat{\mathcal{M}}=\left\{ j:\hat{\rho}_{j}\geq Cn^{-r/\left(2r+q^{\mathsf{c}}+1\right)}\log n\right\}. \label{eq:screeningset}
\end{equation}

\begin{theorem}\label{TH:TPR}
Suppose Assumption \ref{A:signal_strength} holds. Let $h\propto n^{-1/\left(2r+q^{\mathsf{c}}+1\right)}$ and
$\lambda\propto n^{-r/\left(2r+q^{\mathsf{c}}+1\right)}.$ Then, 
\[
\Pr\left(\hat{\mathcal{M}}=\mathcal{M}^{*}\right)\rightarrow1.
\]
\end{theorem}

Theorem \ref{TH:TPR} establishes that, with appropriately chosen thresholds in (\ref{eq:screeningset}), our screening procedure can select the set of relevant covariates with probability approaching one. This result can also be readily extended to the case where $Y$ is a binary random variable, in which the conditional density of $Y$ reduces to the propensity score. We discuss this extension in detail in Section \ref{sec:Propensity-Score}.

\subsection{Thresholds, CV Refinement and Post-Selection Estimation \protect\protect} \label{SEC:post-estimation}
While Theorem \ref{TH:TPR} guarantees that our screening procedure works in theory, the threshold defined in (\ref{eq:screeningset}) depends on an unknown constant $C$ that is difficult to choose in practice. To address this, we move beyond a fixed (subjective) threshold and propose a data-driven second stage to refine the set of selected variables. Our approach utilizes the CV method of \citet{HallRacineLi}, which eliminates irrelevant covariates through kernel bandwidth selection. The refinement proceeds as follows: first, we use our screening method to select a reasonably large set of $\tilde{p}$ candidate covariates, including the pre-selected $\boldsymbol{W}$ and the top covariates in $\boldsymbol{X}$ ranked by $\hat{\rho}_j$. Then, we apply a CV-based refinement procedure to this set of $\tilde{p}$ variables to determine the final conditioning set. A data-driven method for choosing $\tilde{p}$ is provided in Section \ref{sec:summary_procedure}.

One important advantage (and added value) of this refinement step, relative to using a fixed threshold in (\ref{eq:screeningset}), is its ability to further filter out certain
``indirectly relevant'' variables. Consider the following scenario: suppose there exists an $s^{**}<s^{*}$ such that 
\begin{equation}
Y\perp\left(X_{s^{**}+1},...,X_{s^{*}}\right)\left|\left(X_{1},...,X_{s^{**}},\boldsymbol{W}\right)\right.,\label{eq:conditional_refine}
\end{equation}
and such conditional independence is unambiguous.\footnote{Conditional independence can be ambiguous. For example, it is possible for $Y \perp X_1 \mid X_2$, $Y \perp X_2 \mid X_1$, and $Y \not\perp (X_1, X_2)$ to all hold simultaneously. The notion of ``unambiguous'' is formalized in Assumption \ref{A:2}.} In (\ref{eq:conditional_refine}), although $X_{s^{**}+1},...,X_{s^{*}}$ may be dependent on $Y$ conditional on $\boldsymbol{W}$ and thus survive the screening step, they may become conditionally independent of $Y$ once we condition on the subset
\begin{equation}
\boldsymbol{Z}^{*}\equiv\left(X_{1},...,X_{s^{**}},\boldsymbol{W}'\right)'. \label{eq:xstar}
\end{equation}
In this case, the conditional density of $Y$ simplifies as
\[
f\left(Y|X_{1},...,X_{s^{*}},\boldsymbol{W}\right)  =\frac{f\left(Y,X_{s^{**}+1},...,X_{s^{*}}|\boldsymbol{Z}^{*}\right)}{f\left(X_{s^{**}+1},...,X_{s^{*}}|\boldsymbol{Z}^{*}\right)} = \frac{f\left(Y|\boldsymbol{Z}^{*}\right)f\left(X_{s^{**}+1},...,X_{s^{*}}|\boldsymbol{Z}^{*}\right)}{f\left(X_{s^{**}+1},...,X_{s^{*}}|\boldsymbol{Z}^{*}\right)} =f\left(Y|\boldsymbol{Z}^{*}\right).
\]
We refer to $X_{s^{**}+1},..., X_{s^{*}}$ as indirectly relevant covariates and to $\boldsymbol{Z}^{*}$ as directly relevant covariates.

To see that (\ref{eq:conditional_refine}) is a realistic concern, consider the following illustrative example:
\[
Y=g\left(X_{1},\boldsymbol{W},e\right),\text{ }X_{1}=U^{\ast}+U_{1},\text{ }\left(X_{2},...,X_{s}\right)\perp\left(U_{1},\boldsymbol{W},e\right),
\]
for some $s > 1$, where $g$ is an unknown smooth function. In this setup, $Y\perp\left(X_{2},...,X_{s}\right)|\left(X_{1},\boldsymbol{W}\right)$ holds,
and the conditional independence is unambiguous due to the presence
of $U_{1}$. However, it is possible that $\left(X_{2},...,X_{s}\right)\not\perp Y|\boldsymbol{W}$
holds through $U^{\ast}$.

For the refinement stage, we retain $\tilde{p}-q^{\mathsf{c}}-q^{\mathsf{d}}$
(recall $\textrm{dim}(\boldsymbol{W})=q^{\mathsf{c}}+q^{\mathsf{d}}$) covariates from $\boldsymbol{X}$ with the largest values of $\hat{\rho}_j$. Together with $\boldsymbol{W}$, this yields a total of $\tilde{p}$ covariates in the conditioning set for us to work with. For the post-selection estimation and the determination of covariates relevance, we do not distinguish between covariates from $\boldsymbol{X}$ and those in $\boldsymbol{W}$ for simplicity.\footnote{In applications, practitioners can exclude certain or all elements of $\boldsymbol{W}$ from this refinement step, as illustrated in the simulations in Section \ref{sec:simulation}.} Without loss of generality, we denote the surviving covariates after screening (including $\boldsymbol{W}$) as
\[
\boldsymbol{\tilde{X}}\equiv\left(X_{1}^{\mathsf{c}},...,X_{\tilde{s}_{1}}^{\mathsf{c}},X_{1}^{\mathsf{d}},...,X_{\tilde{s}_{2}}^{\mathsf{d}}\right)',
\]
where $\tilde{s}_1$ and $\tilde{s}_2$ denote the number of continuous and discrete covariates, respectively, in $\boldsymbol{\tilde{X}}$.

We then estimate the conditional density of $Y$ given $\boldsymbol{\tilde{X}}$ using the standard Nadaraya–Watson estimator:
\begin{equation}
\hat{f}\left(y|\boldsymbol{\tilde{x}}\right)=\frac{\sum_{i=1}^{n}\left[\Pi_{l=1}^{\tilde{s}_{1}}K_{h_{l}}\left(x_{li}^{\mathsf{c}}-x_{l}^{\mathsf{c}}\right)\right]\left[\Pi_{l=1}^{\tilde{s}_{2}}K_{\lambda_{l}}^{\mathsf{d}}\left(x_{li}^{\mathsf{d}},x_{l}^{\mathsf{d}}\right)\right]K_{h}\left(y_{i}-y\right)}{\sum_{i=1}^{n}\left[\Pi_{l=1}^{\tilde{s}_{1}}K_{h_{l}}\left(x_{li}^{\mathsf{c}}-x_{l}^{\mathsf{c}}\right)\right]\left[\Pi_{l=1}^{\tilde{s}_{2}}K_{\lambda_{l}}^{\mathsf{d}}\left(x_{li}^{\mathsf{d}},x_{l}^{\mathsf{d}}\right)\right]},\label{eq:fhat}
\end{equation}
where $r_l$ denotes the number of support points of the discrete covariate $X_l^{\mathsf{d}}$, $0\leq\lambda_l\leq\left.\left(r_{l}-1\right)\right/r_{l}$, the kernel functions
$K_{h}(\cdot)$, $K_{h_{l}}(\cdot)$, and $K_{\lambda_{l}}^{\mathsf{d}}(\cdot)$ are
defined as in (\ref{eq:CDFhat}).

The logic of using CV for variable selection is based on how kernel bandwidths relate to variable relevance. It is easy to see that when the bandwidth $h_l$ for a continuous covariate $X_{l}^{\mathsf{c}}$ is much larger than the range of its support, the kernel $K_{h_{l}}(\cdot)$ assigns nearly the same weight to all observations, meaning $X_{l}^{\mathsf{c}}$ becomes almost uninformative to the estimate $\hat{f}(y|\boldsymbol{\tilde{x}})$. Similarly, when the bandwidth $\lambda_l$ for a discrete covariate $X_{l}^{\mathsf{d}}$ reaches its upper bound of $(r_l-1)/r_l$, $K_{\lambda_{l}}^{\mathsf{d}}\left(x_{l}^{\mathsf{d}},x^{\mathsf{d}}\right)=1/r_{l}$ for all $\left(x_{l}^{\mathsf{d}},x^{\mathsf{d}}\right)$, effectively removing the variable's influence on the density estimate.

In view of this observation, \citet{HallRacineLi} find that searching for optimal bandwidths to minimize the cross-validated ISE (CV-ISE) will force the bandwidths for irrelevant covariates to these upper extremes. This procedure thus essentially eliminates (``smoothes out'') the influence of irrelevant variables, providing a data-driven method for variable selection. We briefly review the method of \citet{HallRacineLi} in Appendix \ref{APP:hall}.

Formally, the ISE is defined as
\begin{equation}
\text{ISE}\left(h,h_{1},...h_{\tilde{s}_{1}},\lambda_{1},...,\lambda_{\tilde{s}_{2}}\right)=\int\left[\hat{f}\left(y|\boldsymbol{\tilde{x}}\right)-f\left(y|\boldsymbol{\tilde{x}}\right)\right]^{2}f\left(\boldsymbol{\tilde{x}}\right)\,d\boldsymbol{\tilde{x}}\,dy,\label{eq:ISE_define}
\end{equation}
The optimal bandwidths are those that minimize the CV-ISE:
\begin{equation}
(\hat{h},\hat{h}_{1},...\hat{h}_{\tilde{s}_{1}},\hat{\lambda}_{1},...,\hat{\lambda}_{\tilde{s}_{2}})=\arg\min_{h,h_{1},...h_{\tilde{s}_{1}},\lambda_{1},...,\lambda_{\tilde{s}_{2}}}\text{CV}\left(h,h_{1},...h_{\tilde{s}_{1}},\lambda_{1},...,\lambda_{\tilde{s}_{2}}\right),\label{eq:ISE-1}
\end{equation}
and the post-selection conditional density is then obtained using $(\hat{h},\hat{h}_{1},...\hat{h}_{\tilde{s}_{1}},\hat{\lambda}_{1},...,\hat{\lambda}_{\tilde{s}_{2}})$:
\begin{equation}
\hat{f}\left(y|\boldsymbol{\tilde{x}}\right)=\frac{\sum_{i=1}^{n}\left[\Pi_{l=1}^{\tilde{s}_{1}}K_{\hat{h}_{l}}\left(x_{li}^{\mathsf{c}}-x_{l}^{\mathsf{c}}\right)\right]\left[\Pi_{l=1}^{\tilde{s}_{2}}K_{\hat{\lambda}_{l}}^{\mathsf{d}}\left(x_{li}^{\mathsf{d}},x_{l}^{\mathsf{d}}\right)\right]K_{\hat{h}}\left(y_{i}-y\right)}{\sum_{i=1}^{n}\left[\Pi_{l=1}^{\tilde{s}_{1}}K_{\hat{h}_{l}}\left(x_{li}^{\mathsf{c}}-x_{l}^{\mathsf{c}}\right)\right]\left[\Pi_{l=1}^{\tilde{s}_{2}}K_{\hat{\lambda}_{l}}^{\mathsf{d}}\left(x_{li}^{\mathsf{d}},x_{l}^{\mathsf{d}}\right)\right]}.\label{eq:fhat_post}
\end{equation}
For notational simplicity, we abuse the notation to still use $\hat{f}$ to denote the post-selection estimator.

To ensure the validity of $\hat{f}(y|\boldsymbol{\tilde{x}})$ in (\ref{eq:fhat_post}), we impose the following additional assumptions:
\begin{assumption}\label{A:2} 
\begin{enumerate}
\item[(1)] The conditional independence 
\[
Y\perp\left(X_{s^{**}+1},...,X_{s^{*}}\right)\left|\boldsymbol{Z}^{*}\right.
\]
holds and is unambiguous. That is, there do not exist any alternative partition $(\boldsymbol{X}_{1},\boldsymbol{X}_{2})$ of $\left(X_{1},...,X_{s^{*}},\boldsymbol{W}\right)$ with $\boldsymbol{X}_{1}\neq\left(X_{s^{**}+1},...,X_{s^{*}}\right)'$
and $\boldsymbol{X}_{2}\neq\boldsymbol{Z}^{*}$ such that $Y\perp\boldsymbol{X}_{1}\left|\boldsymbol{X}_{2}\right..$
In addition, we assume that $\left(Y,X_{1},...,X_{s^{*}}\right)\perp\left(X_{s^{*}+1},...,X_{p}\right)|\boldsymbol{W}$.

\item[(2)] $f\left(\boldsymbol{Z}^{*}\right)$ is uniformly bounded and bounded
away from zero. 
\item[(3)] $\tilde{p}-q^{\mathsf{c}}-q^{\mathsf{d}}\geq s^{*}$\textcolor{black}{.} 
\end{enumerate}
\end{assumption}

Assumption \ref{A:2}(1) defines the unambiguity of the conditional
independence. $\left(Y,X_{1},...,X_{s^{*}}\right)\perp\left(X_{s^{*}+1},...,X_{p}\right)|\boldsymbol{W}$
is imposed to ensure that the ambiguity does not occur when irrelevant
covariates are included. Assumption \ref{A:2}(2) is a standard regularity
condition required as in \citet{HallRacineLi}. Assumption \ref{A:2}(3) implicitly requires that $s^{*}$
is fixed, so that nonparametric estimation remains feasible. This assumption also guarantees that the initial screening does not mistakenly discard any relevant covariates due to choosing a too small $\tilde{p}$.  

The results in \citet{HallRacineLi} rely on the pure (unconditional) independence
to avoid ambiguity. In the following proposition, we (slightly) generalize
the results of \citet{HallRacineLi} to the case in (\ref{eq:conditional_refine})
when the conditional independence is not ambiguous.

\begin{proposition} \label{Prop:post}
Suppose Assumptions \ref{A:signal_strength}
and \ref{A:2} hold. Let $p^{\mathsf{c}*}$ denote the number of continuous random variables in
$\boldsymbol{Z}^{*}$. Then, 
\begin{enumerate}
\item[(1)] The post-selection estimator $\hat{f}\left(y|\boldsymbol{\tilde{x}}\right)$ in (\ref{eq:fhat_post})
satisfies 
\[
\hat{f}\left(y|\boldsymbol{\tilde{x}}\right)-f\left(y|\boldsymbol{z}^{*}\right)=O_{P}\left(n^{-\frac{r}{p^{\mathsf{c}\ast}+1+2r}}\right).
\]
\item[(2)] The bandwidths from (\ref{eq:ISE-1}) satisfy the
following: for $Y$ and any $X_{l}^{\mathsf{c}}$ and $X_{l}^{\mathsf{d}}$ in $\boldsymbol{Z}^{*}$
(i.e., directly relevant covariates),
\[
n^{\frac{1}{p^{\mathsf{c}\ast}+1+2r}}\hat{h}\overset{P}{\rightarrow}c_{0}, \textrm{ } n^{\frac{1}{p^{\mathsf{c}\ast}+1+2r}}\hat{h}_{l}\overset{P}{\rightarrow}c_{l},\text{ and }n^{\frac{r}{p^{\mathsf{c}\ast}+1+2r}}\hat{\lambda}_{l}\overset{P}{\rightarrow}c_{l}'.
\]
For any $X_{l}^{\mathsf{c}}$ and $X_{l}^{\mathsf{d}}$ in $\boldsymbol{\tilde{X}}\backslash\boldsymbol{Z}^{*}$
(i.e., irrelevant or indirectly relevant covariates), 
\[
\Pr\left(\hat{h}_{l}>C\right)\rightarrow1,\text{ for any positive }C,\text{ and }\hat{\lambda}_{l}\overset{P}{\rightarrow}\frac{r_{l}-1}{r_{l}}.
\]
\end{enumerate}
\end{proposition}

Proposition \ref{Prop:post} shows that the presence of irrelevant or indirectly relevant variables has no impact
on the convergence rate of $\hat{f}\left(y|\boldsymbol{\tilde{x}}\right)$. The post-selection estimator attains the fastest convergence rate achievable as if the estimation were performed using only the conditioning set $\boldsymbol{Z}^{*}$.

\subsubsection*{A Computationally Efficient Modified CV Procedure}
In practice, this procedure tends to be quite time-consuming, and the process can be numerically unstable due to the highly nonlinear nature of the ISE criterion function. Finding the optimal set of bandwidths that minimizes the ISE can be a challenging optimization problem. 

We find that an exhaustive search over the full bandwidth space is unnecessary. Instead, we can achieve the same result by evaluating the ISE only at the theoretically optimal bandwidths corresponding to each possible subset (or ``selection'') of the $\tilde{p}$ candidate covariates and then taking the one that minimizes the ISE. Although conceptually simple, we need to introduce additional notation to formalize this procedure. 

For the $\tilde{p}$ covariates in $\boldsymbol{\tilde{X}}$, there are $2^{\tilde{p}}$ possible selections. Each selection can be represented by a $\tilde{p}\times1$ binary selection vector $\boldsymbol{I}$, with 1 indicating inclusion and 0 indicating exclusion of the corresponding covariate. For example,  
\[
\boldsymbol{I}=(\underset{\tilde{p}}{\underbrace{0,0,\dots,0}})'
\]
indicates that none of the covariates are selected, while 
\[
\boldsymbol{I}^{*} \equiv \left( \underbrace{1, \dots, 1}_{p^{\mathsf{c}*}}, \underbrace{0, \dots, 0}_{\tilde{s}_1 - p^{\mathsf{c}*}}, \underbrace{1, \dots, 1}_{p^{\mathsf{d}*}}, \underbrace{0, \dots, 0}_{\tilde{s}_2 - p^{\mathsf{d}*}} \right)'
\]
corresponds to the correct selection of all directly relevant covariates $\boldsymbol{Z}^{*}$, where $p^{\mathsf{c}*}$ and $p^{\mathsf{d}*}$ denote the numbers of continuous and discrete variables in $\boldsymbol{Z}^{*}$, respectively. We collect all such candidate selection vectors into the set 
\begin{equation}
\mathcal{S}=\left\{ (0,0,\dots,0)',(1,0,\dots,0)',\dots,(1,1,\dots,1)'\right\} .\label{eq:selectionset}
\end{equation}

We define $\boldsymbol{h}_{\boldsymbol{I}}^{\textrm{opt}}\in\mathbb{R}^{\tilde{p}+1}$ as the vector of theoretically optimal bandwidths corresponding to the selection $\boldsymbol{I} \in \mathcal{S}$. For example, when no covariates are selected, i.e., $\boldsymbol{I}=(0,0,\dots,0)'$,
we have
\[
\boldsymbol{h}_{\boldsymbol{I}}^{\textrm{opt}}=\left(c_{0}n^{\frac{-1}{1+2r}},\underset{\tilde{s}_{1}}{\underbrace{\infty,\dots,\infty}},\underset{\tilde{s}_{2}}{\underbrace{\frac{r_{1}-1}{r_{1}},\frac{r_{2}-1}{r_{2}},\dots,\frac{r_{\tilde{s}_{2}}-1}{r_{\tilde{s}_{2}}}}}\right)',
\]
where $c_{0}n^{\frac{-1}{1+2r}}$ is the optimal bandwidth for $Y$ with no conditioning covariates, and the bandwidths for all covariates reach their upper limit. In practice, ``$\infty$'' for a continuous covariate means that it is assigned uniform weight across all observations, which is analogous to using $(r_{l}-1)/r_{l}$ for a discrete covariate $X_{l}^{\mathsf{d}}$. When only $\boldsymbol{Z}^{*}$ is selected, corresponding to $\boldsymbol{I}^{*}$, we obtain 
\[
\boldsymbol{h}_{\boldsymbol{I}^{*}}^{\textrm{opt}}=\left(h_{0}^{*},h_{1}^{*},\dots,h_{p^{\mathsf{c}*}}^{*},\infty,\dots,\infty,\lambda_{1}^{*},\dots,\lambda_{p^{\mathsf{d}*}}^{*},\frac{r_{p^{\mathsf{d}*}+1}-1}{r_{p^{\mathsf{d}*}+1}},\dots,\frac{r_{\tilde{s}_{2}}-1}{r_{\tilde{s}_{2}}}\right)',
\]
where $h_{0}^{*}$, $h_{l}^{*}$, and $\lambda_{l}^{*}$ are the optimal bandwidths
as given in Proposition \ref{Prop:post}.


The following corollary provides the key theoretical justification for our modified CV procedure. Since the corollary follows directly from Proposition \ref{Prop:post},  the proof is omitted. It states that the ISE evaluated at the optimal bandwidths, $\boldsymbol{h}_{\boldsymbol{I}^{*}}^{\textrm{opt}}$, is, with high probability, the minimum among the ISEs evaluated for all other $\boldsymbol{h}_{\boldsymbol{I}}^{\textrm{opt}}$. This result is crucial as it reduces the complex continuous optimization problem to a much faster and more stable discrete search over the $2^{\tilde{p}}$ selection vectors in $\mathcal{S}$.

\begin{corollary}\label{Corollary:post-1}Suppose Assumptions \ref{A:signal_strength}
and \ref{A:2} hold. Then, 
\[
\Pr\left(\text{ISE}\left(\boldsymbol{h}_{\boldsymbol{I}^{*}}^{\textrm{opt}}\right)<\min_{\boldsymbol{I}\in\mathscr{\mathcal{S}}\left\backslash \boldsymbol{I}^{*}\right.}\text{ISE}\left(\boldsymbol{h}_{\boldsymbol{I}}^{\textrm{opt}}\right)\right)\rightarrow1.
\]
\end{corollary}

In practice, we apply the proposed discrete search algorithm to the CV-ISE to obtain an estimate of the optimal selection vector, $\hat{\boldsymbol{I}}$, and its corresponding bandwidths, $\boldsymbol{h}_{\hat{\boldsymbol{I}}}^{\textrm{opt}}$.

\subsection{Propensity Score Estimation \protect} \label{sec:Propensity-Score}
The propensity score plays an important role in many statistical and econometric methods. This section details how our procedure can be applied to its estimation. We denote the binary response (treatment) variable as $D \in \{0, 1\}$. To identify covariates relevant to the propensity score, we adapt the dependence measure from Section \ref{SEC:newMeasure}. Specifically, we modify our KS-based distance measure to compare the distribution of a candidate covariate $X$ for the treated group ($D=1$) with that of the control group ($D=0$), conditional on $\boldsymbol{W}$:
\[ 
\Lambda_{X}\left(\boldsymbol{W}=\boldsymbol{w}\right)  =\sup_{x}\left|F\left(x|D=1,\boldsymbol{w}\right)-F\left(x|D=0,\boldsymbol{w}\right)\right| \equiv\sup_{x}\left|F_{1}\left(x|\boldsymbol{w}\right)-F_{0}\left(x|\boldsymbol{w}\right)\right|.
\]
The overall dependence measure and its sample counterpart are then defined analogously to the general case:
\[
\rho=\mathbb{E}\left[\Lambda_{X}\left(\boldsymbol{W}\right)\right] \quad \text{and} \quad \hat{\rho}=\frac{1}{n}\sum_{i=1}^{n}\hat{\Lambda}_{X}\left(\boldsymbol{W}=\boldsymbol{w}_{i}\right),
\]
where
\[
\hat{\Lambda}_{X}\left(\boldsymbol{W}=\boldsymbol{w}\right)=\max_{i=1,...,n}\left|\hat{F}_{1,n}\left(x_{i}|\boldsymbol{w}\right)-\hat{F}_{0,n}\left(x_{i}|\boldsymbol{w}\right)\right|
\]
with 
\[
\hat{F}_{d,n}\left(x|\boldsymbol{w}\right)=\frac{\sum_{i=1}^{n}\mathbf{1}\left(x_{i}\leq x\right)\mathbf{1}\left(D_{i}=d\right)\left[\Pi_{l=1}^{q^{\mathsf{c}}}K_{h}\left(w_{li}^{\mathsf{c}}-w_{l}^{\mathsf{c}}\right)\right]\left[\Pi_{l=1}^{q^{\mathsf{d}}}K_{\lambda}^{\mathsf{d}}\left(w_{li}^{\mathsf{d}},w_{l}^{\mathsf{d}}\right)\right]}{\sum_{i=1}^{n}\mathbf{1}\left(D_{i}=d\right)\left[\Pi_{l=1}^{q^{\mathsf{c}}}K_{h}\left(w_{li}^{\mathsf{c}}-w_{l}^{\mathsf{c}}\right)\right]\left[\Pi_{l=1}^{q^{\mathsf{d}}}K_{\lambda}^{\mathsf{d}}\left(w_{li}^{\mathsf{d}},w_{l}^{\mathsf{d}}\right)\right]},\textrm{ }d=0,1.
\]

One notable difference here from the general case is that because $D$ is binary, no kernel smoothing is required for it.

\begin{corollary} \label{COR:propensity_score} 
Suppose Assumption \ref{A:signal_strength} holds for the relationship between $\boldsymbol{X}$ and $D$ with the minimum signal strength
\[
\min_{j=1,...,s^{*}}\rho_{j}=\mathbb{E}\left[\Lambda_{X_{j}}\left(\boldsymbol{W}\right)\right]\gtrsim\left(\log n\right)^{1+c}n^{-r/\left(2r+q^{\mathsf{c}}\right)},
\]
for a positive small $c$. Re-define $\hat{\mathcal{M}}=\left\{ j:\hat{\rho}_{j}\geq Cn^{-r/\left(2r+q^{\mathsf{c}}\right)}\log n\right\} $
and set $h\propto n^{-1/\left(2r+q^{\mathsf{c}}\right)}$. Then, 
\[
\Pr\left(\hat{\mathcal{M}}=\mathcal{M}^{*}\right)\rightarrow1.
\]
\end{corollary}

After screening, the post-selection estimate $\hat{f} \left( D = d \mid \tilde{\boldsymbol{x}} \right)$ can be obtained via the refinement procedure outlined in Section \ref{SEC:post-estimation}. The final estimator is given by
\begin{equation}
\hat{f}\left(D=d|\boldsymbol{\tilde{x}}\right)=\frac{\sum_{i=1}^{n}\left[\Pi_{l=1}^{\tilde{s}_{1}}K_{\hat{h}_{l}}\left(x_{li}^{\mathsf{c}}-x_{l}^{\mathsf{c}}\right)\right]\left[\Pi_{l=1}^{\tilde{s}_{2}}K_{\hat{\lambda}_{l}}^{\mathsf{d}}\left(x_{li}^{\mathsf{d}},x_{l}^{\mathsf{d}}\right)\right]\mathbf{1}\left(D_{i}=d\right)}{\sum_{i=1}^{n}\left[\Pi_{l=1}^{\tilde{s}_{1}}K_{\hat{h}_{l}}\left(x_{li}^{\mathsf{c}}-x_{l}^{\mathsf{c}}\right)\right]\left[\Pi_{l=1}^{\tilde{s}_{2}}K_{\hat{\lambda}_{l}}^{\mathsf{d}}\left(x_{li}^{\mathsf{d}},x_{l}^{\mathsf{d}}\right)\right]},\label{eq:propensity_post}
\end{equation}
where $(\hat{h}_{1},...,\hat{h}_{\tilde{s}_{1}},\hat{\lambda}_{1},...,\hat{\lambda}_{\tilde{s}_{2}})$ are the optimal bandwidths that minimize the ISE in the CV-refinement step. 

\begin{corollary} \label{COR:propensity_score-screening} 
Suppose Assumption \ref{A:2} and the conditions in Corollary \ref{COR:propensity_score}
hold. Then, the post-selection estimator $\hat{f}\left(D=d|\tilde{\boldsymbol{x}}\right)$ in  (\ref{eq:propensity_post})
satisfies 
\[
\hat{f}\left(D=d|\tilde{\boldsymbol{x}}\right)-f\left(D=d|\boldsymbol{z}^{*}\right)=O_{P}\left(n^{-\frac{r}{p^{\mathsf{c}\ast}+2r}}\right),\text{ for }d=0,1.
\]
\textcolor{black}{For any }$X_{l}^{\mathsf{c}}$\textcolor{black}{{}
}and\textcolor{black}{{} }$X_{l}^{\mathsf{d}}$ in\textcolor{black}{{}
}$\boldsymbol{Z}^{*},$\textcolor{black}{{} 
\[
n^{\frac{1}{p^{\mathsf{c}\ast}+2r}}\hat{h}\overset{P}{\rightarrow}c_{0},n^{\frac{1}{p^{\mathsf{c}\ast}+2r}}\hat{h}_{l}\overset{P}{\rightarrow}c_{l},\text{ and }n^{\frac{r}{p^{\mathsf{c}\ast}+2r}}\hat{\lambda}_{l}\overset{P}{\rightarrow}c_{l}'.
\]
For any }$X_{l}^{\mathsf{c}}$\textcolor{black}{{} }and\textcolor{black}{{}
}$X_{l}^{\mathsf{d}}$ in\textcolor{black}{{} }$\boldsymbol{\tilde{X}}\backslash\boldsymbol{Z}^{*}$,
\[
\Pr\left(\hat{h}_{l}>C\right)\rightarrow1,\text{ for any positive }C,\text{ and }\hat{\lambda}_{l}\overset{P}{\rightarrow}\frac{r_{l}-1}{r_{l}}.
\]
\end{corollary}

The proofs of Corollaries \ref{COR:propensity_score} and \ref{COR:propensity_score-screening} are straightforward applications of Theorem \ref{TH:TPR} and Proposition \ref{Prop:post} and are thus omitted.

\subsection{Summary of the Proposed Procedure \protect\protect} \label{sec:summary_procedure}
For ease of reference, we summarize the detailed steps of our procedure proposed in Sections \ref{SEC:newMeasure}--\ref{sec:Propensity-Score} below. Recall that the dimension of the pre-selected covariates $\boldsymbol{W}$ is $q^{\mathsf{c}}+q^{\mathsf{d}}$. We assume that $q^{\mathsf{c}}+q^{\mathsf{d}}\leq3$ without loss of generality.
\begin{enumerate}
    \item \textbf{Initialize}: Start by setting an initial number of covariates to screen for, $\tilde{p}$ (e.g., $\tilde{p}=5$).
    \item \textbf{Screen}: Calculate the dependence measure $\hat{\rho}_{j}$ for each candidate covariate $X_{j}$, for $j=1,...,p$, as defined in (\ref{eq:rho_continuousX}). Keep the $\tilde{p}-q^{\mathsf{c}}-q^{\mathsf{d}}$ covariates in $\boldsymbol{X}$ with the largest $\hat{\rho}_{j}$ values.
    \item \textbf{Refine}: For the set of $\tilde{p}$ covariates (the pre-selected $\boldsymbol{W}$ and the $X_j$'s that survive Step 2), evaluate the CV-ISE for each possible subset of variables $\boldsymbol{I}\in\mathcal{S}$ defined in (\ref{eq:selectionset}), using the corresponding optimal bandwidths $\boldsymbol{h}_{\boldsymbol{I}}^{\textrm{opt}}$. Select the subset $\hat{\boldsymbol{I}}$ that minimizes the CV-ISE and take $\boldsymbol{h}_{\hat{\boldsymbol{I}}}^{\textrm{opt}}$ as the working bandwidths.
    \item \textbf{Check and Iterate}: Examine the optimal bandwidths $\boldsymbol{h}_{\hat{\boldsymbol{I}}}^{\textrm{opt}}$ from Step 3. If none of the bandwidths corresponding to the screened covariates diverge to their upper extremes (i.e., if no variables can be “smoothed out”), increase $\tilde{p}$ by 1 and repeat Steps 2 through 4. Otherwise, the procedure terminates. The final conditional density estimate $\hat{f}\left(y|\boldsymbol{\tilde{x}}\right)$ is constructed using the covariates selected by $\hat{\boldsymbol{I}}$ and the bandwidths $\boldsymbol{h}_{\hat{\boldsymbol{I}}}^{\textrm{opt}}$.
\end{enumerate}

It is important to note the limitation of this approach: its performance relies on the sparsity assumption, i.e., the number of truly relevant covariates is fixed. If the procedure continues to iterate until the final selected $\tilde{p}$ is too large for reliable nonparametric estimation (e.g., $\tilde{p}=10$ for $n=1000$), our method may not be suitable.

\section{Estimating Average Treatment Effects}\label{sec:theoryapply}
In this section, we apply the methods proposed in Section \ref{SEC:f(y,x)reduce}
to estimate and conduct inference on ATE
in high-dimensional settings. We use this context to demonstrate how
our screening procedure and dimension-reduced density estimator, combined
with ML methods, enhance the applicability and performance of established
approaches. The identification results presented here may also be
of independent interest. Given the extensive literature on ATE, we
provide only a brief overview of the framework in Section \ref{SEC:ATE_intro}. For a comprehensive review of related methods, readers may refer to \citet{ImbensRubin} and \citet{ding2024first}. Additionally, \citet{chernozhukov2024applied} discuss recent advances in applying ML and AI techniques to causal inference. It is worth emphasizing that the applications of our proposed methods are not limited to ATE estimation; they are broadly applicable and can complement many of the procedures discussed in the aforementioned works and the references therein.


\subsection{Reviewing Average Treatment Effects and Doubly Robust Estimator} \label{SEC:ATE_intro}
In this section, we review the \emph{potential outcomes} framework
and the doubly robust estimator for ATE, providing the necessary background
for the subsequent discussion.

Suppose we observe a random sample of $n$ individuals indexed by
$i=1,...,n$. We consider the classic potential outcomes framework
for observational studies under the \textit{stable unit treatment
value assumption} (\citet{rubin1980randomization}). Let $Y$ denote
an outcome variable and $D$ a binary treatment indicator---$D=0$
for the control group and $D=1$ for the treatment group. Define $Y(0)$
and $Y(1)$ as potential outcomes corresponding to $D=0$ and $D=1$,
respectively. The observed outcome $Y$ is determined by $Y=Y(0)\left(1-D\right)+Y(1)D$.
In addition to $(Y,D)$, we observe a vector of pre-treatment covariates
(control variables) $\boldsymbol{Z}\equiv\left(\boldsymbol{X}',\boldsymbol{W}'\right)'$,
where we use this notation to align with the setting in Section \ref{SEC:f(y,x)reduce}.
We denote the conditional means of the potential outcomes as 
\begin{equation}
g_{0}\left(\boldsymbol{Z}\right)=\mathbb{E}\left[\left.Y(0)\right|\boldsymbol{Z}\right]\text{ and }g_{1}\left(\boldsymbol{Z}\right)=\mathbb{E}\left[\left.Y(1)\right|\boldsymbol{Z}\right].\label{eq:outcome_reg}
\end{equation}
The population ATE to estimate is then defined as 
\[
\psi=\mathbb{E}[g_{1}\left(\boldsymbol{Z}\right)-g_{0}\left(\boldsymbol{Z}\right)].
\]

Let $m\left(\boldsymbol{Z}\right)\equiv\mathbb{E}\left[D|\boldsymbol{Z}\right]=\Pr(D=1|\boldsymbol{Z})$,
commonly referred to as the \emph{propensity score}. The following
assumptions are standard in the literature. Assumption \ref{A:3}(1) is
known as the \emph{ignorability} condition (\citet{rubin1978bayesian} and \citet{rosenbaum1983central}), while Assumption \ref{A:3}(2) is referred to as the \emph{overlap} condition, which ensures the estimability of the ATE (\cite{KhanTamer}).\footnote{The ignorability assumption is also known as the \emph{unconfoundedness}
or \emph{selection on observables} assumption.} \begin{assumption}\label{A:3} 
\begin{enumerate}
\item[(1)] $\left.\left(Y(0),Y(1)\right)\perp D\right|\boldsymbol{Z}$. 
\item[(2)] There exists a small constant $\epsilon\in(0,1/2)$ such that $\epsilon<m\left(\boldsymbol{z}\right)<1-\epsilon$
for all $\boldsymbol{z}\in\textrm{supp}(\boldsymbol{Z})$.\footnote{Assumption \ref{A:3}(2) is stronger than the usual overlap condition
$0<m(\boldsymbol{z})<1$ for all $\boldsymbol{z}\in\textrm{supp}(\boldsymbol{Z})$,
and is therefore referred to as the \emph{strict overlap} condition.
For a detailed discussion of its strong implications for data-generating
processes, see \citet{d2021overlap}.} 
\item[(3)] The observations $\left\{ Y_{i},D_{i},\boldsymbol{Z}_{i}\right\} _{i=1}^{n}$
are i.i.d. across $i$. Additionally, there exists $\delta>0$ such
that $\mathbb{E}(\left|\psi_{i}\right|^{2+\delta})<\infty$, where
$\psi_{i}$ is defined in (\ref{eq:psi}). 
\end{enumerate}
\end{assumption}

Under Assumption \ref{A:3}, we have $g_{0}(\boldsymbol{Z})=\mathbb{E}[Y|D=0,\boldsymbol{Z}]$,
$g_{1}(\boldsymbol{Z})=\mathbb{E}[Y|D=1,\boldsymbol{Z}]$, and the following identification formula for $\psi$: 
\begin{equation}
\psi=\mathbb{E}[\psi_{i}]\textrm{ with }\psi_{i}\equiv[g_{1}(\boldsymbol{Z}_{i})-g_{0}(\boldsymbol{Z}_{i})]+\left[\frac{D_{i}(Y_{i}-g_{1}(\boldsymbol{Z}_{i}))}{m(\boldsymbol{Z}_{i})}-\frac{(1-D_{i})(Y_{i}-g_{0}(\boldsymbol{Z}_{i}))}{1-m(\boldsymbol{Z}_{i})}\right].\label{eq:psi}
\end{equation}
Expression (\ref{eq:psi}) motivates the following moment estimator
for $\psi$: 
\begin{equation}
\hat{\psi}=\frac{1}{n}\sum_{i=1}^{n}\hat{\psi}_{i}\textrm{ with }\hat{\psi}_{i}\equiv\left[\hat{g}_{1}\left(\boldsymbol{z}_{i}\right)-\hat{g}_{0}\left(\boldsymbol{z}_{i}\right)\right]+\left[\frac{D_{i}\left(y_{i}-\hat{g}_{1}\left(\boldsymbol{z}_{i}\right)\right)}{\hat{m}\left(\boldsymbol{z}_{i}\right)}-\frac{\left(1-D_{i}\right)\left(y_{i}-\hat{g}_{0}\left(\boldsymbol{z}_{i}\right)\right)}{1-\hat{m}\left(\boldsymbol{z}_{i}\right)}\right],\label{eq:psi_hat}
\end{equation}
where $(\hat{g}_{0},\hat{g}_{1})$ and $\hat{m}$ are generic estimators for
$(g_{0},g_{1})$ and $m$, respectively. The estimator $\hat{\psi}$
is \emph{doubly robust} (\citet{scharfstein1999adjusting} and
\citet{bang2005doubly}) because it is consistent if either $(\hat{g}_{0},\hat{g}_{1})$
are consistent estimators for $(g_{0},g_{1})$ or $\hat{m}$ is a
consistent estimator for $m$. Similar doubly robust estimators can be constructed
for other treatment effects, such as the average treatment effect
on the treated (ATT) and on the control (ATC).

\subsection{Doubly Robust and De-Biased ATE Estimators with Dimension-Reduced Propensity Score} \label{SEC:debias}
Assumption \ref{A:3}(1) (ignorability) plays a central role in observational
studies. It assumes the absence of unobserved \emph{confounders}---covariates that affect both the treatment and the outcome. However, this assumption is generally \emph{not} testable and is often justified based on the researcher's domain knowledge. To minimize the risk of violating this assumption, practitioners may control for as many pre-treatment
covariates $\boldsymbol{Z}$ as possible (\citet{rubin2007design}
and \citet{vander2011new}). Nevertheless, there is an intrinsic trade-off
between the ignorability assumption and Assumption \ref{A:3}(2) (overlap)
(\citet{d2021overlap}). Controlling for a rich set of covariates
$\boldsymbol{Z}$ makes the former more plausible but simultaneously
reduces the randomness in the propensity score model, making it harder
to satisfy the latter as $D$ becomes more predictable. In applications,
even if the overlap assumption holds, the estimated propensity scores
$\hat{m}(\boldsymbol{z})$ can still be close to 0 or 1, particularly
when overfitting occurs due to the inclusion of many covariates irrelevant
to $D$. This can lead to poor finite-sample performance of estimators involving inverse propensity weighting, such as (\ref{eq:psi_hat}).
Therefore, it is practically important to refine the conditioning
set $\boldsymbol{Z}$ to ensure the ignorability condition while mitigating
its tension with the overlap condition.\footnote{Adjusting for many covariates also increases the risk of \emph{over-adjustment}
problems (\citet{ding2015adjust}). Since this paper focuses on estimation
and inference, we do not discuss their implications for the identification
of the ATE.}

Furthermore, the doubly robust estimator (\ref{eq:psi_hat}) can become inconsistent
(or even ``doubly fragile'') when both the outcome regression model
$(g_{0},g_{1})$ and the propensity score model $m$ are misspecified
(\citet{KangSchafer2007}). To address this issue, recent advancements
in DML methods, such as \citet{farrell2015robust},
\citet{ChernozhukovEtal2018}, and \citet{FarrellEtal2021}, propose
employing (nonparametric) ML tools to estimate $(g_{0},g_{1})$
and $m$. These approaches enhance the robustness of the estimator
(\ref{eq:psi_hat}) by avoiding strong parametric restrictions, such
as logistic or linear regressions. However, the drawback is that the
estimator may suffer from the curse of dimensionality when $\textrm{dim}(\boldsymbol{Z})$ is large or even diverges as the sample size increases. This challenge highlights another practical importance of refining the conditioning set $\boldsymbol{Z}$--dimension reduction for enabling robust DML estimation.

In what follows, we demonstrate how our proposed methods facilitate
such refinement under certain sparsity conditions. The proof of the
following result is provided in Appendix \ref{APP:mainproof}.

\begin{proposition} \label{Prop:refine} Suppose Assumption \ref{A:3}(1)
holds. Let $(\boldsymbol{Z}^{*},\boldsymbol{\underline{Z}})$ be a
partition of $\boldsymbol{Z}$ such that $D\perp\boldsymbol{\underline{Z}}|\boldsymbol{Z}^{*}$. Assume there exists a small constant $\epsilon\in(0,1/2)$ such that
$\epsilon<m(\boldsymbol{z}^{*})<1-\epsilon$ for all $\boldsymbol{z}^{*}\in\textrm{supp}(\boldsymbol{Z}^{*})$.
Then: 
\begin{enumerate}
\item[(1)] $(Y(0),Y(1))\perp D|\boldsymbol{Z}^{*}$ and 
\begin{equation}
\psi=\mathbb{E}\left[g_{1}(\boldsymbol{Z}^{*})-g_{0}(\boldsymbol{Z}^{*})\right]=\mathbb{E}\left[\frac{DY}{m(\boldsymbol{Z}^{*})}-\frac{(1-D)Y}{1-m(\boldsymbol{Z}^{*})}\right]. \label{eq:refined_ipw}
\end{equation} 
\item[(2)] The ATE $\psi$ has the following doubly robust representation: 
\begin{equation}
\psi=\mathbb{E}\left[g_{1}(\boldsymbol{Z}^{*})-g_{0}(\boldsymbol{Z}^{*})\right]+\mathbb{E}\left[\frac{D(Y-g_{1}(\boldsymbol{Z}^{*}))}{m(\boldsymbol{Z}^{*})}-\frac{(1-D)(Y-g_{0}(\boldsymbol{Z}^{*}))}{1-m(\boldsymbol{Z}^{*})}\right].\label{eq:refined_estimand}
\end{equation}
\end{enumerate}
\end{proposition}

Proposition \ref{Prop:refine} ensures that under the additional sparsity
restriction $D\perp\boldsymbol{\underline{Z}}|\boldsymbol{Z}^{*}$, the ignorability condition remains valid when conditioning only on
$\boldsymbol{Z}^{*}$. Consequently, the ATE $\psi$ can be consistently estimated using
(\ref{eq:refined_ipw}) or (\ref{eq:refined_estimand}) with the lower-dimensional subvector $\boldsymbol{Z}^{*}$
instead of the full covariate set $\boldsymbol{Z}$, achieving a ``dual dimension reduction'' for both propensity score and outcome regression models. Moreover, the overlap condition required for these estimators is weaker than Assumption
\ref{A:3}(2), expanding their applicability when the overlap condition
on $m(\boldsymbol{Z})$ fails but holds for $m(\boldsymbol{Z}^{*})$. 

This result is particularly relevant when the sparsity condition in
Assumptions \ref{A:signal_strength} and \ref{A:2} holds. Specifically,
we can apply our approach proposed in Section \ref{SEC:f(y,x)reduce}
to identify $\boldsymbol{Z}^{*}$ with high probability, as ensured
by Theorem \ref{TH:TPR} (Corollary \ref{COR:propensity_score}).
Since $\text{dim}(\boldsymbol{Z}^{*})$ is fixed
and much smaller than $\text{dim}(\boldsymbol{Z})$,
the overlap condition on $\boldsymbol{Z}^{*}$ is both weaker and
more plausible than that on $\boldsymbol{Z}$. 

Furthermore, there exists a complementary result, symmetric to Proposition \ref{Prop:refine}, that relies on a similar sparsity condition on $Y \mid D$. This result is presented in Appendix \ref{App:refine_Y}. Taken together, these results illustrate that our proposed approach can exploit a “double sparsity” structure: when either $D$ or $Y|D$ exhibits a sparsity pattern similar to that in Proposition \ref{Prop:refine}, our method can recover the corresponding relevant covariates and achieve effective dimension reduction. We provide a more detailed discussion of this point in Appendix \ref{App:refine_Y}.

\begin{remark} \label{remark:general} \emph{The doubly robust estimand (\ref{eq:refined_estimand})
can be generalized as follows: 
\[
\psi=\mathbb{E}\left[g_{1}(\boldsymbol{Z}^{*},s(\boldsymbol{\underline{Z}}))-g_{0}(\boldsymbol{Z}^{*},s(\boldsymbol{\underline{Z}}))\right]+\mathbb{E}\left[\frac{D(Y-g_{1}(\boldsymbol{Z}^{*},s(\boldsymbol{\underline{Z}})))}{m(\boldsymbol{Z}^{*})}-\frac{(1-D)(Y-g_{0}(\boldsymbol{Z}^{*},s(\boldsymbol{\underline{Z}})))}{1-m(\boldsymbol{Z}^{*})}\right],
\]
where $s(\boldsymbol{\underline{Z}})$ is a generic measurable function
of $\boldsymbol{\underline{Z}}$. Since $D\perp\boldsymbol{\underline{Z}}\mid\boldsymbol{Z}^{*}$,
the propensity score should be estimated using only $\boldsymbol{Z}^{*}$
for efficiency. However, $\boldsymbol{Z}^{*}$ may not be the optimal
set of covariates for the outcome regressions $(g_{0},g_{1})$. To
achieve potential efficiency gains, we allow $s(\boldsymbol{\underline{Z}})$
to serve as additional covariate adjustments in modeling $(g_{0},g_{1})$.
In practice, $s(\boldsymbol{\underline{Z}})$ can take various forms:
an empty set, a subvector of $\boldsymbol{\underline{Z}}$, or a synthetic
statistic derived from $\boldsymbol{\underline{Z}}$, such as principal
components. Also, there are many data-driven procedures designed for
regression problems that can assist in selecting $s(\boldsymbol{\underline{Z}})$,
including the CV method proposed by \citet{HallEtal2007}.} \end{remark}

To illustrate the usefulness of Proposition \ref{Prop:refine} in
robust causal inference, we replicate the well-established results
of DML developed in \citet{farrell2015robust}, \citet{ChernozhukovEtal2018},
and \citet{FarrellEtal2021}. We can show that the ATE estimator
(\ref{eq:psi_hat}), when using identified relevant covariates, has the following asymptotic properties:

\begin{theorem} \label{TH:debiased} Suppose conditions in Corollary
\ref{COR:propensity_score-screening} and Assumption \ref{A:3} hold.
Denote $\boldsymbol{Z}^{\hat{*}}$ as the set of covariates identified as relevant for $D$ using the procedure described in Section \ref{sec:summary_procedure}. Further, for $d=0,1$, assume the following conditions hold: 
\begin{enumerate}
    \item[(1)] $n^{-1}\sum_{i=1}^{n}\left[\hat{m}\left(\boldsymbol{z}_{i}^{\hat{*}}\right)-m\left(\boldsymbol{z}_{i}^{*}\right)\right]^{2}=o_{P}\left(1\right)$
and $n^{-1}\sum_{i=1}^{n}\left[\hat{g}_{d}\left(\boldsymbol{z}_{i}^{\hat{*}}\right)-g_{d}\left(\boldsymbol{z}_{i}^{*}\right)\right]^{2}=o_{P}\left(1\right)$,
\item[(2)] $\{n^{-1}\sum_{i=1}^{n}\left[\hat{m}\left(\boldsymbol{z}_{i}^{\hat{*}}\right)-m\left(\boldsymbol{z}_{i}^{*}\right)\right]^{2}\}^{1/2}\cdot\{n^{-1}\sum_{i=1}^{n}\left[\hat{g}_{d}\left(\boldsymbol{z}_{i}^{\hat{*}}\right)-g_{d}\left(\boldsymbol{z}_{i}^{*}\right)\right]^{2}\}^{1/2}=o_{P}\left(n^{-1/2}\right)$,
and
\item[(3)] $n^{-1}\sum_{i=1}^{n}\left[\hat{g}_{d}\left(\boldsymbol{z}_{i}^{\hat{*}}\right)-g_{d}\left(\boldsymbol{z}_{i}^{*}\right)\right]\cdot\left[1-\frac{\mathbf{1}\left(D_{i}=d\right)}{m\left(\boldsymbol{z}_{i}^{*}\right)^{d}\left(1-m\left(\boldsymbol{z}_{i}^{*}\right)\right)^{1-d}}\right]=o_{P}\left(n^{-1/2}\right)$.
\end{enumerate}
Then, the doubly robust estimator $\hat{\psi}$ satisfies
\[
[\textrm{Var}\left(\psi_{i}\right)]^{-1/2}\sqrt{n}(\hat{\psi}-\psi)\stackrel{d}{\rightarrow}N\left(0,1\right).
\]
\end{theorem}

Theorem \ref{TH:debiased} states that under standard conditions,
$\hat{\psi}$ has the same asymptotic distribution as it would if
$(g_{0},g_{1})$ and $m$ were known. This result follows directly
from the existing literature and Corollary \ref{COR:propensity_score-screening}, which establishes that our screening procedure can recover $\boldsymbol{Z}^{*}$ with high probability. Thus, we omit the proof. The key implication is that our procedure can achieve dimension reduction not only for $D$ and $m$, but also for $Y\left(d\right)$ and $g_{d}$, $d=0,1$.

We briefly verify conditions (1)--(3) in Theorem \ref{TH:debiased}. For conditions (1) and (2), we assume that $(\hat{g}_{0},\hat{g}_{1})$ are certain estimators
from the literature that satisfy $\{ n^{-1}\sum_{i=1}^{n}\left[\hat{g}_{d}\left(\boldsymbol{z}_{i}^{*}\right)-g_{d}\left(\boldsymbol{z}_{i}^{*}\right)\right]^{2}\} {}^{1/2}=O_{P}(n^{-1/4})$ for
$d=0,1$, under standard regularity conditions.  We show in Lemma \ref{LE:uniform_n} that $\hat{m}\left(\boldsymbol{z}_{i}^{\hat{*}}\right)$ also satisfies the required convergence rate, provided that $r/\left(p^{\mathsf{c}\ast}+2r\right)>1/4$, aligning with
results in \citet{FarrellEtal2021}. Condition (3) can be satisfied through a ``sample splitting'' procedure, as described in \citet{ChernozhukovEtal2018}.

\par\medskip
We end this section with the following remarks.

\begin{remark}\emph{ Our discussion in this section focuses on the
dimension reduction of the propensity score $m(\boldsymbol{Z})$ for the following reasons: (1) the doubly robust estimator is more sensitive to estimation
errors in $\hat{m}(\boldsymbol{Z})$ than in $(\hat{g}_{0}(\boldsymbol{Z}),\hat{g}_{1}(\boldsymbol{Z}))$,
as $\hat{m}(\boldsymbol{Z})$ appears in the denominator; (2) even
under Assumption \ref{A:3}(2), $\hat{m}(\boldsymbol{Z})$ can still
approach 0 or 1 in finite samples, causing numerical instability in
the estimator.}\footnote{For an in-depth discussion on the use of trimming in causal
inference, see \citet{YangDing2018} and references therein.}\emph{ A common practice to address this issue is to trim observations
with extreme $\hat{m}(\boldsymbol{Z})$. However, when many irrelevant
covariates are included, the propensity score model may overfit the
data, leading to a substantial proportion of the sample being trimmed;
and (3) correct specification and efficient estimation of $m(\boldsymbol{Z})$
are the keys for constructing doubly robust estimators for many other weighted
ATEs, where the weights often depend on $m(\boldsymbol{Z})$. See,
e.g., \citet{li2018balancing} and \citet{tao2019doubly}.} \end{remark}

\begin{remark} \emph{As a final remark, our proposed dimension reduction
procedure is model-free. Its applicability extends beyond the specific
model and estimation methods discussed above. Instead, it should be
regarded as a general tool to enhance our understanding of the data
and research problem. For instance, consider a covariate $X_{j}$
satisfying $Y\perp X_{j}\mid D$. In this case, $X_{j}$ can either
be entirely irrelevant to the causal model if $D\perp X_{j}$ also
holds or serves as an instrumental variable (IV) if $D\not\perp X_{j}$.
In both scenarios, including $X_{j}$ in the estimand (\ref{eq:psi})
does not introduce bias, but excluding it could enhance finite-sample
efficiency and reduce the risk of model overfitting. In the latter
scenario, this insight may enable alternative IV regression approaches
for causal inference. Our approaches can be useful in identifying
such cases. }\end{remark}

\section{Simulations}\label{sec:simulation}
This section assesses the finite-sample performance of the procedure proposed in Section \ref{SEC:f(y,x)reduce} through Monte Carlo experiments. All simulations presented in this section are conducted using the R programming language, based on 1,000 independent replications. In Section \ref{sec:simu_1}, we examine dimension reduction in conditional density estimation with a continuous response variable \( Y \). In Section \ref{sec:simu_2}, we study the performance of our procedure in estimating propensity scores for the estimation and inference of ATE.  

\subsection{Simulations with Continuous Response}\label{sec:simu_1}
We explore three distinct simulation designs, each differing in how the response variable \( Y \) depends on covariates $\boldsymbol{X}$, based on the following linear varying coefficient model: 
\begin{equation}
Y = \boldsymbol{X}^{\mathsf{c}}{}' \boldsymbol{\beta}^{\mathsf{c}}(W^{\mathsf{c}}) + \boldsymbol{X}^{\mathsf{d}}{}' \boldsymbol{\beta}^{\mathsf{d}}(W^{\mathsf{c}}) + \epsilon. \label{eq:DGP_1}   
\end{equation} 
Throughout this section, we assume the error term \( \epsilon \sim N(0,1) \) and consider the total number of covariates \( p \in \{20, 50, 100\} \), with \( \boldsymbol{X}^{\mathsf{c}} \) and \( \boldsymbol{X}^{\mathsf{d}} \) each having dimension \( p/2 \). We assume there is no \( \boldsymbol{W}^{\mathsf{d}} \) and only a scalar \( W^{\mathsf{c}} \), i.e., \( \boldsymbol{W}= W^{\mathsf{c}}\). The true parameter functions for the continuous and discrete covariates are specified as follows: 
\[
\underset{(p/2) \times 1}{\boldsymbol{\beta}^{\mathsf{c}}(W^{\mathsf{c}})} = \left( 0.4 W^{\mathsf{c}} + 0.5, \sin(2\pi W^{\mathsf{c}}) + 0.5, 0, \dots, 0 \right)',
\]
and  
\[
\underset{(p/2) \times 1}{\boldsymbol{\beta}^{\mathsf{d}}(W^{\mathsf{c}})} = \left( 0.5 W^{\mathsf{c}} + 0.7, 0, \dots, 0 \right)'.
\]  
Thus, \(\boldsymbol{X}(\mathcal{M}^{*}) \equiv (X_{1}^{\mathsf{c}}, X_{2}^{\mathsf{c}}, X_{1}^{\mathsf{d}})'\) are directly relevant to \( Y \), while the remaining covariates are not.  
\par\medskip

\noindent\textbf{Design 1:} We generate a \(\left(p+1\right) \times 1\) vector \(\boldsymbol{e} = (e_1, \dots, e_{p+1})'\) independently from \( N(0, \boldsymbol{I}_{p+1}) \), where \(\boldsymbol{I}_{p+1}\) denotes the \(\left(p+1\right) \times \left(p+1\right)\) identity matrix. We set \( W^{\mathsf{c}} = e_{p+1} \) and, for \( j = 1, \dots, p/2 \), define $X_{j}^{\mathsf{c}} = e_{j}$, and 
\[
X_{j}^{\mathsf{d}} = (-1) \cdot \boldsymbol{1} \left( \Phi(e_{j+p/2}) \leq 1/3 \right) + \boldsymbol{1} \left(  \Phi(e_{j+p/2})>2/3  \right),
\]
where \(\Phi(\cdot)\) denotes the CDF of the standard normal distribution. Thus, in this design,  
\[
\left(Y, W^{\mathsf{c}}, \boldsymbol{X}(\mathcal{M}^{*})\right) \perp \left(X_{3}^{\mathsf{c}}, \dots, X_{p/2}^{\mathsf{c}}, X_{2}^{\mathsf{d}}, \dots, X_{p/2}^{\mathsf{d}}\right).
\]
\par\medskip

\noindent\textbf{Design 2:} Let \(\boldsymbol{\Sigma}_{k}\) denote a \(k \times k\) covariance matrix with all diagonal elements equal to 1 and all off-diagonal elements equal to 0.25. We generate a \((p-2) \times 1\) vector \(\boldsymbol{e}_{1} = (e_{1,1}, \dots, e_{1,p-2})' \sim N(0, \boldsymbol{\Sigma}_{p-2})\) and a \(3 \times 1\) vector \(\boldsymbol{e}_{2} = (e_{2,1}, e_{2,2}, e_{2,3})' \sim N(0, \boldsymbol{\Sigma}_{3})\) independently. We set \( W^{\mathsf{c}} = e_{1,p-2} \), $X_{j+2}^{\mathsf{c}} = e_{1,j}$, for $j = 1, \dots, p/2 - 2$, and
\[
X_{j+1}^{\mathsf{d}} = (-1) \cdot \boldsymbol{1} \left( \Phi(e_{1,j+p/2-2}) \leq 1/3 \right) + \boldsymbol{1} \left( \Phi(e_{1,j+p/2-2}) > 2/3 \right),
\]  
for \( j = 1, \dots, p/2 - 1 \). For the relevant covariates, we define $X_{j}^{\mathsf{c}} = e_{2,j}$, for $j = 1,2$, and 
\[
X_{1}^{\mathsf{d}} = (-1) \cdot \boldsymbol{1} \left( \Phi(e_{2,3}) \leq 1/3 \right) +   \boldsymbol{1} \left( \Phi(e_{2,3}) > 2/3 \right).
\]  
Thus, in this design,  
\[
Y \not\perp \left(X_{3}^{\mathsf{c}}, \dots, X_{p/2}^{\mathsf{c}}, X_{2}^{\mathsf{d}}, \dots, X_{p/2}^{\mathsf{d}}\right), \textrm{ but } Y \perp \left.\left(X_{3}^{\mathsf{c}}, \dots, X_{p/2}^{\mathsf{c}}, X_{2}^{\mathsf{d}}, \dots, X_{p/2}^{\mathsf{d}}\right) \right| W^{\mathsf{c}}.
\]  
\par\medskip
 
\noindent\textbf{Design 3:} We generate a \((p+1) \times 1\) vector \(\boldsymbol{e} = (e_1, \dots, e_{p+1})' \sim N(0, \boldsymbol{\Sigma}_{p+1})\) and define \(\boldsymbol{X}^{\mathsf{c}}\), \(\boldsymbol{X}^{\mathsf{d}}\), and \(W^{\mathsf{c}}\) following the same formulae based on \(\boldsymbol{e}\) as in Design 1. Thus, in this design,  
\[
Y \not\perp \left.\left(X_{3}^{\mathsf{c}}, \dots, X_{p/2}^{\mathsf{c}}, X_{2}^{\mathsf{d}}, \dots, X_{p/2}^{\mathsf{d}}\right)\right| W^{\mathsf{c}}, \textrm{ but } Y \perp \left.\left(X_{3}^{\mathsf{c}}, \dots, X_{p/2}^{\mathsf{c}}, X_{2}^{\mathsf{d}}, \dots, X_{p/2}^{\mathsf{d}}\right)\right| \left(W^{\mathsf{c}}, \boldsymbol{X}(\mathcal{M}^{*})\right).
\]
\par\medskip

Design 1 is the benchmark design. Design 2 examines the effectiveness of our approach in handling the dependence between \( Y \) and \( (X_{3}^{\mathsf{c}}, \dots, X_{p/2}^{\mathsf{c}}, X_{2}^{\mathsf{d}}, \dots, X_{p/2}^{\mathsf{d}} )'\) through their shared dependence on \(\boldsymbol{W}\). Design 3 assesses the performance of our procedure under a more general dependence structure, as described in (\ref{eq:conditional_refine}), where the dependence between \( Y \) and \(( X_{3}^{\mathsf{c}}, \dots, X_{p/2}^{\mathsf{c}}, X_{2}^{\mathsf{d}}, \dots, X_{p/2}^{\mathsf{d}} )'\) arises from two sources: their common dependence on \(\boldsymbol{W}\) and their relationship with the set \(\boldsymbol{X}(\mathcal{M}^{*})\) of covariates directly relevant to \( Y \).

For each simulation design, we implement the dimension reduction procedure described in Section \ref{sec:summary_procedure} with sample sizes \( n \in \{250, 500, 1000, 2000\} \) and set \(\tilde{p} = 5\) to identify the set $\boldsymbol{X}(\mathcal{M}^{*})$ in the data-generating process (DGP) (\ref{eq:DGP_1}). Let \( \boldsymbol{X}(\hat{\mathcal{M}}^{(1)}) \) denote the set of covariates selected after the first two steps of our procedure (screening), and \(\boldsymbol{X}(\hat{\mathcal{M}}^{(2)}) \) denote the set of covariates not assigned extreme bandwidths in the CV refining steps 3 and 4. In Appendix \ref{appendix:tables_1}, we report two sets of simulation results:
\begin{enumerate}
\item Tables \ref{tab:design1_top4_screening}--\ref{tab:design3_top4_screening} present screening results for Designs 1–3, reporting the probability that each element of \(\boldsymbol{X}(\mathcal{M}^{*})\), as well as the entire set, is contained in \(\boldsymbol{X}(\hat{\mathcal{M}}^{(1)})\). We refer to this probability as the correct recovery rate (CRR) throughout this section. Implementing our proposed screening procedure requires selecting a reference value \(y^{*}\) to calculate the estimate \(\hat{\rho}\) of our conditional dependence measure \(\rho\). In our simulation studies, we consider two options. First, we calculate \(\hat{\rho}\) by simply setting \(y^{*}\) to be the sample median of \(Y\). Second, we compute \(\hat{\rho}\) using the following variant of KS distance:
\begin{equation}
\hat{\Lambda}_X(Y = y, \boldsymbol{W} = \boldsymbol{w}) = \max_{i=1,\ldots,n} \max_{j=1,2,3} \left| \hat{F}_n(x_i \mid y, \boldsymbol{w}) - \hat{F}_n(x_i \mid y_j^*, \boldsymbol{w})\right|, \label{eq:rho_variant}
\end{equation}
where \((y_1^*, y_2^*, y_3^*)\) respectively correspond to the \((0.25, 0.5, 0.75)\) quantiles of \(Y\). Results based on these two approaches are labeled as ``\(\rho\)'' and ``\(\textrm{Quantile-}\rho\)'', respectively,  in Tables \ref{tab:design1_top4_screening}--\ref{tab:design3_top4_screening}. In addition to our method, we conduct variable screening using the conditional distance correlation (labeled as ``CDCSIS'') by \cite{WangEtal2015} and feature ordering by conditional independence (FOCI) by \cite{azadkia2021simple} for comparison. The CDCSIS and FOCI results are obtained using R packages $\texttt{cdcsis}$ and $\texttt{FOCI}$, respectively.

\item The CV refinement results for Designs 1--3 are summarized in Table \ref{tab:prob1_refine}, in which we compute \(\Pr(\boldsymbol{X}(\mathcal{M}^{*}) = \boldsymbol{X}(\hat{\mathcal{M}}^{(2)}))\), the probability that the CV refinement step achieves exact recovery of \(\boldsymbol{X}(\mathcal{M}^{*})\). We compare the performance of our modified CV refinement algorithm introduced in Section \ref{SEC:post-estimation} (labeled as ``Modified-CV'') against that of the original CV-ISE optimization procedure of \citet{HallRacineLi} (labeled as ``CV''), implemented with the R package $\texttt{np}$.
\end{enumerate}

The two variable screening approaches based on our proposed conditional dependence measure can identify relevant covariates with high probability across Designs 1--3, even in high-dimensional settings, when $n \geq 500$. The ``\(\textrm{Quantile-}\rho\)'' approach, which uses the statistic defined in (\ref{eq:rho_variant}) with multiple reference values, generally outperforms the ``$\rho$'' approach that relies solely on the sample median, particularly when screening discrete covariates in small samples. Supplementary simulations suggest that incorporating additional quantiles, such as the $(0.1, \ldots, 0.9)$ quantiles of $Y$, can further enhance the small-sample performance of the ``\(\textrm{Quantile-}\rho\)'' method.

In small samples, our approaches exhibit substantially higher CRRs in variable screening than FOCI and slightly lower CRRs than CDCSIS. As noted in Remark \ref{Remark:compare}, FOCI is the most computationally efficient, with a computational cost of $O(n\log n)$, but our simulation results indicate it requires relatively large sample sizes to attain a desirable CRR. Applying CDCSIS, which has a computational cost of $O(n^3)$, to simulations with $n\geq 2000$ can be quite time-consuming. Thus, we omit the corresponding results in Table \ref{tab:design1_top4_screening}--\ref{tab:design3_top4_screening}, given that CDCSIS already achieved 100\% CRR with $n=1000$. Our two approaches, however, have a computational cost of $O(n^2)$, and we observe that when $n\geq 1000$, they have almost identical CRR to CDCSIS. In other words, our approaches offer significantly more computationally efficient alternatives to CDCSIS for moderately large datasets.

Table \ref{tab:prob1_refine} reports results from the CV refinement step, showing that our ``Modified-CV'' algorithm achieves exact recovery with probability approaching one. It consistently outperforms the original ``CV'' procedure of \cite{HallRacineLi} across all designs and sample sizes. As discussed in Section \ref{SEC:post-estimation}, the ``CV'' procedure minimizes a CV-ISE criterion, which can be numerically unstable due to its high nonlinearity and multiple local minima. Our simulations demonstrate that the ``Modified-CV'' algorithm offers a more computationally efficient and stable post-screening refinement strategy. We omit the results of the original CV for $n=2000$ due to its high computation cost.

\subsection{Simulations with Discrete Response} \label{sec:simu_2}
This section examines the performance of our procedure in propensity score estimation and its application to dimension-reduced doubly robust estimation of ATE, as proposed in Section \ref{sec:theoryapply}.

We consider simulation designs adapted from \cite{FarrellEtal2021}. Specifically, we assume a binary treatment $D$ with the propensity score given by the varying coefficient logistic model:
\begin{equation}
m(\boldsymbol{Z})=\Pr\left(D=1\mid \boldsymbol{Z}\right) = \left\{1+\exp\left[- 2 \cdot \mathcal{S} \left(\boldsymbol{X}^{\mathsf{c}}{}'\boldsymbol{\alpha}^{\mathsf{c}}(W^{\mathsf{c}})+\boldsymbol{X}^{\mathsf{d}}{}'\boldsymbol{\alpha}^{\mathsf{d}}(W^{\mathsf{c}})\right)\right]\right\}^{-1},  \label{eq:pscore_design}   
\end{equation}
where $\boldsymbol{Z}=(\boldsymbol{X}',\boldsymbol{W}')'$ and \(\mathcal{S}(\cdot)\) denotes the standardization operator such that \(\mathcal{S}(X) = [X -\mathbb{E}(X)]/\text{sd}(X)\) for any random variable $X$. As in Section \ref{sec:simu_1}, the dimension of \( \boldsymbol{X} \) is set to \( p \in \{20, 50, 100\} \), where \( \boldsymbol{X}^{\mathsf{c}} \) and \( \boldsymbol{X}^{\mathsf{d}} \) each have dimension \( p/2 \), and \( \boldsymbol{W}=W^{\mathsf{c}} \). The true parameter functions in (\ref{eq:pscore_design}) are specified as:
\[
\boldsymbol{\alpha}^{\mathsf{c}}(W^{\mathsf{c}}) = \left( 0.4 W^{\mathsf{c}} + 0.5, \sin\left(2\pi W^{\mathsf{c}}\right)+0.5, 0, \dots, 0 \right)',
\]
and  
\[
\boldsymbol{\alpha}^{\mathsf{d}}(W^{\mathsf{c}}) = \left( 0.5W^{\mathsf{c}} + 0.7, 0, \dots, 0 \right)'.
\]  
Thus, only \( \boldsymbol{X}(\mathcal{M}^{*}) = (X_{1}^{\mathsf{c}}, X_{2}^{\mathsf{c}}, X_{1}^{\mathsf{d}})'\) are directly relevant to \( D \), and $\boldsymbol{Z}^*=(\boldsymbol{X}(\mathcal{M}^{*})',W^{\mathsf{c}})'$. 

We adopt the same DGPs for \( \boldsymbol{X}\) and \(  W^{\mathsf{c}}\) as in Designs 1–3. These DGPs, when applied to the propensity score model specified in equation (\ref{eq:pscore_design}), establish Designs 4–6, which feature the following dependence structures:
\par\medskip
\noindent\textbf{Design 4:} This serves as our benchmark design, where  
\[
\left(D, W^{\mathsf{c}}, \boldsymbol{X}(\mathcal{M}^{*})\right) \perp \left(X_{3}^{\mathsf{c}}, \dots, X_{p/2}^{\mathsf{c}}, X_{2}^{\mathsf{d}}, \dots, X_{p/2}^{\mathsf{d}}\right).
\]
\noindent\textbf{Design 5:} \( D \) and \( (X_{3}^{\mathsf{c}}, \dots, X_{p/2}^{\mathsf{c}}, X_{2}^{\mathsf{d}}, \dots, X_{p/2}^{\mathsf{d}} )'\) are dependent solely through their shared dependence on \( W^{\mathsf{c}} \), i.e.,
\[
D \not\perp \left(X_{3}^{\mathsf{c}}, \dots, X_{p/2}^{\mathsf{c}}, X_{2}^{\mathsf{d}}, \dots, X_{p/2}^{\mathsf{d}}\right), \text{ but } D \perp \left.\left(X_{3}^{\mathsf{c}}, \dots, X_{p/2}^{\mathsf{c}}, X_{2}^{\mathsf{d}}, \dots, X_{p/2}^{\mathsf{d}}\right)\right| W^{\mathsf{c}}.
\]
\noindent\textbf{Design 6:} \( D \) depends on \( (X_{3}^{\mathsf{c}}, \dots, X_{p/2}^{\mathsf{c}}, X_{2}^{\mathsf{d}}, \dots, X_{p/2}^{\mathsf{d}} )'\) not only through \( W^{\mathsf{c}} \) but also through the covariates \(\boldsymbol{X}(\mathcal{M}^{*})\), which are directly relevant to \( D \), i.e.,
\[
D \not\perp \left.\left(X_{3}^{\mathsf{c}}, \dots, X_{p/2}^{\mathsf{c}}, X_{2}^{\mathsf{d}}, \dots, X_{p/2}^{\mathsf{d}}\right)\right| W^{\mathsf{c}}, \text{ but } D \perp \left.\left(X_{3}^{\mathsf{c}}, \dots, X_{p/2}^{\mathsf{c}}, X_{2}^{\mathsf{d}}, \dots, X_{p/2}^{\mathsf{d}}\right)\right| \left(W^{\mathsf{c}}, \boldsymbol{X}(\mathcal{M}^{*})\right).
\]

Given the covariates and treatment variable, we specify the following DGP for the outcome \( Y \):
\begin{equation}
Y = g_{0}(\boldsymbol{X}(\mathcal{G}^{*}), W^{\mathsf{c}}) + \psi(\boldsymbol{X}(\mathcal{G}^{*}), W^{\mathsf{c}})D + \epsilon, \label{eq:outcome_design}
\end{equation}
where $\boldsymbol{X}(\mathcal{G}^{*})\equiv (X_{1}^{\mathsf{c}}, X_{3}^{\mathsf{c}}, X_{1}^{\mathsf{d}})'$. Notably, the subsets of covariates directly relevant to the outcome and the treatment differ, with $\boldsymbol{X}(\mathcal{G}^{*})= (X_{1}^{\mathsf{c}}, X_{3}^{\mathsf{c}}, X_{1}^{\mathsf{d}})'$ and $\boldsymbol{X}(\mathcal{M}^{*}) = (X_{1}^{\mathsf{c}}, X_{2}^{\mathsf{c}}, X_{1}^{\mathsf{d}})'$. The untreated potential outcome \( g_{0}(\boldsymbol{X}(\mathcal{G}^{*}), W^{\mathsf{c}})\) and the conditional treatment effect \( \psi(\boldsymbol{X}(\mathcal{G}^{*}), W^{\mathsf{c}}) \) are defined as:
\[
g_{0}(\boldsymbol{X}(\mathcal{G}^{*}), W^{\mathsf{c}}) = \varphi(\boldsymbol{X}(\mathcal{G}^{*}))'\beta_{g}(W^{\mathsf{c}}) 
\]
and
\[
\psi(\boldsymbol{X}(\mathcal{G}^{*}), W^{\mathsf{c}}) = \varphi(\boldsymbol{X}(\mathcal{G}^{*}))'\beta_{\psi}(W^{\mathsf{c}}),
\]
where \( \varphi(\boldsymbol{X}(\mathcal{G}^{*})) \) denotes the vector of all 9 second-degree polynomial terms of \( \boldsymbol{X}(\mathcal{G}^{*}) \), including all pairwise interactions. The coefficient functions are given by:
\[
\beta_{g,k}(W^{\mathsf{c}}) = W^{\mathsf{c}} \cdot U_{g,k} \textrm{ with } U_{g,k} \sim N(0.3, 0.7)  
\]
and
\[
\beta_{\psi,k}(W^{\mathsf{c}}) = W^{\mathsf{c}} \cdot U_{\psi,k} \textrm{ with } U_{\psi,k} \sim U(0.1, 0.22),
\]
where the subscript \( k =1,...,9 \) indexes each component of \( \varphi(\boldsymbol{X}(\mathcal{G}^{*})) \).  We integrate the outcome regression model \eqref{eq:outcome_design} into Designs 4–6 to estimate the ATE, defined as \( \psi = \mathbb{E}[\psi(\boldsymbol{X}(\mathcal{G}^{*}), W^{\mathsf{c}})] \). 

We consider sample sizes $n\in\{500, 1000, 2000, 4000\}$ and split each sample into two folds $(I_1,I_2)$, of equal size. Note that the sample splitting is required for condition (3) in Theorem \ref{TH:debiased}. For implementation, we first apply the procedure proposed in Section \ref{sec:summary_procedure} with $\tilde{p}=5$ to sample $I_1$ (with sample size $n_1\in\{250, 500, 1000, 2000\}$) to obtain the dimension-reduced estimator $\hat{m}(\boldsymbol{Z}^{\hat{*}})$ of the propensity score \( m(\boldsymbol{Z})\), where \(\boldsymbol{Z}^{\hat{*}} \equiv (\boldsymbol{X}(\hat{\mathcal{M}}^{(2)}), W^{\mathsf{c}}) \), following the notation in Sections \ref{sec:theoryapply} and \ref{sec:simu_1}. This step aims to identify the components of \( \boldsymbol{X} \) that are directly relevant to the treatment \( D \), namely \( \boldsymbol{X}(\mathcal{M}^{*})\), with high probability. It is worth noting that, by Proposition \ref{Prop:refine}, consistent estimation of $\psi$ does not require recovering $\boldsymbol{X}(\mathcal{G}^{*})$. 

Tables \ref{tab:design4_top4_screening}--\ref{tab:design6_top4_screening} and Table \ref{tab:prob1_refine_d456} in Appendix \ref{appendix:tables_2} present the dimension reduction results for Designs 4--6. These tables parallel Tables \ref{tab:design1_top4_screening}--\ref{tab:design3_top4_screening} and Table \ref{tab:prob1_refine} from Designs 1--3, respectively, reporting the same set of metrics. The only exception is the absence of the ``$\textrm{Quantile-}\rho$'' results, as $D$ is a binary variable in Designs 4--6. In these designs, our proposed screening method and CDCSIS demonstrate finite-sample performance similar to that observed in Designs 1--3 in almost all aspects. However, it appears that $n_1=2000$ is not always sufficient for FOCI to attain a desirable CRR in some designs. Taken together with our earlier discussion on computational efficiency, these findings underscore the advantage of our approach in nonparametric estimation involving discrete response variables, such as propensity score estimation. The CV refinement results in Table \ref{tab:prob1_refine_d456} reveal similar patterns to those in Table \ref{tab:prob1_refine}. In particular, while our ``Modified-CV'' algorithm continues to outperform the ``CV'' procedure of \citet{HallRacineLi}, the performance gap narrows in these designs. 

Following \citet{YangDing2018}, we focus on the subpopulation whose propensity scores lie within the interval \(\left[0.1,0.9\right]\). This practice helps avoid issues with extreme estimated propensity scores by trimming the sample. To estimate the ATE \( \psi \), we apply the doubly robust estimator defined in (\ref{eq:psi_hat}) and implement the multi-layer perceptron (MLP) procedure employed in \cite{FarrellEtal2021}. Specifically, we define \(\hat{\omega} = \mathbf{1}\left(0.1\leq \hat{m}\left(\cdot\right)\leq 0.9\right)\) and consider the following variants: 

\begin{enumerate}
\item Estimate $\hat{m}(\boldsymbol{Z})$ and $(\hat{g}_{0}(\boldsymbol{Z}), \hat{g}_{1}(\boldsymbol{Z}))$ via MLP using the full set $\boldsymbol{Z}$ in sample \(I_{1}\), then compute the estimator \(\hat{\psi}_{1}\) in sample \(I_{2}\): 
\[\hat{\psi}_1 = \frac{1}{\underset{i\in I_{2}}{\sum}\hat{\omega}_{i}}
\sum_{i \in I_{2}} \hat{\omega}_{i}
\left\{
\hat{g}_{1}\left(\boldsymbol{Z}_{i}\right) - \hat{g}_{0}\left(\boldsymbol{Z}_{i}\right)
+ \frac{D_{i}\left(Y_{i} - \hat{g}_{1}\left(\boldsymbol{Z}_{i}\right)\right)}{\hat{m}\left(\boldsymbol{Z}_{i}\right)}
- \frac{\left(1 - D_{i}\right)\left(Y_{i} - \hat{g}_{0}\left(\boldsymbol{Z}_{i}\right)\right)}{1 - \hat{m}\left(\boldsymbol{Z}_{i}\right)}
\right\}.
\]
\item Estimate $\hat{m}(\boldsymbol{Z}^{\hat{*}})$ via the kernel method and \(\left(\hat{g}_{0}\left(\boldsymbol{Z})\right), \hat{g}_{1}\left(\boldsymbol{Z}\right)\right)\) on the full set $\boldsymbol{Z}$ via MLP in sample \(I_{1}\), then compute the estimator \(\hat{\psi}_{2}\) in sample \(I_{2}\):  
\[
\hat{\psi}_2 = \frac{1}{\underset{i\in I_{2}}{\sum}\hat{\omega}_{i}}
\sum_{i \in I_{2}} \hat{\omega}_{i}
\left\{
\hat{g}_{1}\left(\boldsymbol{Z}_{i}\right) - \hat{g}_{0}\left(\boldsymbol{Z}_{i}\right)
+ \frac{D_{i}\left(Y_{i} - \hat{g}_{1}\left(\boldsymbol{Z}_{i}\right)\right)}{\hat{m}(\boldsymbol{Z}_{i}^{\hat{*}})}
- \frac{\left(1 - D_{i}\right)\left(Y_{i} - \hat{g}_{0}\left(\boldsymbol{Z}_{i}\right)\right)}{1 - \hat{m}(\boldsymbol{Z}_{i}^{\hat{*}})}
\right\}.
\]
\item Estimate $\hat{m}(\boldsymbol{Z}^{\hat{*}})$ via the kernel method and \((\hat{g}_{0}(\boldsymbol{Z}^{\hat{*}}),\hat{g}_{1}(\boldsymbol{Z}^{\hat{*}}))\) via MLP in sample \(I_{1}\), then compute the estimator \(\hat{\psi}_{3}\) in sample \(I_{2}\): 
\[
\hat{\psi}_3 = \frac{1}{\underset{i\in I_{2}}{\sum}\hat{\omega}_{i}}
\sum_{i \in I_{2}} \hat{\omega}_{i}
\left\{
\hat{g}_{1}(\boldsymbol{Z}_{i}^{\hat{*}}) - \hat{g}_{0}(\boldsymbol{Z}_{i}^{\hat{*}})
+ \frac{D_{i}(Y_{i} - \hat{g}_{1}(\boldsymbol{Z}_{i}^{\hat{*}}))}{\hat{m}(\boldsymbol{Z}_{i}^{\hat{*}})}
- \frac{\left(1 - D_{i}\right)(Y_{i} - \hat{g}_{0}(\boldsymbol{Z}_{i}^{\hat{*}}))}{1 - \hat{m}(\boldsymbol{Z}_{i}^{\hat{*}})}
\right\}.
\]
\item Estimate both \(\hat{m}(\boldsymbol{Z}^{\hat{*}})\) and \((\hat{g}_{0}(\boldsymbol{Z}^{\hat{*}}),\hat{g}_{1}(\boldsymbol{Z}^{\hat{*}}))\) via MLP in sample \(I_{1}\), then compute the estimator \(\hat{\psi}_{4}\) in sample \(I_{2}\): 
\[
\hat{\psi}_4 = \frac{1}{\underset{i\in I_{2}}{\sum}\hat{\omega}_{i}}
\sum_{i \in I_{2}} \hat{\omega}_{i}
\left\{
\hat{g}_{1}(\boldsymbol{Z}_{i}^{\hat{*}}) - \hat{g}_{0}(\boldsymbol{Z}_{i}^{\hat{*}})
+ \frac{D_{i}(Y_{i} - \hat{g}_{1}(\boldsymbol{Z}_{i}^{\hat{*}}))}{\hat{m}(\boldsymbol{Z}_{i}^{\hat{*}})}
- \frac{\left(1 - D_{i}\right)(Y_{i} - \hat{g}_{0}(\boldsymbol{Z}_{i}^{\hat{*}}))}{1 - \hat{m}(\boldsymbol{Z}_{i}^{\hat{*}})}
\right\}.
\]
\end{enumerate}

When implementing MLP to estimate the nuisance functions, the network architecture consists of two hidden layers with 64 and 32 neurons, respectively, each followed by a ReLU activation function. Training is performed using the \texttt{Adam} optimizer with a learning rate of $10^{-3}$, a mini-batch size of 32, and 100 training epochs. We use the mean squared error loss function for outcome regression and the binary cross-entropy loss function for propensity score estimation. All computations are conducted using the \texttt{PyTorch} framework.

For all these estimators, we report their mean bias (Bias), root mean squared error (RMSE), and the length (IL) and coverage probability (CP) of the 95\% confidence interval. We summarize these results in Tables \ref{tab:design4_ate}--\ref{tab:design6_ate} in Appendix \ref{appendix:tables_2}, corresponding to Designs 4--6, respectively. Estimators $\hat{\psi}_3$ and $\hat{\psi}_4$ exhibit similarly satisfactory performance across all designs. As $n$ increases, both estimators display vanishing biases, RMSEs that decline at approximately $\sqrt{n}$ rate, and shrinking confidence intervals with coverage probabilities approaching the target value of 0.95. Their performance, as expected, slightly deteriorates as $p$ increases. Between the two, $\hat{\psi}_4$ tends to have slightly higher RMSE and wider confidence interval than $\hat{\psi}_3$. In contrast, $\hat{\psi}_1$ performs poorly, yielding larger bias, higher RMSE, and wider confidence interval compared to the other estimators. Estimator $\hat{\psi}_2$ also suffers from relatively large bias and, more critically, produces narrow confidence intervals when $p$ is large, resulting in poor coverage. Overall, we recommend that practitioners employ $\hat{\psi}_3$ or $\hat{\psi}_4$ in empirical applications, as we will do in the next section.

\section{Empirical Illustration} \label{sec:application}
There is considerable debate in empirical studies on whether 401(k) eligibility increases (``crowd-in'') or decreases (``crowd-out'') the accumulation of other assets (savings). A key challenge in identifying this effect is that 401(k) eligibility may be correlated with an individual's unobserved savings preferences, potentially leading to selection bias. For example, individuals with strong savings preferences may be more likely to seek employment at firms that offer 401(k) plans. \cite{benjamin2003does} investigates this question using regression within propensity score subclasses, with the validity of his analysis relying on the absence of unobserved confounding factors. \cite{gelber2011401} aims to address this challenge by employing a difference-in-differences (DID) approach using matched observations based on stratified propensity scores.\footnote{\cite{gelber2011401}'s DID approach estimates the ATT of 401(k) eligibility.} However, both the linear DID regression and the Probit propensity score model employed in his analysis may suffer from model misspecification. In this section, we re-analyze the effect of 401(k) eligibility on savings by applying our proposed dimension-reduced propensity score estimator (Section \ref{sec:Propensity-Score}) and the doubly robust ATE estimator leveraging Proposition \ref{Prop:refine} (Section \ref{SEC:debias}) to the rich longitudinal dataset used by \cite{gelber2011401}, attempting to address the methodological limitations in existing studies.

Broadly, evaluating the effect of tax-advantaged retirement savings plans, such as the 401(k) in the US or pension systems in other countries, on private saving or debt, along with its influence channels, has attracted long-standing interest. Related topics have been extensively studied in recent years using various empirical strategies (e.g., DID, regression discontinuity (kink) designs), as illustrated by works such as \cite{chetty2014active}, \cite{andersen2018tax}, \cite{messacar2018crowd}, \cite{goodman2020catching}, \cite{beshears2022borrowing}, and \cite{garcia2024public}. Identifying and estimating the corresponding causal effects also contributes to a better understanding of individual responses to policy incentives for private retirement savings (\cite{chan2022income}) and the broader macroeconomic implications of pension policy for the structure of the financial system (\cite{scharfstein2018presidential}). The related literature is vast. We refer interested readers to the cited studies and the references therein for a more comprehensive review.

\subsubsection*{Data}
\begin{figure}[htbp]
    \centering
    \includegraphics[width=0.9\textwidth]{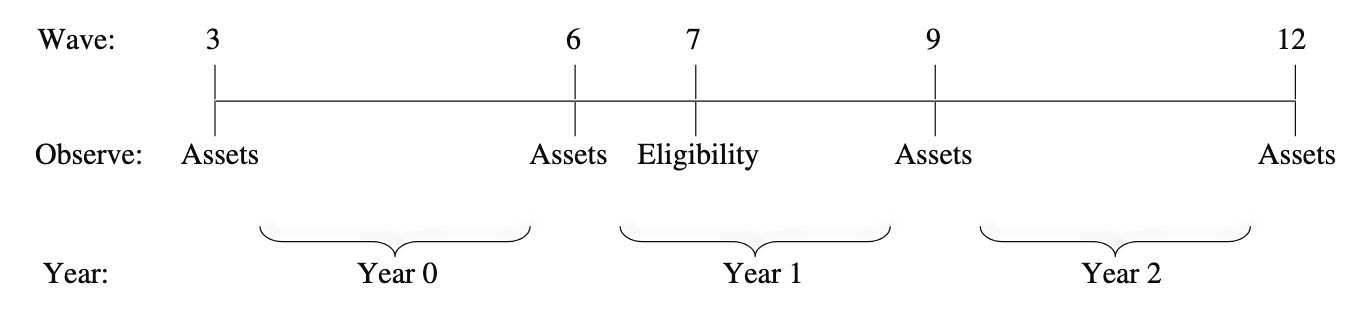}
    \caption{Timing of Observations}
    \label{fig:timing}  
\end{figure}

We use data from Waves 3, 6, 7, 9, and 12 of the 1996 SIPP.\footnote{The raw data and accompanying do-files used for processing were obtained from \citet{gelber2011401}'s replication package, which is publicly available at \url{https://www.openicpsr.org/openicpsr/project/116534/version/V1/view}.} 401(k) eligibility is observed in Wave 7, and asset (or liability) information is collected in the other waves.\footnote{For simplicity, we treat all liabilities as assets and represent them in positive terms, so that a larger value indicates a larger liability.} Following \citet{gelber2011401}, we define the period from Wave 3 to Wave 6 as ``Year 0'', the period from Wave 6 to Wave 9 as ``Year 1'', and the period from Wave 9 to Wave 12 as ``Year 2'', which approximately correspond to calendar years 1997, 1998, and 1999, respectively. The data timeline is illustrated in Figure \ref{fig:timing}, sourced from \citet{gelber2011401}. Following prior studies, our sample is restricted to reference persons in each household who are 22–64 years old and work at a for-profit firm in Year 1. 

As in \citet{benjamin2003does} and \citet{gelber2011401}, our treatment variable $D$ equals 1 if the individual is eligible for a 401(k) plan and 0 otherwise. In the literature, eligible individuals with positive 401(k) balances are defined as 401(k) participants, while those with a zero balance are considered non-participants. It is well documented in the literature that firms decide whether to offer employees 401(k) plans, whereas participation is the employee's choice. Therefore, eligibility, rather than participation, is a more plausible treatment variable that is conditionally independent of individual-level potential outcomes, given observed individual, household, and firm characteristics, satisfying Assumption \ref{A:3}(1) (ignorability).

We follow \citet{gelber2011401} in examining the following asset categories: 401(k) assets, IRA assets, other financial assets, secured debt, unsecured debt, and car value. Let $A_{t}^{k}$ denote the balance of asset category $k$ in Wave $t$. Our outcome variables are the first differences of asset balances, defined as $\Delta A_{2}^{k} \equiv A_{12}^{k} - A_{9}^{k}$, measuring the change in the balance of asset $k$ in Year 2. This transformation helps control for time-invariant unobserved heterogeneity, such as individual saving preferences. To mitigate the influence of large outliers commonly observed in asset data, we follow \citet{gelber2011401} and winsorize outcome variables at the 5th and 95th percentiles. Summary statistics for these outcome variables are presented in Panel A of Table \ref{tab:app_summary}. The reported $p$-values, from regressions of each $\Delta A_{2}^{k}$ on $D$ without other covariates, indicate that eligible and ineligible individuals differ significantly in their accumulation of 401(k) and IRA assets. The latter suggests a ``crowd-in'' effect, a finding consistent with that of \citet{gelber2011401}.

To strengthen the plausibility of Assumption \ref{A:3}(1), we include a comprehensive set of covariates, encompassing all those used by \citet{benjamin2003does} and \citet{gelber2011401}, along with many additional variables.\footnote{These include the number of children under 18, indicators for home ownership, gender, race, metropolitan residence, state of residence, private and public health insurance coverage, financial aid receipt, union or employee association membership, lifetime armed forces service, among others.} Leveraging the longitudinal nature of the data, we also control for changes in 401(k) and other asset balances in Year 0. In total, we include 93 covariates in our analysis.\footnote{These are all stand-alone variables, not including any transformation such as polynomial or interaction terms.} All monetary variables are measured in 1996 US dollars (in units of \$1,000). After excluding observations with missing values, our working data comprises $n = 6741$ observations, of whom 59.98\% are 401(k) eligible ($D=1$) and 40.02\% are ineligible ($D=0$).

\subsubsection*{Empirical Model}
Given a set of covariates $\boldsymbol{Z}$, we consider the following empirical model for a representative individual's (potential) asset holdings in wave $t \in \{3, 6, 9, 12\}$:
\begin{equation}
A_{t}^{k}(d) = G_{t,d}^{k}(\boldsymbol{Z}) + \alpha_{d}^{k} + u_{t,d}^{k}, \label{eq:app_eq_1}
\end{equation}
where $A_{t}^{k}(d)$ denotes an individual's holdings of asset $k$ in Wave $t$, given the 401(k) eligibility status $D = d \in \{0, 1\}$. The term $\alpha_{d}^{k}$ captures unobserved time-invariant confounding factors, $u_{t,d}^{k}$ is an idiosyncratic error, and $G_{t,d}^{k}(\cdot)$ is an unknown smooth function. \cite{benjamin2003does} estimates model (\ref{eq:app_eq_1}) using cross-sectional SIPP data, assuming a linear specification for $G_{t,d}^{k}(\boldsymbol{Z})$ and excluding the unobserved confounder $\alpha_{d}^{k}$. Taking first differences between Waves 9 and 12 removes $\alpha_{d}^{k}$ and yields
\begin{equation}
\Delta A_{2}^{k}(d) = A_{12}^{k}(d) - A_{9}^{k}(d) = \left( G_{12,d}^{k}(\boldsymbol{Z}) - G_{9,d}^{k}(\boldsymbol{Z}) \right) + \left(u_{12,d}^{k} - u_{9,d}^{k} \right), \label{eq:app_eq_2}
\end{equation}
which defines the potential outcome for our analysis. Then, the observed outcome can be expressed as $\Delta A_{2}^{k} = \Delta A_{2}^{k}(1) D + \Delta A_{2}^{k}(0) (1 - D)$, following the standard potential outcomes framework.

Suppose Assumption \ref{A:3} holds in this application. We apply the approaches proposed in Sections \ref{SEC:f(y,x)reduce} and \ref{sec:theoryapply} to the observed data $(\Delta A_{2}^{k}, D, \boldsymbol{Z})$ to estimate the ATE:
\begin{align}
\psi^{k}\equiv & \mathbb{E}[\Delta A_{2}^{k}(1)-\Delta A_{2}^{k}(0)] \nonumber \\
=&\mathbb{E}[g_{1}^{k}(\boldsymbol{Z})-g_{0}^{k}(\boldsymbol{Z})]+\mathbb{E}\left[\frac{D\left(\Delta A_{2}^{k}-g_{1}^{k}(\boldsymbol{Z})\right)}{m(\boldsymbol{Z})}-\frac{\left(1-D\right)\left(\Delta A_{2}^{k}-g_{0}^{k}(\boldsymbol{Z})\right)}{1-m(\boldsymbol{Z})}\right],  \label{eq:app_eq_3} 
\end{align}
where $m(\boldsymbol{Z}) = \Pr(D | \boldsymbol{Z})$ and $g_{d}^{k}(\boldsymbol{Z}) = \mathbb{E}[\Delta A_{2}^{k}(d) | \boldsymbol{Z}]$ for $d\in\{0,1\}$, consistent with the notation in previous sections.

\subsubsection*{Variable Selection}
We begin by applying the variable selection procedure from Section \ref{sec:summary_procedure} to evaluate the dependence between each pre-treatment covariate and $D$. For the procedure's implementation, we set $\tilde{p}=8$ and use the change in an individual's 401(k) balance in Year 0, i.e., $\Delta A_{0}^{\textrm{401(k)}}\equiv A_{6}^{\textrm{401(k)}}-A_{3}^{\textrm{401(k)}}$, as $\boldsymbol{W}$. This variable serves as a proxy for 401(k) participation in Year 0 and is therefore expected to be highly relevant for predicting subsequent eligibility. By capturing prior participation behavior, this variable serves as a sufficient surrogate for many observed and unobserved characteristics that influence 401(k) eligibility, thereby simplifying the dimension reduction task. To minimize subjective judgment, no other covariates are included in $\boldsymbol{W}$.

To ensure a robust selection, we apply our procedure to the full sample as well as to two equal-sized (randomly) split samples ($I_1, I_2$), which yields slightly different sets of selected covariates conditional on $\Delta A_{0}^{\textrm{401(k)}}$. In accordance with the principle of conservatism, our final set of relevant covariates for subsequent ATE estimation, denoted by $\boldsymbol{Z}^{\hat{*}}$, is the union of the variables selected from all three analyses. Definitions and summary statistics for these covariates are provided in Panel B of Table \ref{tab:app_summary}. The reported $p$-values show significant imbalances between the eligible and ineligible groups for most of these characteristics. Notably, our choice of $\tilde{p}=8$ proves to be sufficiently large, as there are always some covariates ``smoothed out'' in the CV-refining step of our procedure across all three analyses. 

Our procedure selects eight covariates as relevant to one’s 401(k) eligibility. As expected, firm characteristics are important. Larger firms and those in certain industries\footnote{According to the 1996 SIPP data documentation, $\texttt{Industry4}$ corresponds to firms engaged in transportation, communication, and utilities.} are more likely to offer 401(k) plans. Among individual-level variables, prior changes in the balances of 401(k) accounts, other financial assets, and unsecured debt may serve as plausible proxies for unobserved saving preferences. Gender and income are also selected, likely because they are important determinants of occupation and employment status. These selected covariates are also used in \cite{benjamin2003does} and \cite{gelber2011401} for propensity score estimation. Notably, our method identifies private health insurance as a strong predictor of 401(k) eligibility, a factor that has been overlooked in both prior studies. This finding highlights the importance of data-driven variable selection in uncovering economic relationships and enhancing model specification.

\subsubsection*{ATE Estimation and Balance Tests}
We first estimate $\psi^{k}$ by applying the estimators $\hat{\psi}_{3}$ and $\hat{\psi}_{4}$, as defined in Section \ref{sec:simu_2}, using the full sample. For all procedures, we standardize all continuous covariates and use the trimming rule $\hat{\omega} = \mathbf{1}(0.05 \leq \hat{m}(\boldsymbol{Z}^{\hat{*}}) \leq 0.95)$ to handle extreme propensity scores. For the MLP implementation, we use a neural network with three hidden layers containing $(64, 32, 16)$ neurons, respectively. The training is conducted with a learning rate of $10^{-5}$ and a batch size of 64. All other settings follow those in Section \ref{sec:simu_2}. These estimation results are reported in Table \ref{tab:app_estimation}. In addition, we consider two alternative estimators. First, we implement the H{\'a}jek form of the doubly robust estimator proposed in \cite{robins2007comment}. Second, we apply the DML procedure of \cite{ChernozhukovEtal2018} using a two-fold data partition $(I_{1}, I_{2})$. These results are presented in Tables \ref{tab:app_estimation_Hajek} and \ref{tab:app_estimation_split}, respectively, for comparison. 
 
A correctly specified propensity score, $m(\boldsymbol{Z}^{*})$, should balance the distribution of covariates between the treatment and control groups, i.e., $D\perp \boldsymbol{Z}^{*}|m(\boldsymbol{Z}^{*})$. Accordingly, a reliable estimator should yield approximately balanced samples within subclasses defined by the estimated propensity scores. Table \ref{tab:app_balance} reports the $p$-values from two-sample $t$-tests used to assess covariate balance within given subclasses for both the kernel and MLP propensity score estimators. Subclasses corresponding to intervals $[0, 0.1]$ and $[0.9, 1.0]$ are omitted due to trimming. An entry of ``NA'' indicates that the subclass contains too few treated observations to perform a meaningful test. The results show that the kernel estimator passes the balance test for most covariates and subclasses, and substantially outperforms the MLP estimator in this regard. In view of this, our subsequent interpretation of the empirical results focuses on the ATE estimates obtained using the kernel-based propensity score, as presented in Panel A of Tables \ref{tab:app_estimation}, \ref{tab:app_estimation_Hajek}, and \ref{tab:app_estimation_split}. 

\subsubsection*{Empirical Results and Concluding Remarks}
Our ATE estimates across different estimators yield largely similar results. As expected, 401(k) eligibility significantly increases contributions to 401(k) accounts. In addition, we find weak evidence suggesting that eligibility leads to increases in IRA savings, other financial assets, and secured debt, which indicates potential ``crowd-in'' effects. The estimated ATEs for these outcomes are only marginally significant. However, due to large standard errors and consequently wide confidence intervals, we cannot rule out the possibility of economically meaningful effects in these categories. Finally, we find no evidence that 401(k) eligibility significantly affects unsecured debt or car values in either a statistical or an economic sense. A practical limitation of our analysis is the imprecision of the ATE estimates, which may stem from the large noise in asset (liability) data, as reflected by the large standard deviations reported in Table \ref{tab:app_summary}.

The approach we illustrate here allows for unobserved heterogeneity and flexibly accommodates pre-treatment covariates to account for observed heterogeneity. This offers a more robust framework for estimating treatment effects compared to the common practice of using linear or logistic regressions with homogeneous coefficients. The results presented in Panel A of Table \ref{tab:app_summary} provide a ready example. The reported $p$-values essentially summarize estimation results from two-way fixed effects DID regressions without covariates. These regressions do not provide suggestive evidence for the potential ``crowd-in'' effects of 401(k) eligibility on other financial assets and secured debt that are detected by our proposed methods.\footnote{A direct comparison should be made with caution, as DID estimates the ATT, while our analysis focuses on the ATE. A more direct comparison would involve integrating our methods into doubly robust ATT estimation (e.g., \cite{shinozaki2015brief}) or modern doubly robust DID frameworks (e.g., \cite{sant2020doubly}). This is beyond the scope of this paper, and we leave it for future research.} This highlights the value of adopting a more robust, data-driven approach for uncovering important causal effects.

As a final remark, while our procedure is data-driven, it should be applied under the supervision of researchers with domain knowledge and a degree of conservatism. Once the assumed sparsity structure (like in Proposition \ref{Prop:refine}) in the data can be justified, as in our application, nonparametric estimation in fact becomes more straightforward to implement than many commonly used parametric approaches. Prior studies, such as \citet{benjamin2003does} and \citet{gelber2011401}, expend significant effort determining and justifying the best specification for their empirical models. This challenge is doubled in causal inference, which often requires specifying both a propensity score and an outcome regression model. A key advantage of our approach is therefore clear: by focusing on identifying the relevant variables, our method enables us to sidestep the need to argue for a specific functional form, thereby simplifying the empirical analysis while maintaining robustness.

\section{Conclusion} \label{SEC:conclusion} 
This paper introduces a novel method for nonparametric conditional density estimation in high-dimensional settings, addressing the curse of dimensionality through a computationally efficient approach. Rather than imposing strong distributional or functional form assumptions, our dimension reduction approach leverages a sparsity assumption that only a small subset of covariates is truly relevant. The core of our methodology is a screening procedure based on a new measure of conditional dependence, which achieves a favorable balance between small-sample statistical power and computational efficiency. Simulation studies show that it offers a substantial improvement over existing approaches in terms of recovery rates and processing time. Our method is particularly well-suited for applications with moderate sample sizes, where alternative methods may be either computationally costly (e.g., CDSIS by \cite{WangEtal2015}) or lack desirable small sample power (e.g., FOCI by \cite{azadkia2021simple}).

We further propose to refine the screening process using a modified CV algorithm that provides a data-driven approach to effectively eliminate irrelevant covariates through bandwidth selection, thereby refining the conditioning set for subsequent density estimation. Simulations demonstrate that our algorithm consistently outperforms the original CV procedure by \cite{HallRacineLi} in terms of exact recovery probability and numerical stability. Combined with the screening step as a whole, our proposed variable selection procedure provides a practical and theoretically grounded solution to the curse of dimensionality, which often hinders the application of nonparametric methods in complex, high-dimensional settings.

We demonstrate the practical utility of our dimension-reduced density estimator by applying it to the estimation of ATE with doubly robust methods. We show that our dimension reduction procedure offers a dual benefit in this context: it not only refines the set of variables for estimating the propensity score, mitigating the practical challenges posed by the trade-off between ignorability and overlap conditions, but also, through a refined ignorability condition, effectively reduces the dimensionality of the outcome regression models. Simulations confirm our method's potential to enhance the robustness and precision of causal inference in the presence of many covariates. 

We further illustrate the real-world applicability of our approach through an empirical analysis of the effect of 401(k) eligibility on savings. Aiming to address unobserved confounders and potential model misspecification issues in prior literature, this application underscores how our proposed dimension-reduced propensity score estimator and doubly robust ATE estimator can lead to more credible empirical findings.  

In conclusion, this paper provides a robust and efficient framework for nonparametric conditional density estimation in high-dimensional contexts. Beyond improving the performance and applicability of established causal inference methods, our approach can enhance practitioners’ ability to uncover data structures and address research questions. While we focus on ATE estimation, the proposed procedure offers a general-purpose toolkit for advancing a broad class of econometric and statistical models where understanding conditional distributions is essential.

\begin{center}
{\large\textbf{Acknowledgments }}{\large\par}
\par\end{center}
We are grateful for the valuable feedback and discussions provided by seminar participants at the University of Queensland, as well as conference attendees at the 19th International Symposium on Econometric Theory and Applications (SETA 2025). All errors are our responsibility.
\par\bigskip

\bibliography{references}
\newpage{}

\appendix

\begin{center}
{\LARGE\textbf{{}{}{}{}{}Supplementary Appendix}}{\LARGE{}{}{}{} }{\LARGE\par}
\par\end{center}


\section{Additional Results and Introduction of Some Existing Results\protect\protect\protect\protect\protect\protect\protect\protect\protect} \label{APP:Additional-Results}

\subsection{Dimension Reduction via Outcome Regression} \label{App:refine_Y}
The following results, based on a sparsity restriction on $Y$, are
symmetric to Proposition \ref{Prop:refine}. \begin{proposition}
\label{Prop:refine_2} Suppose Assumption \ref{A:3}(1) holds. Let
$(\boldsymbol{Z}^{*},\boldsymbol{\underline{Z}})$ be a partition
of $\boldsymbol{Z}$ such that $Y\perp\boldsymbol{\underline{Z}}|(D,\boldsymbol{Z}^{*})$.\footnote{As demonstrated in the proof, the results presented in Proposition
\ref{Prop:refine_2} rely essentially only on the sparsity condition
\[
\mathbb{E}[Y\mid D,\boldsymbol{Z}^{*},\boldsymbol{\underline{Z}}]=\mathbb{E}[Y\mid D,\boldsymbol{Z}^{*}],
\]
which states that $Y$ is mean independent of $\boldsymbol{\underline{Z}}$
conditional on $(D,\boldsymbol{Z}^{*})$. This condition is strictly
weaker than the full stochastic conditional independence assumption
$Y\perp\boldsymbol{\underline{Z}}\mid(D,\boldsymbol{Z}^{*})$.} Assume there exists a small constant $\epsilon\in(0,1/2)$ such that
$\epsilon<m(\boldsymbol{z}^{*})<1-\epsilon$ for all $\boldsymbol{z}^{*}\in\textrm{supp}(\boldsymbol{Z}^{*})$.
Then: 
\begin{enumerate}
\item[(1)] $g_{0}(\boldsymbol{Z})=\mathbb{E}[Y|D=0,\boldsymbol{Z}^{*}]=g_{0}(\boldsymbol{Z}^{*})$,
$g_{1}(\boldsymbol{Z})=\mathbb{E}[Y|D=1,\boldsymbol{Z}^{*}]=g_{1}(\boldsymbol{Z}^{*})$,
and 
\[
\psi=\mathbb{E}\left[g_{1}(\boldsymbol{Z}^{*})-g_{0}(\boldsymbol{Z}^{*})\right]=\mathbb{E}\left[\frac{DY}{m(\boldsymbol{Z}^{*})}-\frac{(1-D)Y}{1-m(\boldsymbol{Z}^{*})}\right],
\]
where $(g_{0},g_{1})$ are defined in (\ref{eq:outcome_reg}). 
\item[(2)] The ATE $\psi$ has the following doubly robust representation: 
\[
\psi=\mathbb{E}\left[g_{1}(\boldsymbol{Z}^{*})-g_{0}(\boldsymbol{Z}^{*})\right]+\mathbb{E}\left[\frac{D(Y-g_{1}(\boldsymbol{Z}^{*}))}{m(\boldsymbol{Z}^{*})}-\frac{(1-D)(Y-g_{0}(\boldsymbol{Z}^{*}))}{1-m(\boldsymbol{Z}^{*})}\right].
\]
\end{enumerate}
\end{proposition}

Proposition \ref{Prop:refine_2} is particularly relevant when there
are only a few covariates $\boldsymbol{Z}^{*}$ such that $Y\perp\boldsymbol{\underline{Z}}\mid(D,\boldsymbol{Z}^{*})$.
Practitioners can follow a procedure similar to that in Section \ref{sec:summary_procedure}
to achieve dimension reduction in estimating $\psi$. Specifically,
one can first apply our screening method to identify $\boldsymbol{Z}^{*}$
in the conditional independence condition $Y\perp\boldsymbol{\underline{Z}}\mid(D,\boldsymbol{Z}^{*})$
and then use the CV method proposed by \citet{HallEtal2007} for post-selection
estimation of $(g_{0},g_{1})$. \citet{HallEtal2007} parallels \citet{HallRacineLi},
employing the same CV approach to eliminate the impact of irrelevant
covariates in local polynomial estimations.\footnote{See also \citet{hall2015infinite}.}
It can be shown that the local polynomial estimators $(\hat{g}_{0},\hat{g}_{1})$
obtained through this procedure exhibit the same asymptotic properties
as estimators based solely on the truly relevant covariates $\boldsymbol{Z}^{*}$,
analogous to Proposition \ref{Prop:post} for kernel density estimation.
The proof follows similar arguments to those for Proposition \ref{Prop:post}
and is omitted for brevity.

We conclude this section by presenting the proof of Proposition \ref{Prop:refine_2}.

\noindent\textbf{Proof of Proposition \ref{Prop:refine_2}.} We begin
by noting that 
\begin{align*}
g_{1}(\boldsymbol{Z}) & =\mathbb{E}[Y(1)|\boldsymbol{Z}]=\mathbb{E}[Y(1)|D=1,\boldsymbol{Z}]=\mathbb{E}[Y|D=1,\boldsymbol{Z}]=\mathbb{E}[Y|D=1,\boldsymbol{Z}^{*},\boldsymbol{\underline{Z}}]\\
 & =\mathbb{E}[Y|D=1,\boldsymbol{Z}^{*}]=g_{1}(\boldsymbol{Z}^{*}),
\end{align*}
where the second equality follows from Assumption \ref{A:3}(1), the
third equality follows from $Y=Y(1)D+Y(0)(1-D)$, and the fifth equality
uses $Y\perp\boldsymbol{\underline{Z}}\mid(D,\boldsymbol{Z}^{*})$.
By analogous arguments, we can show $g_{0}(\boldsymbol{Z})=\mathbb{E}[Y|D=0,\boldsymbol{Z}^{*}]=g_{0}(\boldsymbol{Z}^{*})$.
Taking expectations, we then have 
\begin{equation}
\psi=\mathbb{E}[g_{1}(\boldsymbol{Z})-g_{0}(\boldsymbol{Z})]=\mathbb{E}[g_{1}(\boldsymbol{Z}^{*})-g_{0}(\boldsymbol{Z}^{*})].\label{eq:COR_refine_1}
\end{equation}
Next, given the overlap condition $0<m(\boldsymbol{Z}^{*})<1$, we
derive 
\begin{align}
 & \mathbb{E}[Y\mid D=1,\boldsymbol{Z}^{*}]=\frac{\mathbb{E}[DY\mid\boldsymbol{Z}^{*}]}{m(\boldsymbol{Z}^{*})}=\mathbb{E}\left[\left.\frac{D(Y-g_{1}(\boldsymbol{Z}^{*}))}{m(\boldsymbol{Z}^{*})}\right|\boldsymbol{Z}^{*}\right]+\mathbb{E}\left[\left.\frac{Dg_{1}(\boldsymbol{Z}^{*})}{m(\boldsymbol{Z}^{*})}\right|\boldsymbol{Z}^{*}\right]\nonumber \\
= & \mathbb{E}\left[\left.\frac{D(Y-g_{1}(\boldsymbol{Z}^{*}))}{m(\boldsymbol{Z}^{*})}\right|\boldsymbol{Z}^{*}\right]+\mathbb{E}\left[g_{1}(\boldsymbol{Z}^{*})\mid\boldsymbol{Z}^{*}\right]=\mathbb{E}\left[\left.g_{1}(\boldsymbol{Z}^{*})+\frac{D(Y-g_{1}(\boldsymbol{Z}^{*}))}{m(\boldsymbol{Z}^{*})}\right|\boldsymbol{Z}^{*}\right],\label{eq:COR_refine_2}
\end{align}
and similarly, 
\begin{equation}
\mathbb{E}[Y\mid D=0,\boldsymbol{Z}^{*}]=\frac{\mathbb{E}[(1-D)Y\mid\boldsymbol{Z}^{*}]}{1-m(\boldsymbol{Z}^{*})}=\mathbb{E}\left[\left.g_{0}(\boldsymbol{Z}^{*})+\frac{(1-D)(Y-g_{0}(\boldsymbol{Z}^{*}))}{1-m(\boldsymbol{Z}^{*})}\right|\boldsymbol{Z}^{*}\right].\label{eq:COR_refine_3}
\end{equation}
Substituting (\ref{eq:COR_refine_2}) and (\ref{eq:COR_refine_3})
into (\ref{eq:COR_refine_1}) and taking expectations, we obtain the
IPW and doubly robust estimands. \hfill{}$\square$

\subsection{An Introduction to \citet{HallRacineLi} \protect\protect\protect\protect\protect\protect\protect}

\label{APP:hall}

In this section, we provide more details on \citet{HallRacineLi},
specifically, how to implement (\ref{eq:ISE-1}) and its justification.
Using the notation from Section \ref{SEC:post-estimation}, we define the following technical terms: 
\begin{align*}
\hat{f}\left(\boldsymbol{x},y\right)= & n^{-1}\sum_{i=1}^{n}\Pi_{l=1}^{\tilde{s}_{1}}K_{h_{l}}\left(x_{li}^{\mathsf{c}}-x_{l}^{\mathsf{c}}\right)\cdot\Pi_{l=1}^{\tilde{s}_{2}}K_{\lambda_{l}}^{\mathsf{d}}\left(x_{li}^{\mathsf{d}},x_{l}^{\mathsf{d}}\right)\cdot K_{h}\left(y_{i}-y\right),\\
\hat{f}\left(\boldsymbol{x}\right)= & n^{-1}\sum_{i=1}^{n}\Pi_{l=1}^{\tilde{s}_{1}}K_{h_{l}}\left(x_{li}^{\mathsf{c}}-x_{l}^{\mathsf{c}}\right)\cdot\Pi_{l=1}^{\tilde{s}_{2}}K_{\lambda_{l}}^{\mathsf{d}}\left(x_{li}^{\mathsf{d}},x_{l}^{\mathsf{d}}\right),\text{ and}\\
\hat{G}\left(\boldsymbol{x}\right)= & n^{-2}\sum_{i_{1}=1}^{n}\sum_{i_{2}=1}^{n}\left\{ \Pi_{l=1}^{\tilde{s}_{1}}K_{h_{l}}\left(x_{li_{1}}^{\mathsf{c}}-x_{l}^{\mathsf{c}}\right)\cdot\Pi_{l=1}^{\tilde{s}_{2}}K_{\lambda_{l}}^{\mathsf{d}}\left(x_{li_{1}}^{\mathsf{d}},x_{l}^{\mathsf{d}}\right)\cdot\Pi_{l=1}^{\tilde{s}_{1}}K_{h_{l}}\left(x_{li_{2}}^{\mathsf{c}}-x_{l}^{\mathsf{c}}\right)\right.\\
 & \left.\cdot\Pi_{l=1}^{\tilde{s}_{2}}K_{\lambda_{l}}^{\mathsf{d}}\left(x_{li_{2}}^{\mathsf{d}},x_{l}^{\mathsf{d}}\right)\times\int K_{h}\left(y_{i_{1}}-y\right)K_{h}\left(y_{i_{2}}-y\right)dy\right\} .
\end{align*}
The cross-validation criterion is defined as 
\begin{equation}
\text{CV}\left(\hat{h},\hat{h}_{1},...\hat{h}_{\tilde{s}_{1}},\hat{\lambda}_{1},...,\hat{\lambda}_{\tilde{s}_{2}}\right)=\hat{I}_{n1}\left(\hat{h},\hat{h}_{1},...\hat{h}_{\tilde{s}_{1}},\hat{\lambda}_{1},...,\hat{\lambda}_{\tilde{s}_{2}}\right)-2\hat{I}_{n2}\left(\hat{h},\hat{h}_{1},...\hat{h}_{\tilde{s}_{1}},\hat{\lambda}_{1},...,\hat{\lambda}_{\tilde{s}_{2}}\right),\label{EQ:CV}
\end{equation}
where $\hat{I}_{n1}$ and $\hat{I}_{n2}$ are computed as 
\begin{align*}
\hat{I}_{n1}\left(\hat{h},\hat{h}_{1},...\hat{h}_{\tilde{s}_{1}},\hat{\lambda}_{1},...,\hat{\lambda}_{\tilde{s}_{2}}\right) & =\frac{1}{n}\sum_{i=1}^{n}\frac{\hat{G}_{-i}\left(\boldsymbol{x}_{i}\right)}{\hat{f}_{-i}\left(\boldsymbol{x}_{i}\right)^{2}}\text{ and }\\
\hat{I}_{n2}\left(\hat{h},\hat{h}_{1},...\hat{h}_{\tilde{s}_{1}},\hat{\lambda}_{1},...,\hat{\lambda}_{\tilde{s}_{2}}\right) & =\frac{1}{n}\sum_{i=1}^{n}\frac{\hat{f}_{-i}\left(\boldsymbol{x}_{i},y_{i}\right)}{\hat{f}_{-i}\left(\boldsymbol{x}_{i}\right)},
\end{align*}
with $\hat{G}_{-i}$ and $\hat{f}_{-i}$ being the leave-one-out (leaving
the $i$-th observation out) estimators. Weighting functions used
in \citet{HallRacineLi} to address issues with near-zero denominators
are omitted for conciseness.

In practice, the optimal bandwidth is calculated as 
\begin{equation}
\left(\hat{h},\hat{h}_{1},...\hat{h}_{\tilde{s}_{1}},\hat{\lambda}_{1},...,\hat{\lambda}_{\tilde{s}_{2}}\right)=\arg\min_{h,h_{1},...h_{\tilde{s}_{1}},\lambda_{1},...,\lambda_{\tilde{s}_{2}}}\text{CV}\left(h,h_{1},...h_{\tilde{s}_{1}},\lambda_{1},...,\lambda_{\tilde{s}_{2}}\right).\label{EQ:bandwidth}
\end{equation}
The CV serves as a valid proxy for the ISE defined in (\ref{eq:ISE_define}).
To illustrate the connection between the CV criterion and the ISE,
we can decompose the ISE as follows: 
\[
\text{ISE}=I_{1n}-2I_{2n}+I_{3n},
\]
where 
\begin{align*}
I_{1n} & =\int\hat{f}\left(y|\boldsymbol{x}\right)^{2}f\left(\boldsymbol{x}\right)d\boldsymbol{x}dy,\\
 & =\int\left[\int\hat{f}\left(y,\boldsymbol{x}\right)^{2}dy\right]\frac{f\left(\boldsymbol{x}\right)}{\hat{f}\left(\boldsymbol{x}\right)^{2}}d\boldsymbol{x}=\int\frac{\hat{G}\left(\boldsymbol{x}\right)}{\hat{f}\left(\boldsymbol{x}\right)^{2}}f\left(\boldsymbol{x}\right)d\boldsymbol{x},\\
I_{2n} & =\int\hat{f}\left(y|\boldsymbol{x}\right)f\left(y,\boldsymbol{x}\right)d\boldsymbol{x}dy=\int\frac{\hat{f}\left(y,\boldsymbol{x}\right)}{\hat{f}\left(\boldsymbol{x}\right)}f\left(y,\boldsymbol{x}\right)d\boldsymbol{x}dy,
\end{align*}
and $I_{3n}$ is irrelevant due to its independence of the tuning
parameters$.$

Clearly, $\hat{I}_{1n}$ and $\hat{I}_{2n}$ are the leave-one-out
sample analogs of $I_{1n}$ and $I_{2n}$, respectively. $I_{3n}$
can be disregarded because it does not depend on tuning parameters.

\section{Algorithm: Computation of $\rho$-values}\label{APP:compute_Rho}

We present an efficient computation algorithm for $\rho$ in this appendix. We assume $X$ is continuous without loss of generality, as the case where $X$ is discrete can be handled similarly. The key innovation lies in Steps 2 and 3, which reduce the computational cost of building the conditional CDFs from a naive $O(n^3)$ to a more manageable $O(n^2)$.

\subsection*{Step 1: Compute Kernel Weights (Cost: $O(n^2)$)}
For each pair of observations $(i, j)$, we compute $b_{ji}$ and $b_{ji}^*$ defined as:
\[
b_{ji}=K_{h}(y_{j}-y_{i})\left[\prod_{l=1}^{q^{\mathsf{c}}}K_{h}(w_{lj}^{\mathsf{c}}-w_{li}^{\mathsf{c}})\right]\left[\prod_{l=1}^{q^{\mathsf{d}}}K_{\lambda}^{\mathsf{d}}(w_{lj}^{\mathsf{d}},w_{li}^{\mathsf{d}})\right]
\]
and
\[
b_{ji}^{*}=K_{h}(y^{*}-y_{i})\left[\prod_{l=1}^{q^{\mathsf{c}}}K_{h}(w_{lj}^{\mathsf{c}}-w_{li}^{\mathsf{c}})\right]\left[\prod_{l=1}^{q^{\mathsf{d}}}K_{\lambda}^{\mathsf{d}}(w_{lj}^{\mathsf{d}},w_{li}^{\mathsf{d}})\right].
\]
Next, for each conditioning observation $j=1,...,n$, we compute the normalizing sums $b_{j\cdot}=\sum_{i=1}^{n}b_{ji}$ and $b_{j\cdot}^{*}=\sum_{i=1}^{n}b_{ji}^{*}$. Finally, we construct two $n \times n$ weight matrices, $\boldsymbol{\omega}$ and $\boldsymbol{\omega}^*$, with elements $\omega_{ji} = b_{ji}/b_{j\cdot}$ and $\omega_{ji}^{*} = b_{ji}^{*}/b_{j\cdot}^{*}$, respectively. The total computation cost for this step is $O(n^2)$. 

\subsection*{Step 2: Sorting (Cost: $O(n \log n)$)}
We sort the realized values of $X$ in ascending order: $x_{(1)} < x_{(2)} < \dots < x_{(n)}$. We assume no ties for simplicity as $X$ is continuous.\footnote{The cases where ``ties'' occur or $X$ is a discrete variable can be similarly handled but with more tedious
notation.} Recall that the $i$-th column in the matrices $\boldsymbol{\omega}$ and $\boldsymbol{\omega}^{*}$
corresponds to observation $i$. We then reorder the columns of the matrices $\boldsymbol{\omega}$ and $\boldsymbol{\omega}^*$ according to this sort order of $X$, yielding the following column-shuffled matrix, denoted by $\boldsymbol{\omega}_{()}$: 
\[
\boldsymbol{\omega}_{\left(\right)}=\left(\begin{array}{cccc}
\omega_{1\left(1\right)} & \omega_{1\left(2\right)} & \cdots & \omega_{1\left(n\right)}\\
\omega_{2\left(1\right)} & \omega_{2\left(2\right)} & \cdots & \omega_{2\left(n\right)}\\
\vdots & \vdots & \ddots & \vdots\\
\omega_{n\left(1\right)} & \omega_{n\left(2\right)} & \cdots & \omega_{n\left(n\right)}
\end{array}\right).
\]
Note that the element in the $j$-th row and $i$-th column of the matrix $\boldsymbol{\omega}_{()}$ is $\omega_{j\left(i\right)}=b_{j\left(i\right)}/b_{j\cdot}$, where the subscript $(i)$ refers to the observation corresponding to the $i$-th sorted value, $x_{(i)}$. We similarly define $\boldsymbol{\omega}_{\left(\right)}^{*}$
with $\omega_{j\left(i\right)}^{*}=b_{j\left(i\right)}^{*}/b_{j\cdot}^{*}$. The computation cost of sorting is proportional to $n\log n$. 

\subsection*{Step 3: Calculate $\hat{F}_{n}$ for Each Observation (Cost: $O(n^2)$)}
With the columns sorted by $X$, we can efficiently compute the conditional cumulative distribution functions (CDFs) for each conditioning observation $j=1,...,n$ by taking the cumulative sum (cumsum) across the rows of the shuffled weight matrices. For each row $j$, the CDF of $X$ conditional on $(y_j, \boldsymbol{w}_j)$ evaluated at the sorted points $x_{(i)}$ is:
\[
\hat{F}_{n}(x_{(i)} | y_j, \boldsymbol{w}_j) = \sum_{l=1}^{i} \omega_{j(l)}.
\]
Take the first row $(\omega_{1(1)}\textrm{ }\omega_{1(2)}\textrm{ }\cdots\textrm{ }\omega_{1(n)})$ of $\boldsymbol{\omega}_{\left(\right)}$ as an example. We can obtain 
\begin{align*}
\hat{F}_{n}\left(x_{\left(1\right)}|y_{1},\boldsymbol{w}_{1}\right) & =\frac{b_{1\left(1\right)}}{b_{1\cdot}}=\omega_{1\left(1\right)},\\
\hat{F}_{n}\left(x_{\left(2\right)}|y_{1},\boldsymbol{w}_{1}\right) & =\frac{b_{1\left(1\right)}+b_{1\left(2\right)}}{b_{1\cdot}}=\omega_{1\left(1\right)}+\omega_{1\left(2\right)},\\
 & \vdots\\
\hat{F}_{n}\left(x_{\left(n\right)}|y_{1},\boldsymbol{w}_{1}\right) & =\frac{b_{1\left(1\right)}+b_{1\left(2\right)}+...+b_{1\left(n\right)}}{b_{1\cdot}}=\omega_{1\left(1\right)}+\omega_{1\left(2\right)}+...+\omega_{1\left(n\right)}.
\end{align*}
Since $\hat{F}_{n}(x_{(i+1)} | y_j, \boldsymbol{w}_j) = \hat{F}_{n}(x_{(i)} | y_j, \boldsymbol{w}_j) + \omega_{j(i+1)}$, the CDF for each conditioning observation $j$ (corresponding to the $j$-th row) can be computed in $O(n)$ operations. Repeating this for all $n$ rows gives a total cost of $O(n^2)$. We similarly compute $\hat{F}_{n}(x_{(i)} | y^*, \boldsymbol{w}_j)$ for all $(j, i)$ using the matrix $\boldsymbol{\omega}^*_{()}$.

\subsection*{Step 4: Compute Empirical Measure $\hat{\rho}$ (Cost: $O(n^2)$)}
Finally, for each conditioning observation $j=1,...,n$, we compute
\[
\hat{\Lambda}_{X}\left(y_{j},\boldsymbol{w}_{j}\right)=\max_{i=1,...,n}\left|\hat{F}_{n}\left(x_{\left(i\right)}|y_{j},\boldsymbol{w}_{j}\right)-\hat{F}_{n}\left(x_{\left(i\right)}|y^{*},\boldsymbol{w}_{j}\right)\right|,
\]
and then
\[
\hat{\rho}=\frac{1}{n}\sum_{j=1}^{n}\hat{\Lambda}_{X}\left(y_{j},\boldsymbol{w}_{j}\right).
\]
The computational cost of this final step is proportional to $n^{2}$.

\section{Main Proofs} \label{APP:mainproof}

\noindent\textbf{Proof of Theorem \ref{TH:TPR}.} Lemma \ref{LE:boundf}
underpins the proof of this theorem. It suffices to show that 
\[
\Pr\left(\cap_{j=1}^{s^{*}}\left\{ \hat{\rho}_{j}\geq Cn^{-\frac{r}{2r+q^{\mathsf{c}}+1}}\log n\right\} \cap_{j=s^{*}+1}^{p}\left\{ \hat{\rho}_{j}<Cn^{-\frac{r}{2r+q^{\mathsf{c}}+1}}\log n\right\} \right)\rightarrow1.
\]

For a relevant $X_{j}$, $j=1,...,s^{*}$, with $\rho_{j}\gtrsim\left(\log n\right)^{1+c}n^{-\frac{r}{2r+q^{\mathsf{c}}+1}}$,
Lemma \ref{LE:boundf} implies that for sufficiently large $n$ 
\begin{align}
\Pr\left(\hat{\rho}_{j}<Cn^{-\frac{r}{2r+q^{\mathsf{c}}+1}}\log n\right) & \leq\Pr\left(\hat{\rho}_{j}<\frac{\rho_{j}}{2}\right)\leq\Pr\left(\left|\hat{\rho}_{j}-\rho_{j}\right|>\frac{\rho_{j}}{2}\right)\nonumber \\
 & \leq\Pr\left(\left|\hat{\rho}_{j}-\rho_{j}\right|>Cn^{-\frac{r}{2r+q^{\mathsf{c}}+1}}\log n\right)\apprle n^{-M},\label{eq:rho_bound1}
\end{align}
for a fixed large $M,$ where the above derivation repeatedly uses
$\rho_{j}\gg n^{-\frac{r}{2r+q^{\mathsf{c}}+1}}\log n.$

For an irrelevant $X_{j}$, $j=s^{*}+1,...,p$, with $\rho_{j}=0,$
Lemma \ref{LE:boundf} implies that 
\begin{equation}
\Pr\left(\hat{\rho}_{j}\geq Cn^{-\frac{r}{2r+q^{\mathsf{c}}+1}}\log n\right)=\Pr\left(\left|\hat{\rho}_{j}-\rho_{j}\right|\geq Cn^{-\frac{r}{2r+q^{\mathsf{c}}+1}}\log n\right)\apprle n^{-M}.\label{eq:rho_bound2}
\end{equation}

Using the results in (\ref{eq:rho_bound1}) and (\ref{eq:rho_bound2})
by taking $M>B_{p}$ and Assumption \ref{A:signal_strength}(2), we
obtain 
\begin{align*}
 & \Pr\left(\cap_{j=1}^{s^{*}}\left\{ \hat{\rho}_{j}\geq Cn^{-\frac{r}{2r+q^{\mathsf{c}}+1}}\log n\right\} \cap_{j=s^{*}+1}^{p}\left\{ \hat{\rho}_{j}<Cn^{-\frac{r}{2r+q^{\mathsf{c}}+1}}\log n\right\} \right)\\
= & 1-\Pr\left(\cup_{j=1}^{s^{*}}\left\{ \hat{\rho}_{j}<Cn^{-\frac{r}{2r+q^{\mathsf{c}}+1}}\log n\right\} \cup_{j=s^{*}+1}^{p}\left\{ \hat{\rho}_{j}\geq Cn^{-\frac{r}{2r+q^{\mathsf{c}}+1}}\log n\right\} \right)\\
\geq & 1-\sum_{j=1}^{s^{*}}\Pr\left(\hat{\rho}_{j}<Cn^{-\frac{r}{2r+q^{\mathsf{c}}+1}}\log n\right)-\sum_{j=s^{*}+1}^{p}\Pr\left(\hat{\rho}_{j}\geq Cn^{-\frac{r}{2r+q^{\mathsf{c}}+1}}\log n\right)\\
= & 1-O\left(n^{-M}p\right)\rightarrow1,
\end{align*}
as desired. 

\hfill{}$\square$

\noindent\textbf{Proof of Proposition \ref{Prop:post}.} Under the
unambiguity condition in Assumption \ref{A:2}(1), the ISE does not
possess multiple local minima corresponding to alternative instances
of conditional independence (e.g., $Y\perp\boldsymbol{X}_{1}\left|\boldsymbol{X}_{2}\right.$
for any $\boldsymbol{X}_{1}\neq\left(X_{s^{**}+1},...,X_{\tilde{p}}\right)$
and $\boldsymbol{X}_{2}\neq\boldsymbol{Z}^{*}$). Consequently, the
population minimum of the $\text{ISE}$ is attained at the desired
bandwidths.

Theorem \ref{TH:TPR} implies that the following event 
\[
\hat{E}=\left\{ \min_{1\leq j\leq s^{*}}\hat{\rho}_{j}\geq Cn^{-r/\left(2r+q^{\mathsf{c}}+1\right)}\log n\geq\max_{s^{*}+1\leq j\leq p}\hat{\rho}_{j}\right\} ,
\]
occurs with probability approaching 1 as $n\rightarrow\infty,$ for
a fixed positive $C$. Consequently, given that $\tilde{p}-q^{\mathsf{c}}-q^{\mathsf{d}}\geq s^{*}$
as per Assumption \ref{A:2}(2), our first-stage screening retains
all relevant covariates in $\boldsymbol{\tilde{X}}$ with high probability,
since their $\hat{\rho}$ values exceed those of irrelevant covariates,
as stated in $\hat{E}$.

 In what follows, we assume the occurrence of event $\hat{E}$. This
allows us to proceed under conditions almost identical to those employed
in \citet{HallRacineLi} and to temporarily disregard the ``variable
selection'' issue, which will be revisited at the end of the proof.

Without loss of generality, let $(X_{1}^{\mathsf{c}},...,X_{p^{\mathsf{c}*}}^{\mathsf{c}})$
and $(X_{1}^{\mathsf{d}},...,X_{p^{\mathsf{d}*}}^{\mathsf{d}})$
denote the continuous and discrete relevant covariates in $\boldsymbol{Z}^{*}$.
Thus, \(\boldsymbol{Z}^{*}\equiv(X_{1}^{\mathsf{c}},...,X_{p^{\mathsf{c}*}}^{\mathsf{c}},X_{1}^{\mathsf{d}},...,X_{p^{\mathsf{d}*}}^{\mathsf{d}})'\).

Our proof employs arguments analogous to those utilized in the proofs
of Theorems 2 and 3 of \citet{HallRacineLi}\footnote{A more detailed proof can be found in the \href{https://citeseerx.ist.psu.edu/document?repid=rep1&type=pdf&doi=d70d7990a2165f78456b28d4cb04cbf9188dcd9d}{working paper} version
of \citet{HallRacineLi}.}, with only two discrepancies necessitating adjustments. Firstly, the primary distinction (aside from the first-stage screening) lies in that \citet{HallRacineLi}'s proofs rely on an independence assumption
different from our Assumption \ref{A:2}(1). Resolving this discrepancy
necessitates the establishment of (\ref{eq:condition_f_equi}) and
(\ref{eq:mu_g}) presented below. Another minor difference is the
use of a second-order kernel in \citet{HallRacineLi}, compared to
our use of an $r$-th order kernel. This extension is standard and
straightforward, and thus omitted for brevity. Upon addressing these
two issues, the remaining arguments from \citet{HallRacineLi} proceed.
In summary, it suffices to prove 
\begin{equation}
f\left(Y|X_{1},...,X_{\tilde{p}}\right)=f\left(Y|X_{1},...,X_{s^{**}},\boldsymbol{W}\right)\equiv f\left(Y|\boldsymbol{Z}^{*}\right),\label{eq:condition_f_equi}
\end{equation}
and 
\begin{align}
\mu_{g}\left(y\left|x_{1},...,x_{\tilde{p}}\right.\right)-\mu_{g}\left(y\left|x_{1},...,x_{s^{**}},\boldsymbol{w}\right.\right) & \equiv\mu_{g}\left(y\left|x_{1},...,x_{\tilde{p}}\right.\right)-\mu_{g}\left(y\left|\boldsymbol{z}^{*}\right.\right)=O\left(\left(nh_{1}...h_{p^{\mathsf{c}*}}\right)^{-1}\right),\label{eq:mu_g}
\end{align}
for 
\begin{equation}
h_{1}^{r},...,h_{p^{\mathsf{c}*}}^{r}\lesssim\left(nh_{1}...h_{p^{\mathsf{c}*}}\right)^{-1}\text{ and }\lambda_{1},...,\lambda_{p^{\mathsf{d}*}}\lesssim\left(nh_{1}...h_{p^{\mathsf{c}*}}\right)^{-1},\label{eq:hlamda_restrict}
\end{equation}
where 
\begin{align*}
\mu_{g}\left(y\left|x_{1},...,x_{\tilde{p}}\right.\right) & \equiv\frac{\mu_{f}\left(x_{1},...,x_{\tilde{p}},y\right)}{\mu_{m}\left(x_{1},...,x_{\tilde{p}}\right)},\\
\mu_{f}\left(x_{1},...,x_{\tilde{p}},y\right) & \equiv\mathbb{E}\left[\Pi_{l=1}^{\tilde{s}_{1}}K_{h_{l}}\left(X_{l}^{\mathsf{c}}-x_{l}^{\mathsf{c}}\right)\cdot\Pi_{l=1}^{\tilde{s}_{2}}K_{\lambda_{l}}^{\mathsf{d}}\left(X_{l}^{\mathsf{d}},x_{l}^{\mathsf{d}}\right)\cdot K_{h}\left(Y-y\right)\right],\\
\mu_{m}\left(x_{1},...,x_{\tilde{p}}\right) & \equiv\mathbb{E}\left[\Pi_{l=1}^{\tilde{s}_{1}}K_{h_{l}}\left(X_{l}^{\mathsf{c}}-x_{l}^{\mathsf{c}}\right)\cdot\Pi_{l=1}^{\tilde{s}_{2}}K_{\lambda_{l}}^{\mathsf{d}}\left(X_{l}^{\mathsf{d}},x_{l}^{\mathsf{d}}\right)\right],
\end{align*}
and $\mu_{g}\left(y\left|\boldsymbol{z}^{*}\right.\right)$ is similarly
defined.

First, note that (\ref{eq:condition_f_equi}) follows from 
\[
f\left(Y|X_{1},...,X_{\tilde{p}}\right)=f\left(Y|X_{1},...,X_{s^{*}},\boldsymbol{W}\right)
\]
by $\left(Y,X_{1},...,X_{s^{*}},\boldsymbol{W}\right)\perp\left(X_{s^{*}+1},...,X_{p}\right)$,
and 
\begin{align*}
f\left(Y|X_{1},...,X_{s^{*}},\boldsymbol{W}\right) & =\frac{f\left(Y,X_{s^{**}+1},...,X_{s^{*}}|X_{1},...,X_{s^{**}},\boldsymbol{W}\right)}{f\left(X_{s^{**}+1},...,X_{s^{*}}|X_{1},...,X_{s^{**}},\boldsymbol{W}\right)},\\
 & =\frac{f\left(Y|X_{1},...,X_{s^{**}},\boldsymbol{W}\right)f\left(X_{s^{**}+1},...,X_{s^{*}}|X_{1},...,X_{s^{**}},\boldsymbol{W}\right)}{f\left(X_{s^{**}+1},...,X_{s^{*}}|X_{1},...,X_{s^{**}},\boldsymbol{W}\right)}\\
 & =f\left(Y|\boldsymbol{Z}^{*}\right),
\end{align*}
due to $Y\perp\left(X_{s^{**}+1},...,X_{s^{*}}\right)\left|\left(X_{1},...,X_{s^{**}},\boldsymbol{W}\right)\right.$
and $\boldsymbol{Z}^{*}=\left(X_{1},...,X_{s^{**}},\boldsymbol{W}\right)'.$

To establish (\ref{eq:mu_g}), we first define 
\[
Q\left(\boldsymbol{z}^{*}\right)\equiv\mathbb{E}\left[\left.\Pi_{l=p^{\mathsf{c}*}+1}^{\tilde{s}_{1}}K_{h_{l}}\left(X_{l}^{\mathsf{c}}-x_{l}^{\mathsf{c}}\right)\cdot\Pi_{l=p^{\mathsf{d}*}+1}^{\tilde{s}_{2}}K_{\lambda_{l}}^{\mathsf{d}}\left(X_{l}^{\mathsf{d}},x_{l}^{\mathsf{d}}\right)\right|\boldsymbol{Z}^{*}=\boldsymbol{z}^{*}\right].
\]
Then, applying the law of iterated expectations and $Y\perp\left(X_{s^{**}+1},...,X_{s^{*}}\right)\left|\boldsymbol{Z}^{*}\right.$
yields 
\begin{align*}
\mu_{f}\left(x_{1},...,x_{\tilde{p}},y\right) & =\mathbb{E}\left\{ \Pi_{l=1}^{p^{\mathsf{c}*}}K_{h_{l}}\left(X_{l}^{\mathsf{c}}-x_{l}^{\mathsf{c}}\right)\cdot\Pi_{l=1}^{p^{\mathsf{d}*}}K_{\lambda_{l}}^{\mathsf{d}}\left(X_{l}^{\mathsf{d}},x_{l}^{\mathsf{d}}\right)\cdot\mathbb{E}\left[\left.K_{h}\left(Y-y\right)\right|\boldsymbol{Z}^{*}\right]Q\left(\boldsymbol{Z}^{*}\right)\right\} \\
 & =\mathbb{E}\left[\left.K_{h}\left(Y-y\right)\right|\boldsymbol{Z}^{*}=\boldsymbol{z}^{*}\right]Q\left(\boldsymbol{z}^{*}\right)f\left(\boldsymbol{z}^{*}\right)+O\left(\sum_{l=1}^{p^{\mathsf{c}*}}h_{l}^{r}+\sum_{l=1}^{p^{\mathsf{d}*}}\lambda_{l}\right),
\end{align*}
where the second equality is obtained through standard Taylor expansion.
Similarly, we have 
\[
\mu_{m}\left(x_{1},...,x_{\tilde{p}}\right)=Q\left(\boldsymbol{z}^{*}\right)f\left(\boldsymbol{z}^{*}\right)+O\left(\sum_{l=1}^{p^{\mathsf{c}*}}h_{l}^{r}+\sum_{l=1}^{p^{\mathsf{d}*}}\lambda_{l}\right).
\]
Combining these results, and given that $f\left(\boldsymbol{z}^{*}\right)$ is uniformly bounded away from 0, we deduce 
\begin{align*}
\mu_{g}\left(y\left|x_{1},...,x_{\tilde{p}}\right.\right) & =\frac{\mu_{f}\left(x_{1},...,x_{\tilde{p}},y\right)}{\mu_{m}\left(x_{1},...,x_{\tilde{p}},y\right)}\\
 & =\mathbb{E}\left[\left.K_{h}\left(Y-y\right)\right|\boldsymbol{Z}^{*}=\boldsymbol{z}^{*}\right]+O\left(\sum_{l=1}^{p^{\mathsf{c}*}}h_{l}^{r}+\sum_{l=1}^{p^{\mathsf{d}*}}\lambda_{l}\right).
\end{align*}
Similar calculations yield 
\begin{align*}
\mu_{g}\left(y\left|\boldsymbol{z}^{*}\right.\right) & =\mathbb{E}\left[\left.K_{h}\left(Y-y\right)\right|\boldsymbol{Z}^{*}=\boldsymbol{z}^{*}\right]+O\left(\sum_{l=1}^{p^{\mathsf{c}*}}h_{l}^{r}+\sum_{l=1}^{p^{\mathsf{d}*}}\lambda_{l}\right).
\end{align*}
Therefore, (\ref{eq:hlamda_restrict}) guarantees 
\[
\mu_{g}\left(y\left|x_{1},...,x_{\tilde{p}}\right.\right)-\mu_{g}\left(y\left|\boldsymbol{z}^{*}\right.\right)=O\left(\sum_{l=1}^{p^{\mathsf{c}*}}h_{l}^{r}+\sum_{l=1}^{p^{\mathsf{d}*}}\lambda_{l}\right)=O\left(\left(nh_{1}...h_{p^{\mathsf{c}*}}\right)^{-1}\right),
\]
as desired.

Finally, the results in \citet{HallRacineLi} hold in probability.
Specifically, let $A$ denote the desired results established in \citet{HallRacineLi}.
They show $\Pr\left(A\right)\rightarrow1$. We have shown above that
\[
\Pr\left(A\left|\hat{E}\right.\right)\rightarrow1,
\]
conditional on the occurrence of event $\hat{E}$. Since $\Pr(\hat{E})\rightarrow1$
by Theorem \ref{TH:TPR}, we can remove the conditional set as follows:
\[
\Pr\left(A\right)=\Pr\left(A\left|\hat{E}\right.\right)\Pr\left(\hat{E}\right)+\Pr\left(A\left|\hat{E}^{c}\right.\right)\Pr\left(\hat{E}^{c}\right)\geq\Pr\left(A\left|\hat{E}\right.\right)\Pr\left(\hat{E}\right)\rightarrow1,
\]
as desired. 

\hfill{}$\square$

\noindent\textbf{Proof of Proposition \ref{Prop:refine}.} Let $F_{\boldsymbol{\underline{Z}}|\boldsymbol{Z}^{*}=\boldsymbol{z}^{*}}$
denote the CDF of $\boldsymbol{\underline{Z}}$ conditional on $\boldsymbol{Z}^{*}=\boldsymbol{z}^{*}$.
Assume without loss of generality that $Y(0)$ and $Y(1)$ are discrete random variables
to simplify the exposition. For any $(y_{0},y_{1})\in\textrm{supp}((Y(0),Y(1)))$,
$d\in\{0,1\}$, and $\boldsymbol{z}^{*}\in\textrm{supp}(\boldsymbol{Z}^{*})$, we can write 
\begin{align*}
 & \Pr((Y(0),Y(1))=(y_{0},y_{1}),D=d|\boldsymbol{Z}^{*}=\boldsymbol{z}^{*})\\
= & \int\Pr((Y(0),Y(1))=(y_{0},y_{1}),D=d|\boldsymbol{Z}^{*}=\boldsymbol{z}^{*},\boldsymbol{\underline{Z}}=\boldsymbol{\underline{z}})dF_{\boldsymbol{\underline{Z}}|\boldsymbol{Z}^{*}=\boldsymbol{z}^{*}}(\boldsymbol{\underline{z}})\\
= & \int\Pr((Y(0),Y(1))=(y_{0},y_{1})|\boldsymbol{Z}^{*}=\boldsymbol{z}^{*},\boldsymbol{\underline{Z}}=\boldsymbol{\underline{z}})\Pr(D=d|\boldsymbol{Z}^{*}=\boldsymbol{z}^{*},\boldsymbol{\underline{Z}}=\boldsymbol{\underline{z}})dF_{\boldsymbol{\underline{Z}}|\boldsymbol{Z}^{*}=\boldsymbol{z}^{*}}(\boldsymbol{\underline{z}})\\
= & \int\Pr((Y(0),Y(1))=(y_{0},y_{1})|\boldsymbol{Z}^{*}=\boldsymbol{z}^{*},\boldsymbol{\underline{Z}}=\boldsymbol{\underline{z}})\Pr(D=d|\boldsymbol{Z}^{*}=\boldsymbol{z}^{*})dF_{\boldsymbol{\underline{Z}}|\boldsymbol{Z}^{*}=\boldsymbol{z}^{*}}(\boldsymbol{\underline{z}})\\
= & \left(\int\Pr((Y(0),Y(1))=(y_{0},y_{1})|\boldsymbol{Z}^{*}=\boldsymbol{z}^{*},\boldsymbol{\underline{Z}}=\boldsymbol{\underline{z}})dF_{\boldsymbol{\underline{Z}}|\boldsymbol{Z}^{*}=\boldsymbol{z}^{*}}(\boldsymbol{\underline{z}})\right)\Pr(D=d|\boldsymbol{Z}^{*}=\boldsymbol{z}^{*})\\
= & \Pr((Y(0),Y(1))=(y_{0},y_{1})|\boldsymbol{Z}^{*}=\boldsymbol{z}^{*})\Pr(D=d|\boldsymbol{Z}^{*}=\boldsymbol{z}^{*}),
\end{align*}
where the second equality follows from Assumption \ref{A:3}(1) and
the third equality follows from $D\perp\boldsymbol{\underline{Z}}|\boldsymbol{Z}^{*}$.
This establishes that $(Y(0),Y(1))\perp D|\boldsymbol{Z}^{*}$. When
$(Y(0),Y(1))\perp D\mid\boldsymbol{Z}^{*}$, combined with the overlap
condition $0<m(\boldsymbol{Z}^{*})<1$, the outcome regression and
IPW estimands in (\ref{eq:refined_ipw}) can be derived using standard
arguments from the literature. Thus, we omit the details here.

In what follows, we establish the doubly robust estimand in Remark \ref{remark:general}:
\begin{align}
\psi= & \mathbb{E}\left[g_{1}(\boldsymbol{Z}^{*})-g_{0}(\boldsymbol{Z}^{*})\right]\label{eq:refine_estimand_ext}\\
= & \mathbb{E}\left[g_{1}(\boldsymbol{Z}^{*},s(\boldsymbol{\underline{Z}}))-g_{0}(\boldsymbol{Z}^{*},s(\boldsymbol{\underline{Z}}))\right]+\mathbb{E}\left[\frac{D(Y-g_{1}(\boldsymbol{Z}^{*},s(\boldsymbol{\underline{Z}})))}{m(\boldsymbol{Z}^{*})}-\frac{(1-D)(Y-g_{0}(\boldsymbol{Z}^{*},s(\boldsymbol{\underline{Z}})))}{1-m(\boldsymbol{Z}^{*})}\right],\nonumber 
\end{align}
which generalizes the estimand (\ref{eq:refined_estimand}) ($s(\boldsymbol{\underline{Z}})=\emptyset$)
by allowing for a nontrivial function $s(\boldsymbol{\underline{Z}})$.

By definition and the law of iterated expectations (LIE), we obtain
\[
g_{1}(\boldsymbol{Z}^{*})=\mathbb{E}[Y|D=1,\boldsymbol{Z}^{*}]=\mathbb{E}\left\{ \mathbb{E}[Y|D=1,\boldsymbol{Z}^{*},s(\boldsymbol{\underline{Z}})]|D=1,\boldsymbol{Z}^{*}\right\} =\mathbb{E}\left[g_{1}(\boldsymbol{Z}^{*},s(\boldsymbol{\underline{Z}}))|D=1,\boldsymbol{Z}^{*}\right].
\]
Taking expectations on both sides yields $\mathbb{E}[g_{1}(\boldsymbol{Z}^{*})]=\mathbb{E}\left[g_{1}(\boldsymbol{Z}^{*},s(\boldsymbol{\underline{Z}}))\right]$.
Next, by LIE and $\Pr(D=1|\boldsymbol{Z}^{*},s(\boldsymbol{\underline{Z}}))=\Pr(D=1|\boldsymbol{Z}^{*})=m(\boldsymbol{Z}^{*})$
due to $D\perp\boldsymbol{\underline{Z}}|\boldsymbol{Z}^{*}$, it
follows that 
\begin{align}
 & \mathbb{E}\left[\frac{D(Y-g_{1}(\boldsymbol{Z}^{*},s(\boldsymbol{\underline{Z}})))}{m(\boldsymbol{Z}^{*})}\right]=\mathbb{E}\left\{ \mathbb{E}\left[\left.\frac{D(Y-g_{1}(\boldsymbol{Z}^{*},s(\boldsymbol{\underline{Z}})))}{m(\boldsymbol{Z}^{*})}\right|\boldsymbol{Z}^{*},s(\boldsymbol{\underline{Z}})\right]\right\} \nonumber \\
= & \mathbb{E}\left\{ \frac{\mathbb{E}\left[D(Y-g_{1}(\boldsymbol{Z}^{*},s(\boldsymbol{\underline{Z}})))|\boldsymbol{Z}^{*},s(\boldsymbol{\underline{Z}})\right]}{m(\boldsymbol{Z}^{*})}\right\} \nonumber \\
= & \mathbb{E}\left\{ \frac{\mathbb{E}\left[Y-g_{1}(\boldsymbol{Z}^{*},s(\boldsymbol{\underline{Z}}))|D=1,\boldsymbol{Z}^{*},s(\boldsymbol{\underline{Z}})\right]\Pr(D=1|\boldsymbol{Z}^{*},s(\boldsymbol{\underline{Z}}))}{m(\boldsymbol{Z}^{*})}\right\} \nonumber \\
= & \mathbb{E}\left\{ \mathbb{E}\left[Y-g_{1}(\boldsymbol{Z}^{*},s(\boldsymbol{\underline{Z}}))|D=1,\boldsymbol{Z}^{*},s(\boldsymbol{\underline{Z}})\right]\right\} =0.\label{eq:refined_dr1}
\end{align}
Similar arguments yield $\mathbb{E}[g_{0}(\boldsymbol{Z}^{*})]=\mathbb{E}\left[g_{0}(\boldsymbol{Z}^{*},s(\boldsymbol{\underline{Z}}))\right]$
and 
\begin{equation}
\mathbb{E}\left[\frac{(1-D)(Y-g_{0}(\boldsymbol{Z}^{*},s(\boldsymbol{\underline{Z}})))}{1-m(\boldsymbol{Z}^{*})}\right]=0.\label{eq:refined_dr2}
\end{equation}
Combining these results establishes the estimand (\ref{eq:refine_estimand_ext}).

The remaining task is to verify the double robustness of the estimand
(\ref{eq:refine_estimand_ext}). Let $(\tilde{g}_{0},\tilde{g}_{1})$
and $\tilde{m}$ be generic working models for $(g_{0},g_{1})$ and
$m$, respectively. Using $D\perp\boldsymbol{\underline{Z}}|\boldsymbol{Z}^{*}$,
we obtain 
\begin{align*}
 & \mathbb{E}\left[\tilde{g}_{1}(\boldsymbol{Z}^{*},s(\boldsymbol{\underline{Z}}))\right]+\mathbb{E}\left[\frac{D(Y-\tilde{g}_{1}(\boldsymbol{Z}^{*},s(\boldsymbol{\underline{Z}})))}{\tilde{m}(\boldsymbol{Z}^{*})}\right]-\mathbb{E}\left[g_{1}(\boldsymbol{Z}^{*},s(\boldsymbol{\underline{Z}}))\right]\\
= & \mathbb{E}\left[\tilde{g}_{1}(\boldsymbol{Z}^{*},s(\boldsymbol{\underline{Z}}))-g_{1}(\boldsymbol{Z}^{*},s(\boldsymbol{\underline{Z}}))\right]+\mathbb{E}\left\{ \mathbb{E}\left[\left.\frac{D(Y-\tilde{g}_{1}(\boldsymbol{Z}^{*},s(\boldsymbol{\underline{Z}})))}{\tilde{m}(\boldsymbol{Z}^{*})}\right|\boldsymbol{Z}^{*},s(\boldsymbol{\underline{Z}})\right]\right\} \\
= & \mathbb{E}\left[\tilde{g}_{1}(\boldsymbol{Z}^{*},s(\boldsymbol{\underline{Z}}))-g_{1}(\boldsymbol{Z}^{*},s(\boldsymbol{\underline{Z}}))\right]+\mathbb{E}\left\{ \mathbb{E}\left[\left.\frac{Y-\tilde{g}_{1}(\boldsymbol{Z}^{*},s(\boldsymbol{\underline{Z}}))}{\tilde{m}(\boldsymbol{Z}^{*})}\right|D=1,\boldsymbol{Z}^{*},s(\boldsymbol{\underline{Z}})\right]m(\boldsymbol{Z}^{*})\right\} \\
= & \mathbb{E}\left[\tilde{g}_{1}(\boldsymbol{Z}^{*},s(\boldsymbol{\underline{Z}}))-g_{1}(\boldsymbol{Z}^{*},s(\boldsymbol{\underline{Z}}))\right]+\mathbb{E}\left[\left(g_{1}(\boldsymbol{Z}^{*},s(\boldsymbol{\underline{Z}}))-\tilde{g}_{1}(\boldsymbol{Z}^{*},s(\boldsymbol{\underline{Z}}))\right)\frac{m(\boldsymbol{Z}^{*})}{\tilde{m}(\boldsymbol{Z}^{*})}\right]\\
= & \mathbb{E}\left[\left(g_{1}(\boldsymbol{Z}^{*},s(\boldsymbol{\underline{Z}}))-\tilde{g}_{1}(\boldsymbol{Z}^{*},s(\boldsymbol{\underline{Z}}))\right)\left(\frac{m(\boldsymbol{Z}^{*})}{\tilde{m}(\boldsymbol{Z}^{*})}-1\right)\right].
\end{align*}
Thus, if either $\tilde{g}_{1}(\boldsymbol{Z}^{*},s(\boldsymbol{\underline{Z}}))=g_{1}(\boldsymbol{Z}^{*},s(\boldsymbol{\underline{Z}}))$
or $\tilde{m}(\boldsymbol{Z}^{*})=m(\boldsymbol{Z}^{*})$, we have
\begin{equation}
\mathbb{E}\left[g_{1}(\boldsymbol{Z}^{*},s(\boldsymbol{\underline{Z}}))\right]=\mathbb{E}\left[\tilde{g}_{1}(\boldsymbol{Z}^{*},s(\boldsymbol{\underline{Z}}))\right]+\mathbb{E}\left[\frac{D(Y-\tilde{g}_{1}(\boldsymbol{Z}^{*},s(\boldsymbol{\underline{Z}})))}{\tilde{m}(\boldsymbol{Z}^{*})}\right].\label{eq:refined_dr3}
\end{equation}
A similar derivation gives 
\begin{equation}
\mathbb{E}\left[g_{0}(\boldsymbol{Z}^{*},s(\boldsymbol{\underline{Z}}))\right]=\mathbb{E}\left[\tilde{g}_{0}(\boldsymbol{Z}^{*},s(\boldsymbol{\underline{Z}}))\right]+\mathbb{E}\left[\frac{(1-D)(Y-\tilde{g}_{0}(\boldsymbol{Z}^{*},s(\boldsymbol{\underline{Z}})))}{1-\tilde{m}(\boldsymbol{Z}^{*})}\right],\label{eq:refined_dr4}
\end{equation}
if either $\tilde{g}_{0}(\boldsymbol{Z}^{*},s(\boldsymbol{\underline{Z}}))=g_{0}(\boldsymbol{Z}^{*},s(\boldsymbol{\underline{Z}}))$
or $\tilde{m}(\boldsymbol{Z}^{*})=m(\boldsymbol{Z}^{*})$. Combining
(\ref{eq:refine_estimand_ext}) and (\ref{eq:refined_dr1})--(\ref{eq:refined_dr4}),
we conclude that for any working models $(\tilde{g}_{0},\tilde{g}_{1})$
and $\tilde{m}$, 
\[
\psi=\mathbb{E}\left[\tilde{g}_{1}(\boldsymbol{Z}^{*},s(\boldsymbol{\underline{Z}}))-\tilde{g}_{0}(\boldsymbol{Z}^{*},s(\boldsymbol{\underline{Z}}))\right]+\mathbb{E}\left[\frac{D(Y-\tilde{g}_{1}(\boldsymbol{Z}^{*},s(\boldsymbol{\underline{Z}})))}{\tilde{m}(\boldsymbol{Z}^{*})}-\frac{(1-D)(Y-\tilde{g}_{0}(\boldsymbol{Z}^{*},s(\boldsymbol{\underline{Z}})))}{1-\tilde{m}(\boldsymbol{Z}^{*})}\right],
\]
when either $(\tilde{g}_{0},\tilde{g}_{1})=(g_{0},g_{1})$ or $\tilde{m}=m$,
completing the proof. 

\hfill{}$\square$

\section{Technical Lemmas}

\label{APP:lemmasProofs}

We briefly explain the roles of the following Lemmas. Lemma \ref{LE:bound0}
is an existing result in the literature (e.g., \citet{Concentration}).
Lemmas \ref{LE:bound3} and \ref{LE:bound4} are used to derive the
probability bound for $\hat{\rho}$ in Lemma \ref{LE:boundf}.  Lemma
\ref{LE:uniform_n} serves to establish Theorem \ref{TH:debiased}.

\begin{lemma}{(Bernstein's Inequality)}\label{LE:bound0} For an
i.i.d. series $\{z_{i}\}_{i=1}^{\infty}$ with zero mean and bounded
support $\left[-L,L\right]$, we have 
\[
\Pr\left(\left|\sum_{i=1}^{n}z_{i}\right|>t\right)\leq2\exp\left\{ \frac{-t^{2}}{2\left(v+Lt/3\right)}\right\} ,
\]
for any $t>0$ and $v\geq\text{Var}\left(\sum_{i=1}^{n}z_{i}\right)$.
\end{lemma}

\begin{lemma}\label{LE:bound3} Suppose Assumption \ref{A:signal_strength}
holds and set $h\propto n^{-1/\left(2r+q^{\mathsf{c}}+1\right)}$
and $\lambda\propto n^{-r/\left(2r+q^{\mathsf{c}}+1\right)}$, then
\[
\sup_{x,y,\boldsymbol{w}}\Pr\left(\left|\hat{F}_{n}\left(x|y,\boldsymbol{w}\right)-F\left(x|y,\boldsymbol{w}\right)\right|>\frac{C\log n}{n^{r/\left(2r+q^{\mathsf{c}}+1\right)}}\right)\lesssim n^{-M},
\]
for any fixed positive $C$ and any fixed large positive $M.$ \end{lemma}

\begin{lemma}\label{LE:bound4}Under the same conditions in Lemma
\ref{LE:bound3}, 
\[
\sup_{y,\boldsymbol{w}}\Pr\left(\left|\hat{\Lambda}_{X}\left(Y=y,\boldsymbol{W}=w\right)-\Lambda_{X}\left(Y=y,\boldsymbol{W}=\boldsymbol{w}\right)\right|>\frac{C\log n}{n^{r/\left(2r+q^{\mathsf{c}}+1\right)}}\right)\lesssim n^{-M},
\]
for any fixed $C$ and any fixed large positive $M.$ \end{lemma}

\begin{lemma}\label{LE:boundf}Under the same conditions in Lemma
\ref{LE:bound3}, 
\[
\Pr\left(\left|\hat{\rho}-\rho\right|>\frac{C\log n}{n^{r/\left(2r+q^{\mathsf{c}}+1\right)}}\right)\lesssim n^{-M},
\]
for any fixed large positive $M.$ \end{lemma}

\begin{lemma}\label{LE:uniform_n}Suppose the conditions in Corollary
\ref{COR:propensity_score-screening} hold. In addition, $r\left/\left(p^{\mathsf{c}\ast}+2r\right)\right.>1/4$.
Write $\hat{m}\left(\boldsymbol{z}_{i}^{\hat{*}}\right)$= $\hat{f}\left(D=1|\boldsymbol{\tilde{x}}_{i}\right)$
in (\ref{eq:propensity_post}). Then, 
\[
n^{-1}\sum_{i=1}^{n}\left[\hat{m}\left(\boldsymbol{z}_{i}^{\hat{*}}\right)-m\left(\boldsymbol{z}_{i}\right)\right]^{2}=o_{P}\left(n^{-1/2}\right),
\]
\end{lemma}

\noindent\textbf{Proof of Lemma \ref{LE:bound3}.} We establish the
result using the following decomposition: 
\begin{align*}
 & \Pr\left(\left|\hat{F}_{n}\left(x|y,\boldsymbol{w}\right)-F\left(x|y,\boldsymbol{w}\right)\right|\geq\frac{C\log n}{n^{r/\left(2r+q^{\mathsf{c}}+1\right)}}\right)\\
= & \Pr\left(\left|\frac{\left[\hat{F}_{n}\left(x|y,\boldsymbol{w}\right)-F\left(x|y,\boldsymbol{w}\right)\right]\hat{f}(y,\boldsymbol{w})}{\hat{f}(y,\boldsymbol{w})}\right|\geq\frac{C\log n}{n^{r/\left(2r+q^{\mathsf{c}}+1\right)}}\right)\\
\leq & \Pr\left(\left|\left[\hat{F}_{n}\left(x|y,\boldsymbol{w}\right)-F\left(x|y,\boldsymbol{w}\right)\right]\hat{f}(y,\boldsymbol{w})\right|\geq\frac{B_{1}C\log n}{2n^{r/\left(2r+q^{\mathsf{c}}+1\right)}}\right)+\Pr\left(\frac{1}{\hat{f}(y,\boldsymbol{w})}\geq\frac{2}{B_{1}}\right)\\
\lesssim & n^{-M}+\exp\left(-C_{1}n^{2/3}\right)\propto n^{-M},
\end{align*}
where the last line is shown in the following Parts 1 and 2, specifically,
the bound on $T_{n}$ and $T_{n}'$ defined in (\ref{eq:Tn}) and
(\ref{eq:Tn'}), respectively.

Furthermore, we demonstrate below that $C_{1}$ and $M$ are independent
of $x,y,$ or $\boldsymbol{w}$. Thus, 
\[
\sup_{x,y,\boldsymbol{w}}\Pr\left(\left|\hat{F}_{n}\left(x|y,\boldsymbol{w}\right)-F\left(x|y,\boldsymbol{w}\right)\right|>\frac{C\log n}{n^{r/\left(2r+q^{\mathsf{c}}+1\right)}}\right)\lesssim n^{-M}.
\]
For notational convenience, we use $C$ and $M$ to denote generic
constants that may vary across the lines below.

\noindent\textbf{Part 1.} We first conduct the following decomposition.
\begin{align}
T_{n}= & \Pr\left(\left|\left[\hat{F}_{n}\left(x|y,\boldsymbol{w}\right)-F\left(x|y,\boldsymbol{w}\right)\right]\hat{f}\left(y,\boldsymbol{w}\right)\right|\geq\frac{C\log n}{n^{r/\left(2r+q^{\mathsf{c}}+1\right)}}\right)\nonumber \\
\leq & \text{\ensuremath{\Pr}\ensuremath{\left(\left|\left[\hat{F}_{n}\left(x|y,\boldsymbol{w}\right)-F\left(x|y,\boldsymbol{w}\right)\right]\hat{f}\left(y,\boldsymbol{w}\right)-\mathbb{E}\left\{ \left[\hat{F}_{n}\left(x|y,\boldsymbol{w}\right)-F\left(x|y,\boldsymbol{w}\right)\right]\hat{f}\left(y,\boldsymbol{w}\right)\right\} \right|\geq\frac{C\log n}{2n^{r/\left(2r+q^{\mathsf{c}}+1\right)}}\right)}}\nonumber \\
 & +\Pr\left(\left|\mathbb{E}\left\{ \left[\hat{F}_{n}\left(x|y,\boldsymbol{w}\right)-F\left(x|y,\boldsymbol{w}\right)\right]\hat{f}\left(y,\boldsymbol{w}\right)\right\} \right|\geq\frac{C\log n}{2n^{r/\left(2r+q^{\mathsf{c}}+1\right)}}\right)\nonumber \\
\equiv & T_{n1}+T_{n2}.\label{eq:Tn}
\end{align}
We now derive bounds for $T_{n1}$ and $T_{n2}$.

By Assumption \ref{A:signal_strength}(6) and (7), we have 
\[
\mathbb{E}\left\{ \left[\hat{F}_{n}\left(x|y,\boldsymbol{w}\right)-F\left(x|y,\boldsymbol{w}\right)\right]\hat{f}\left(y,\boldsymbol{w}\right)\right\} =O\left(h^{r}+\lambda\right),
\]
due to Theorem 6.1 of \citet{LiRacine}. Since $h^{r}\propto\lambda\propto n^{-r/\left(2r+q^{\mathsf{c}}+1\right)}\ll n^{-r/\left(2r+q^{\mathsf{c}}+1\right)}\log n,$
\begin{align*}
T_{n2}= & \Pr\left(\left|\mathbb{E}\left\{ \left[\hat{F}_{n}\left(x|y,\boldsymbol{w}\right)-F\left(x|y,\boldsymbol{w}\right)\right]\hat{f}\left(y,\boldsymbol{w}\right)\right\} \right|\geq\frac{C\log n}{2n^{r/\left(2r+q^{\mathsf{c}}+1\right)}}\right)=0
\end{align*}
for sufficiently large $n$.

$T_{n1}$ can be expressed as 
\begin{align*}
T_{n1}= & \ensuremath{\Pr}\left(\left|\left[\hat{F}_{n}\left(x|y,\boldsymbol{w}\right)-F\left(x|y,\boldsymbol{w}\right)\right]\hat{f}\left(y,\boldsymbol{w}\right)-\mathbb{E}\left\{ \left[\hat{F}_{n}\left(x|y,\boldsymbol{w}\right)-F\left(x|y,\boldsymbol{w}\right)\right]\hat{f}\left(y,\boldsymbol{w}\right)\right\} \right|\geq\frac{C\log n}{2n^{r/\left(2r+q^{\mathsf{c}}+1\right)}}\right)\\
= & \Pr\left(\left|n^{-1}\sum_{i=1}^{n}\left\{ \left[\boldsymbol{1}\left(x_{i}<x\right)-F\left(x|y,\boldsymbol{w}\right)\right]K_{h}\left(y_{i}-y\right)\Pi_{l=1}^{q^{\mathsf{c}}}K_{h}\left(w_{li}^{\mathsf{c}}-w_{l}^{\mathsf{c}}\right)\Pi_{l=1}^{q^{\mathsf{d}}}K_{\lambda}^{\mathsf{d}}\left(w_{li}^{\mathsf{d}},w_{l}^{\mathsf{d}}\right)\right.\right.\right.\\
 & \left.\left.\left.-\mathbb{E}\left[\left(\boldsymbol{1}\left(x_{i}<x\right)-F\left(x|y,\boldsymbol{w}\right)\right)K_{h}\left(y_{i}-y\right)\Pi_{l=1}^{q^{\mathsf{c}}}K_{h}\left(w_{li}^{\mathsf{c}}-w_{l}^{\mathsf{c}}\right)\Pi_{l=1}^{q^{\mathsf{d}}}K_{\lambda}^{\mathsf{d}}\left(w_{li}^{\mathsf{d}},w_{l}^{\mathsf{d}}\right)\right]\right\} \right|\geq\frac{C\log n}{2n^{r/\left(2r+q^{\mathsf{c}}+1\right)}}\right)\\
= & \Pr\left(n^{-1}\left|\sum_{i=1}^{n}z_{n,i}\right|\geq\frac{C\log n}{2n^{r/\left(2r+q^{\mathsf{c}}+1\right)}}\right),
\end{align*}
with 
\begin{align*}
z_{n,i}\equiv & \left[\boldsymbol{1}\left(x_{i}<x\right)-F\left(x|y,\boldsymbol{w}\right)\right]K_{h}\left(y_{i}-y\right)\Pi_{l=1}^{q^{\mathsf{c}}}K_{h}\left(w_{li}^{\mathsf{c}}-w_{l}^{\mathsf{c}}\right)\Pi_{l=1}^{q^{\mathsf{d}}}K_{\lambda}^{\mathsf{d}}\left(w_{li}^{\mathsf{d}},w_{l}^{\mathsf{d}}\right)\\
 & -\mathbb{E}\left[\left(\boldsymbol{1}\left(x_{i}<x\right)-F\left(x|y,\boldsymbol{w}\right)\right)K_{h}\left(y_{i}-y\right)\Pi_{l=1}^{q^{\mathsf{c}}}K_{h}\left(w_{li}^{\mathsf{c}}-w_{l}^{\mathsf{c}}\right)\Pi_{l=1}^{q^{\mathsf{d}}}K_{\lambda}^{\mathsf{d}}\left(w_{li}^{\mathsf{d}},w_{l}^{\mathsf{d}}\right)\right].
\end{align*}

Applying Lemma \ref{LE:bound0}, we have 
\begin{align*}
T_{n1} & =\Pr\left(\left|\sum_{i=1}^{n}z_{n,i}\right|>\frac{Cn^{\left(r+q^{\mathsf{c}}+1\right)/\left(2r+q^{\mathsf{c}}+1\right)}\log n}{2}\right)\\
 & \leq2\exp\left\{ -\frac{\left.C^{2}n^{2\left(r+q^{\mathsf{c}}+1\right)/\left(2r+q^{\mathsf{c}}+1\right)}\left(\log n\right)^{2}\right/4}{2\left(C_{1}nh^{-q^{\mathsf{c}}-1}+C_{2}h^{-q^{\mathsf{c}}-1}Cn^{\left(r+q^{\mathsf{c}}+1\right)/\left(2r+q^{\mathsf{c}}+1\right)}\log n/6\right)}\right\} ,\\
 & =2\exp\left\{ -\frac{\left.C^{2}n^{2\left(r+q^{\mathsf{c}}+1\right)/\left(2r+q^{\mathsf{c}}+1\right)}\left(\log n\right)^{2}\right/4}{2\left(C_{3}n^{2\left(r+q^{\mathsf{c}}+1\right)/\left(2r+q^{\mathsf{c}}+1\right)}+C_{4}n^{\left(r+2q^{\mathsf{c}}+2\right)/\left(2r+q^{\mathsf{c}}+1\right)}\log n\right)}\right\} \lesssim n^{-M},
\end{align*}
for some positive $C_{1},...,C_{4}$ and any fixed $M>0$, where we
use $\text{\text{Var}\ensuremath{\left(\sum_{i=1}^{n}z_{n,i}\right)}\ensuremath{\ensuremath{\propto}n\ensuremath{h^{-q^{\mathsf{c}}-1}}}}$
due to Theorem 6.1 of \citet{LiRacine}, $h\propto n^{-1/\left(2r+q^{\mathsf{c}}+1\right)}$,
and $\max_{i=1,...,n}\left|z_{n,i}\right|\leq C_{2}h^{-q^{\mathsf{c}}-1}$.

Combining these results, we conclude that $T_{n}\lesssim n^{-M}.$
Note that the values of those fixed constants $(C\text{ or }M)$ are
only related to those upper bounds in Assumption \ref{A:signal_strength}.
Thus, they do not depend on the values of $x,y,$ and $\boldsymbol{w}$.

\textbf{Part 2.} We first conduct the following decomposition: 
\begin{align}
T_{n}'= & \Pr\left(\left|\hat{f}(y,\boldsymbol{w})-f(y,\boldsymbol{w})\right|\geq B_{1}/2\right)\nonumber \\
\leq & \Pr\left(\left|\hat{f}(y,\boldsymbol{w})-\mathbb{E}\left[\hat{f}(y,\boldsymbol{w})\right]\right|\geq B_{1}/4\right)+\Pr\left(\left|\mathbb{E}\left[\hat{f}(y,\boldsymbol{w})\right]-f(y,\boldsymbol{w})\right|\geq B_{1}/4\right)\nonumber \\
\equiv & T_{n1}'+T_{n2}'.\label{eq:Tn'}
\end{align}
Analogous to the process of bounding $T_{n2}$ in Part 1, we can show
\[
T_{n2}'=\Pr\left(\left|\mathbb{E}\left[\hat{f}(y,\boldsymbol{w})\right]-f(y,\boldsymbol{w})\right|\geq B_{1}/4\right)=0,
\]
for sufficiently large $n.$

Let 
\[
z_{n,i}'\equiv K_{h}\left(y_{i}-y\right)\Pi_{l=1}^{q^{\mathsf{c}}}K_{h}\left(w_{li}^{\mathsf{c}}-w_{l}^{\mathsf{c}}\right)\Pi_{l=1}^{q^{\mathsf{d}}}K_{\lambda}^{\mathsf{d}}\left(w_{li}^{\mathsf{d}},w_{l}^{\mathsf{d}}\right)-\mathbb{E}\left[K_{h}\left(y_{i}-y\right)\Pi_{l=1}^{q^{\mathsf{c}}}K_{h}\left(w_{li}^{\mathsf{c}}-w_{l}^{\mathsf{c}}\right)\Pi_{l=1}^{q^{\mathsf{d}}}K_{\lambda}^{\mathsf{d}}\left(w_{li}^{\mathsf{d}},w_{l}^{\mathsf{d}}\right)\right].
\]
Analogously to how we addressed $T_{n2}$ in Part 1, 
\begin{align*}
 & T_{n1}'=\Pr\left(\left|\hat{f}(y,\boldsymbol{w})-\mathbb{E}\left[\hat{f}(y,\boldsymbol{w})\right]\right|\geq B_{1}/4\right)\\
= & \Pr\left(\left|\sum_{i=1}^{n}z_{n,i}'\right|\geq nB_{1}/4\right)\leq2\exp\left\{ -\frac{n^{2}B_{1}^{2}/16}{2\left(C_{1}nh^{-q^{\mathsf{c}}-1}+C_{2}h^{-q^{\mathsf{c}}-1}nB_{1}/12\right)}\right\} \\
= & 2\exp\left\{ -Cn^{\frac{2r}{2r+q^{\mathsf{c}}+1}}\right\} ,
\end{align*}
holds for some fixed positive $C$, $C_{1}$, and $C_{2}.$

Substituting the results for $T_{n1}'$ and $T_{n2}'$ into $T_{n}'$,
we obtain 
\[
T_{n}'\leq2\exp\left\{ -Cn^{2/3}\right\} ,
\]
for sufficiently large $n$. For the same reasoning as in Part 1,
$C$ does not depend on $y\text{ and }\boldsymbol{w}$.

Finally, note that $\left\{ \hat{f}(y,\boldsymbol{w})\leq B_{1}/2\right\} \subset\left\{ \left|\hat{f}(y,\boldsymbol{w})-f(y,\boldsymbol{w})\right|\geq B_{1}/2\right\} $
because of the condition $f(y,\boldsymbol{w})\geq B_{1}/2$ imposed
in Assumption \ref{A:signal_strength}(4). Therefore, for sufficiently
large $n$, 
\[
\Pr\left(\frac{1}{\hat{f}(y,\boldsymbol{w})}\geq\frac{2}{B_{1}}\right)\leq T_{n}'\leq2\exp\left\{ -Cn^{2/3}\right\} .
\]
\hfill{}$\square$

\noindent\textbf{Proof of Lemma \ref{LE:bound4}.} Given any $(y,\boldsymbol{w})$,
because we evaluate $\hat{F}_{n}(x|y,\boldsymbol{w})-\hat{F}_{n}(x|y^{*},\boldsymbol{w})$
solely at the sample points $x_{1},...,x_{n}$, there exists an $x_{i^{*}}\in\{x_{1},...,x_{n}\}$
such that $\hat{\Lambda}_{X}(Y=y,\boldsymbol{W}=\boldsymbol{w})=\hat{F}_{n}(x_{i^{*}}|y,\boldsymbol{w})-F(x_{i^{*}}|y^{*},\boldsymbol{w})$.
However, $\Lambda_{X}(Y=y,\boldsymbol{W}=\boldsymbol{w})$ is generally
not attained at any point in $\{x_{1},...,x_{n}\}$. This lemma addresses
this issue. In our proof, we assume $X$ is a continuous random variable,
as a discrete $X$ does not have this concern.

Suppose $\Lambda_{X}\left(Y=y,\boldsymbol{W}=\boldsymbol{w}\right)$
is attained at $\tilde{x},$ that is, 
\begin{equation}
F\left(\tilde{x}|y^{*},\boldsymbol{w}\right)-F\left(\tilde{x}|y,\boldsymbol{w}\right)=\sup_{x}\left|F\left(x|y,\boldsymbol{w}\right)-F\left(x|y^{*},\boldsymbol{w}\right)\right|=\Lambda_{X}\left(Y=y,\boldsymbol{W}=\boldsymbol{w}\right).\label{eq:xtide_define}
\end{equation}
The case when $F\left(\tilde{x}|y,\boldsymbol{w}\right)>F\left(\tilde{x}|y^{*},\boldsymbol{w}\right)$
is symmetric.

Since $\hat{F}_{n}$ is a nonparametric estimator, the number of effectively
observations is proportional to $nh^{q^{\mathsf{c}}+1}$, given that
$f\left(y,\boldsymbol{w}\right)$ is bounded away from zero. Without
loss of generality, suppose we use a uniform kernel and the number
of effectively observations is $\left\lfloor nh^{q^{\mathsf{c}}+1}\right\rfloor $,
where $\left\lfloor \cdot\right\rfloor $ denotes the integer part
of $\cdot$.

Further, let 
\[
x_{0,\left(1\right)},x_{0,\left(2\right)},...,x_{0,\left(\left\lfloor nh^{q^{\mathsf{c}}+1}\right\rfloor \right)}\text{ and }x_{1,\left(1\right)},x_{1,\left(2\right)},...,x_{1,\left(\left\lfloor nh^{q^{\mathsf{c}}+1}\right\rfloor \right)}
\]
be the effective observations for obtaining $\hat{\Lambda}_{X}\left(Y=y,\boldsymbol{W}=\boldsymbol{w}\right)$
and $\hat{\Lambda}_{X}\left(Y=y^{*},\boldsymbol{W}=\boldsymbol{w}\right)$,
respectively, and $x_{d,\left(1\right)}\leq...\leq x_{d,\left(\left\lfloor nh^{q^{\mathsf{c}}+1}\right\rfloor \right)}$
for $d=0,1$. Denote $\tilde{i}_{d}$ such that 
\[
x_{d,\left(\tilde{i}_{d}\right)}\leq\tilde{x}\leq x_{d,\left(\tilde{i}_{d}+1\right)},
\]
for $d=0,1$. The case where $\tilde{x}$ is smaller or greater than any observation can be handled similarly with slightly more details.
By definition, 
\begin{equation}
\hat{F}_{n}\left(x_{0,\left(\tilde{i}_{0}+1\right)}|y,\boldsymbol{w}\right)-\hat{F}_{n}\left(x_{0,\left(\tilde{i}_{0}\right)}|y,\boldsymbol{w}\right)=\frac{1}{\left\lfloor nh^{q^{\mathsf{c}}+1}\right\rfloor },\label{eq:quantile_result}
\end{equation}
and 
\begin{equation}
\hat{F}_{n}\left(x_{0,\left(\tilde{i}_{0}\right)}|y,\boldsymbol{w}\right)\leq\hat{F}_{n}\left(\tilde{x}|y,\boldsymbol{w}\right)\leq\hat{F}_{n}\left(x_{0,\left(\tilde{i}_{0}+1\right)}|y,\boldsymbol{w}\right).\label{eq:quantile_result2}
\end{equation}
The above holds analogously for the case where $d=1$ or $Y=y^{*}$.

We use $A$ to denote the following event 
\begin{align}
A= & \left\{ \left|\hat{F}_{n}\left(\tilde{x}|y,\boldsymbol{w}\right)-F\left(\tilde{x}|y,\boldsymbol{w}\right)\right|\leq\frac{C\log n}{3n^{r/\left(2r+q^{\mathsf{c}}+1\right)}}\right\} \label{eq:eventA}\\
 & \cap\left\{ \left|\hat{F}_{n}\left(\tilde{x}|y^{*},\boldsymbol{w}\right)-F\left(\tilde{x}|y^{*},\boldsymbol{w}\right)\right|\leq\frac{C\log n}{3n^{r/\left(2r+q^{\mathsf{c}}+1\right)}}\right\} .\nonumber 
\end{align}
Without loss of generality, assume that $x_{0,\left(\tilde{i}_{0}\right)}\geq x_{1,\left(\tilde{i}_{1}\right)},$
implying $x_{0,\left(\tilde{i}_{0}\right)}$ is closer to $\tilde{x}$
than $x_{1,\left(\tilde{i}_{1}\right)}$. Then, conditional on event
$A,$ 
\begin{align*}
 & \hat{F}_{n}\left(x_{0,\left(\tilde{i}_{0}\right)}|y^{*},\boldsymbol{w}\right)-\hat{F}_{n}\left(x_{0,\left(\tilde{i}_{0}\right)}|y,\boldsymbol{w}\right)\geq\hat{F}_{n}\left(x_{1,\left(\tilde{i}_{1}\right)}|y^{*},\boldsymbol{w}\right)-\hat{F}_{n}\left(x_{0,\left(\tilde{i}_{0}\right)}|y,\boldsymbol{w}\right)\\
= & \hat{F}_{n}\left(\tilde{x}|y^{*},\boldsymbol{w}\right)-\hat{F}_{n}\left(\tilde{x}|y,\boldsymbol{w}\right)+\left[\hat{F}_{n}\left(x_{1,\left(\tilde{i}_{1}\right)}|y^{*},\boldsymbol{w}\right)-\hat{F}_{n}\left(\tilde{x}|y^{*},\boldsymbol{w}\right)\right]-\left[\hat{F}_{n}\left(x_{0,\left(\tilde{i}_{0}\right)}|y,\boldsymbol{w}\right)-\hat{F}_{n}\left(\tilde{x}|y,\boldsymbol{w}\right)\right]\\
= & F\left(\tilde{x}|y^{*},\boldsymbol{w}\right)-F\left(\tilde{x}|y,\boldsymbol{w}\right)+\left[\hat{F}_{n}\left(\tilde{x}|y^{*},\boldsymbol{w}\right)-F\left(\tilde{x}|y^{*},\boldsymbol{w}\right)\right]-\left[\hat{F}_{n}\left(\tilde{x}|y,\boldsymbol{w}\right)-F\left(\tilde{x}|y,\boldsymbol{w}\right)\right]\\
 & +\left[\hat{F}_{n}\left(x_{1,\left(\tilde{i}_{1}\right)}|y^{*},\boldsymbol{w}\right)-\hat{F}_{n}\left(\tilde{x}|y^{*},\boldsymbol{w}\right)\right]-\left[\hat{F}_{n}\left(x_{0,\left(\tilde{i}_{0}\right)}|y,\boldsymbol{w}\right)-\hat{F}_{n}\left(\tilde{x}|y,\boldsymbol{w}\right)\right]\\
\geq & \Lambda_{X}\left(Y=y,\boldsymbol{W}=\boldsymbol{w}\right)-\frac{2C\log n}{3n^{r/\left(2r+q^{\mathsf{c}}+1\right)}}-\frac{2}{\left\lfloor nh^{q^{\mathsf{c}}+1}\right\rfloor }\\
\geq & \Lambda_{X}\left(Y=y,\boldsymbol{W}=\boldsymbol{w}\right)-\frac{C\log n}{n^{r/\left(2r+q^{\mathsf{c}}+1\right)}},
\end{align*}
for sufficiently large $n,$ given $h\propto n^{-1/\left(2r+q^{\mathsf{c}}+1\right)}$.
The above inequalities follow from the definition in (\ref{eq:xtide_define}),
the results in (\ref{eq:quantile_result}) and (\ref{eq:quantile_result2}),
and the assumption that event $A$ in (\ref{eq:eventA}) occurs. Using
the above, 
\begin{align*}
\hat{\Lambda}_{X}\left(Y=y,\boldsymbol{W}=\boldsymbol{w}\right) & =\max_{i=1,...,n}\left|\hat{F}_{n}\left(x_{i}|y,\boldsymbol{w}\right)-\hat{F}_{n}\left(x_{i}|y^{*},\boldsymbol{w}\right)\right|\\
 & \geq\hat{F}_{n}\left(x_{0,\left(\tilde{i}_{0}\right)}|y^{*},\boldsymbol{w}\right)-\hat{F}_{n}\left(x_{0,\left(\tilde{i}_{0}\right)}|y,\boldsymbol{w}\right)\\
 & \geq\Lambda_{X}\left(Y=y,\boldsymbol{W}=\boldsymbol{w}\right)-\frac{C\log n}{n^{r/\left(2r+q^{\mathsf{c}}+1\right)}},
\end{align*}
which is equivalent to 
\[
\hat{\Lambda}_{X}\left(Y=y,\boldsymbol{W}=\boldsymbol{w}\right)-\Lambda_{X}\left(Y=y,\boldsymbol{W}=\boldsymbol{w}\right)\geq-\frac{C\log n}{n^{r/\left(2r+q^{\mathsf{c}}+1\right)}}.
\]

We now proceed to establish the reverse inequality. We denote event
$A'$ as 
\begin{align*}
A'= & \left\{ \max_{i=1,...,n}\left|\hat{F}_{n}\left(x_{i}|y,\boldsymbol{w}\right)-F\left(x_{i}|y,\boldsymbol{w}\right)\right|\leq\frac{C\log n}{2n^{r/\left(2r+q^{\mathsf{c}}+1\right)}}\right\} \\
 & \cap\left\{ \max_{i=1,...,n}\left|\hat{F}_{n}\left(x_{i}|y^{*},\boldsymbol{w}\right)-F\left(x_{i}|y^{*},\boldsymbol{w}\right)\right|\leq\frac{C\log n}{2n^{r/\left(2r+q^{\mathsf{c}}+1\right)}}\right\} .
\end{align*}
Then, conditional on $A',$ 
\begin{align*}
 & \hat{\Lambda}_{X}\left(Y=y,\boldsymbol{W}=\boldsymbol{w}\right)=\max_{i=1,...,n}\left|\hat{F}_{n}\left(x_{i}|y,\boldsymbol{w}\right)-\hat{F}_{n}\left(x_{i}|y^{*},\boldsymbol{w}\right)\right|\\
\leq & \max_{i=1,...,n}\left\{ \left|\hat{F}_{n}\left(x_{i}|y,\boldsymbol{w}\right)-F\left(x_{i}|y,\boldsymbol{w}\right)\right|+\left|F\left(x_{i}|y,\boldsymbol{w}\right)-F\left(x_{i}|y^{*},\boldsymbol{w}\right)\right|+\left|\hat{F}_{n}\left(x_{i}|y^{*},\boldsymbol{w}\right)-F\left(x_{i}|y^{*},\boldsymbol{w}\right)\right|\right\} \\
\leq & \frac{C\log n}{n^{r/\left(2r+q^{\mathsf{c}}+1\right)}}+\Lambda_{X}\left(Y=y,\boldsymbol{W}=\boldsymbol{w}\right),
\end{align*}
which implies that 
\[
\hat{\Lambda}_{X}\left(Y=y,\boldsymbol{W}=\boldsymbol{w}\right)-\Lambda_{X}\left(Y=y,\boldsymbol{W}=\boldsymbol{w}\right)\leq\frac{C\log n}{n^{r/\left(2r+q^{\mathsf{c}}+1\right)}}.
\]

To summarize, the preceding analysis demonstrates that the event $A\cap A'$
implies 
\[
\left|\hat{\Lambda}_{X}\left(Y=y,\boldsymbol{W}=\boldsymbol{w}\right)-\Lambda_{X}\left(Y=y,\boldsymbol{W}=\boldsymbol{w}\right)\right|\leq\frac{C\log n}{n^{r/\left(2r+q^{\mathsf{c}}+1\right)}}.
\]
Lemma \ref{LE:bound3} implies that 
\begin{align*}
\Pr(A'{}^{c})= & \Pr\left(\max_{i=1,...,n}\max_{a=y,y^{*}}\left|\hat{F}_{n}\left(x_{i}|Y=a,\boldsymbol{w}\right)-F\left(x_{i}|Y=a,\boldsymbol{w}\right)\right|>\frac{C\log n}{2n^{r/\left(2r+q^{\mathsf{c}}+1\right)}}\right)\\
\leq & \sum_{i=1}^{n}\sum_{a=y,y^{*}}\Pr\left(\left|\hat{F}_{n}\left(x_{i}|a,\boldsymbol{w}\right)-F\left(x_{i}|a,\boldsymbol{w}\right)\right|>\frac{C\log n}{2n^{r/\left(2r+q^{\mathsf{c}}+1\right)}}\right)\lesssim n^{-M'+1},
\end{align*}
for any fixed large positive $M'$. Similarly, 
\[
\Pr(A^{c})=\Pr\left(\max_{a=y,y^{*}}\left|\hat{F}_{n}\left(\tilde{x}|Y=a,\boldsymbol{w}\right)-F\left(\tilde{x}|Y=a,\boldsymbol{w}\right)\right|>\frac{C\log n}{3n^{r/\left(2r+q^{\mathsf{c}}+1\right)}}\right)\lesssim n^{-M}.
\]
Therefore, by letting $M=M'-1$, we have 
\begin{align*}
 & \Pr\left(\left|\hat{\Lambda}_{X}\left(Y=y,\boldsymbol{W}=\boldsymbol{w}\right)-\Lambda_{X}\left(Y=y,\boldsymbol{W}=\boldsymbol{w}\right)\right|>\frac{C\log n}{n^{r/\left(2r+q^{\mathsf{c}}+1\right)}}\right)\\
= & 1-\Pr\left(\left|\hat{\Lambda}_{X}\left(Y=y,\boldsymbol{W}=\boldsymbol{w}\right)-\Lambda_{X}\left(Y=y,\boldsymbol{W}=\boldsymbol{w}\right)\right|\leq\frac{C\log n}{n^{r/\left(2r+q^{\mathsf{c}}+1\right)}}\right)\\
\leq & 1-\Pr\left(A\cap A'\right)=\Pr\left(A^{c}\cup A'{}^{c}\right)\leq\Pr(A^{c})+\Pr(A'{}^{c})\lesssim n^{-M},
\end{align*}
as desired. 

\hfill{}$\square$

\noindent\textbf{Proof of Lemma \ref{LE:boundf}.} We denote $\Lambda_{X}\left(Y=y_{i},\boldsymbol{W}=\boldsymbol{w}_{i}\right)$
as $\Lambda_{i}$ for brevity. The term $\hat{\Lambda}_{i}$ is defined
analogously. By the i.i.d. assumption, we obtain the following decomposition:
\[
\hat{\rho}-\rho=\frac{1}{n}\sum_{i=1}^{n}\left[\hat{\Lambda}_{i}-E\left(\Lambda_{i}\right)\right]=\frac{1}{n}\sum_{i=1}^{n}\left[\Lambda_{i}-E\left(\Lambda_{i}\right)\right]+\frac{1}{n}\sum_{i=1}^{n}\left(\hat{\Lambda}_{i}-\Lambda_{i}\right).
\]
From this decomposition, we have 
\begin{align*}
\Pr\left(\left|\hat{\rho}-\rho\right|\geq\frac{C\log n}{n^{r/\left(2r+q^{\mathsf{c}}+1\right)}}\right)\leq & \Pr\left(\left|\frac{1}{n}\sum_{i=1}^{n}\left[\Lambda_{i}-E\left(\Lambda_{i}\right)\right]\right|\geq\frac{C\log n}{2n^{r/\left(2r+q^{\mathsf{c}}+1\right)}}\right)\\
 & +\Pr\left(\left|\frac{1}{n}\sum_{i=1}^{n}\left(\hat{\Lambda}_{i}-\Lambda_{i}\right)\right|\geq\frac{C\log n}{2n^{r/\left(2r+q^{\mathsf{c}}+1\right)}}\right).
\end{align*}
We now proceed to bound each term on the right-hand side.

Since $\left|\Lambda_{i}\right|\leq1$ by definition, Hoeffding's inequality implies that 
\[
\Pr\left(\left|\frac{1}{n}\sum_{i=1}^{n}\left[\Lambda_{i}-E\left(\Lambda_{i}\right)\right]\right|\geq\frac{C\log n}{2n^{r/\left(2r+q^{\mathsf{c}}+1\right)}}\right)\leq\exp\left\{ -\frac{C^{2}\left(\log n\right)^{2}n^{\left(q^{\mathsf{c}}+1\right)/\left(2r+q^{\mathsf{c}}+1\right)}}{8}\right\} \lesssim n^{-M}.
\]
For the second term, Lemma \ref{LE:bound4} implies that 
\begin{align*}
\Pr\left(\left|\frac{1}{n}\sum_{i=1}^{n}\left(\hat{\Lambda}_{i}-\Lambda_{i}\right)\right|\geq\frac{C\log n}{2n^{r/\left(2r+q^{\mathsf{c}}+1\right)}}\right) & \leq\sum_{i=1}^{n}\Pr\left(\left|\hat{\Lambda}_{i}-\Lambda_{i}\right|\geq\frac{C\log n}{2n^{r/\left(2r+q^{\mathsf{c}}+1\right)}}\right)\\
 & \leq n\cdot n^{-M'}=n^{-M'+1},
\end{align*}
for any fixed large positive $M'$. Setting $M=M'-1$ completes the
proof. 

\hfill{}$\square$

\noindent\textbf{Proof of Lemma \ref{LE:uniform_n}.} It is sufficient
to show that 
\[
\max_{i=1,...,n}\left|\hat{m}\left(\boldsymbol{z}_{i}^{\hat{*}}\right)-m\left(\boldsymbol{z}_{i}^{*}\right)\right|=O_{P}\left(\frac{\log n}{n^{r\left/\left(p^{\mathsf{c}\ast}+2r\right)\right.}}\right)=o_{P}\left(n^{-1/4}\right).
\]

As established in the proofs of Theorem \ref{TH:TPR} and Proposition
\ref{Prop:post}, the following two events occur with probability
approaching 1: 
\[
\hat{E}=\left\{ \min_{1\leq j\leq s^{*}}\hat{\rho}_{j}\geq Cn^{-r/\left(2r+q^{\mathsf{c}}+1\right)}\log n\geq\max_{s^{*}+1\leq j\leq p}\hat{\rho}_{j}\right\} 
\]
and 
\begin{align*}
\hat{F} & =\left\{ n^{\frac{1}{p^{\mathsf{c}\ast}+2r}}\hat{h}_{l}\rightarrow c_{l},\text{ and }n^{\frac{r}{p^{\mathsf{c}\ast}+2r}}\hat{\lambda}_{l'}\rightarrow c_{l'},\text{ for any \ensuremath{X_{l}^{\mathsf{c}}} and \ensuremath{X_{l'}^{\mathsf{d}}} in \ensuremath{\boldsymbol{Z}^{*}}},\right.\\
 & \left.\hat{h}_{l}>C,\text{ for any positive }C,\text{ and }\hat{\lambda}_{l'}\rightarrow\frac{r_{l'}-1}{r_{l'}},\text{ for any \ensuremath{X_{l}^{\mathsf{c}}} and \ensuremath{X_{l'}^{\mathsf{d}}} in \ensuremath{\boldsymbol{\tilde{X}}\backslash\boldsymbol{Z}^{*}}}\right\} .
\end{align*}
Consequently, 
\[
\Pr\left(\hat{E}\cap\hat{F}\right)\rightarrow1.
\]

Recall that 
\[
\hat{f}\left(D=1|\boldsymbol{\tilde{x}}\right)=\frac{n^{-1}\sum_{i=1}^{n}\Pi_{l=1}^{\tilde{s}_{1}}K_{\hat{h}_{l}}\left(x_{li}^{\mathsf{c}}-x_{l}^{\mathsf{c}}\right)\cdot\Pi_{l=1}^{\tilde{s}_{2}}K_{\hat{\lambda}_{l}}^{\mathsf{d}}\left(x_{li}^{\mathsf{d}},x_{l}^{\mathsf{d}}\right)\cdot\mathbf{1}\left(D_{i}=1\right)}{n^{-1}\sum_{i=1}^{n}\Pi_{l=1}^{\tilde{s}_{1}}K_{\hat{h}_{l}}\left(x_{li}^{\mathsf{c}}-x_{l}^{\mathsf{c}}\right)\cdot\Pi_{l=1}^{\tilde{s}_{2}}K_{\hat{\lambda}_{l}}^{\mathsf{d}}\left(x_{li}^{\mathsf{d}},x_{l}^{\mathsf{d}}\right)}\equiv\frac{\hat{f}\left(1,\boldsymbol{\tilde{x}}\right)}{\hat{f}\left(\boldsymbol{\tilde{x}}\right)}.
\]
Conditional on $\hat{E}\cap\hat{F}$, and using the same proof as
that of Lemma \ref{LE:bound3}, we obtain 
\begin{equation}
\Pr\left(\left|\hat{f}\left(1,\boldsymbol{\tilde{x}}\right)-\mathbb{E}\left[\hat{f}\left(1,\boldsymbol{\tilde{x}}\right)\right]\right|\geq\frac{C_{1}\log n}{\sqrt{nh_{1}...h_{p^{\mathsf{c}*}}}}\right)=\Pr\left(\left|\hat{f}\left(1,\boldsymbol{\tilde{x}}\right)-\mathbb{E}\left[\hat{f}\left(1,\boldsymbol{\tilde{x}}\right)\right]\right|\geq\frac{C_{2}\log n}{n^{r\left/\left(p^{\mathsf{c}\ast}+2r\right)\right.}}\right)\lesssim n^{-M}\label{eq:f1xbound}
\end{equation}
and 
\begin{equation}
\Pr\left(\left|\hat{f}\left(\boldsymbol{\tilde{x}}\right)-\mathbb{E}\left[\hat{f}\left(\boldsymbol{\tilde{x}}\right)\right]\right|\geq\frac{C_{1}\log n}{\sqrt{nh_{1}...h_{p^{\mathsf{c}*}}}}\right)=\Pr\left(\left|\hat{f}\left(\boldsymbol{\tilde{x}}\right)-\mathbb{E}\left[\hat{f}\left(\boldsymbol{\tilde{x}}\right)\right]\right|\geq\frac{C_{2}\log n}{n^{r\left/\left(p^{\mathsf{c}\ast}+2r\right)\right.}}\right)\lesssim n^{-M},\label{eq:fxbound}
\end{equation}
for some positive $C_{1},C_{2}$, and some positive $M>1.$ Note that
the above utilizes $\hat{h}_{l}>C$ for $l=p^{\mathsf{c}*}+1,...,\tilde{s}_{1}$
and hence, these bandwidths have no impact on the rate.

$\mathbb{E}\left[\hat{f}\left(1,\boldsymbol{\tilde{x}}\right)\right]$
and $\mathbb{E}\left[\hat{f}\left(\boldsymbol{\tilde{x}}\right)\right]$
are essentially $\mu_{f}$ and $\mu_{g}$ in the proof of Proposition
\ref{Prop:post}. Based on the results for $\mu_{f}$ and $\mu_{g}$
and conditional on $\hat{E}\cap\hat{F},$ 
\begin{align}
\frac{\mathbb{E}\left[\hat{f}\left(1,\boldsymbol{\tilde{x}}\right)\right]}{\mathbb{E}\left[\hat{f}\left(\boldsymbol{\tilde{x}}\right)\right]} & =\mathbb{E}\left(D=1|\boldsymbol{Z}^{*}=\boldsymbol{z}^{*}\right)+O\left(\left(nh_{1}...h_{p^{\mathsf{c}*}}\right)^{-1}\right)\equiv m\left(\boldsymbol{z}^{*}\right)+O\left(\left(nh_{1}...h_{p^{\mathsf{c}*}}\right)^{-1}\right)\nonumber \\
 & =m\left(\boldsymbol{z}^{*}\right)+O\left(n^{-2r\left/\left(p^{\mathsf{c}\ast}+2r\right)\right.}\right).\label{eq:f1xfx_E}
\end{align}
Assumption \ref{A:2}(2) ensures that $\mathbb{E}\left[\hat{f}\left(\boldsymbol{\tilde{x}}\right)\right]$,
conditional on $\hat{E}\cap\hat{F}$ (and thus $\hat{h}_{l}>C$ for
$l=p^{\mathsf{c}*}+1,...,\tilde{s}_{1}$), is uniformly bounded away
from zero, i.e., 
\begin{equation}
\mathbb{E}\left[\hat{f}\left(\boldsymbol{\tilde{x}}\right)\right]>\varrho_{1}>0.\label{eq:deno_baway0}
\end{equation}
Further, (\ref{eq:fxbound}) implies that there exists a $\varrho_{2}>0$
such that 
\begin{equation}
\Pr\left(\hat{f}\left(\boldsymbol{\tilde{x}}\right)\geq\varrho_{2}\right)\lesssim n^{-M}.\label{eq:deno_baway0_^}
\end{equation}
Combining (\ref{eq:f1xbound}) through (\ref{eq:deno_baway0_^}),
and conditional on $\hat{E}\cap\hat{F}$, we conclude that 
\begin{align}
 & \Pr\left(\left|\hat{f}\left(D=1|\boldsymbol{\tilde{x}}\right)-m\left(\boldsymbol{z}^{*}\right)\right|\geq\frac{C_{3}\log n}{n^{r\left/\left(p^{\mathsf{c}\ast}+2r\right)\right.}}\right)\nonumber \\
\leq & \Pr\left(\left|\frac{\hat{f}\left(1,\boldsymbol{\tilde{x}}\right)}{\hat{f}\left(\boldsymbol{\tilde{x}}\right)}-\frac{\mathbb{E}\left[\hat{f}\left(1,\boldsymbol{\tilde{x}}\right)\right]}{\mathbb{E}\left[\hat{f}\left(\boldsymbol{\tilde{x}}\right)\right]}\right|\geq\frac{1}{2}\frac{C_{3}\log n}{n^{r\left/\left(p^{\mathsf{c}\ast}+2r\right)\right.}}\right)\nonumber \\
\leq & \Pr\left(\left|\hat{f}\left(1,\boldsymbol{\tilde{x}}\right)-\mathbb{E}\left[\hat{f}\left(1,\boldsymbol{\tilde{x}}\right)\right]\right|\geq\frac{C_{4}\log n}{n^{r\left/\left(p^{\mathsf{c}\ast}+2r\right)\right.}}\right)+\Pr\left(\left|\hat{f}\left(\boldsymbol{\tilde{x}}\right)-\mathbb{E}\left[\hat{f}\left(\boldsymbol{\tilde{x}}\right)\right]\right|\geq\frac{C_{5}\log n}{n^{r\left/\left(p^{\mathsf{c}\ast}+2r\right)\right.}}\right)\nonumber \\
 & +\Pr\left(\hat{f}\left(\boldsymbol{\tilde{x}}\right)\geq\varrho_{2}\right)\nonumber \\
\lesssim & n^{-M},\label{eq:bound_last}
\end{align}
for some positive $C_{3},C_{4},\text{ and }C_{5}$, where the second
inequality uses Assumption \ref{A:3}(2) and the following decomposition:
\begin{align*}
\frac{\hat{f}\left(1,\tilde{\boldsymbol{x}}\right)}{\hat{f}\left(\tilde{\boldsymbol{x}}\right)}-\frac{\mathbb{E}\left[\hat{f}\left(1,\tilde{\boldsymbol{x}}\right)\right]}{\mathbb{E}\left[\hat{f}\left(\tilde{\boldsymbol{x}}\right)\right]} & =\frac{\hat{f}\left(1,\tilde{\boldsymbol{x}}\right)-\mathbb{E}\left[\hat{f}\left(1,\tilde{\boldsymbol{x}}\right)\right]}{\hat{f}\left(\tilde{\boldsymbol{x}}\right)}-\frac{\hat{f}\left(\tilde{\boldsymbol{x}}\right)-\mathbb{E}\left[\hat{f}\left(\tilde{\boldsymbol{x}}\right)\right]}{\hat{f}\left(\tilde{\boldsymbol{x}}\right)}\frac{\mathbb{E}\left[\hat{f}\left(1,\tilde{\boldsymbol{x}}\right)\right]}{\mathbb{E}\left[\hat{f}\left(\tilde{\boldsymbol{x}}\right)\right]}\\
 & =\frac{\hat{f}\left(1,\tilde{\boldsymbol{x}}\right)-\mathbb{E}\left[\hat{f}\left(1,\tilde{\boldsymbol{x}}\right)\right]}{\hat{f}\left(\tilde{\boldsymbol{x}}\right)}-\frac{\hat{f}\left(\tilde{\boldsymbol{x}}\right)-\mathbb{E}\left[\hat{f}\left(\tilde{\boldsymbol{x}}\right)\right]}{\hat{f}\left(\tilde{\boldsymbol{x}}\right)}\left(m(\boldsymbol{z}^{*})+o\left(1\right)\right).
\end{align*}

We now proceed to show the claim stated at the beginning of the proof.
The result (\ref{eq:bound_last}) implies 
\begin{align*}
 & \Pr\left(\left.\max_{i=1,...,n}\left|\hat{m}\left(\boldsymbol{z}_{i}^{\hat{*}}\right)-m\left(\boldsymbol{z}_{i}^{*}\right)\right|\geq\frac{C_{2}\log n}{n^{r\left/\left(p^{\mathsf{c}\ast}+2r\right)\right.}}\right|\hat{E}\cap\hat{F}\right)\\
= & \Pr\left(\left.\max_{i=1,...,n}\left|\hat{f}\left(D=1|\boldsymbol{\tilde{x}}_{i}\right)-m\left(\boldsymbol{z}_{i}^{*}\right)\right|\geq\frac{C_{2}\log n}{n^{r\left/\left(p^{\mathsf{c}\ast}+2r\right)\right.}}\right|\hat{E}\cap\hat{F}\right)\\
\leq & \sum_{i=1}^{n}\Pr\left(\left.\left|\hat{f}\left(D=1|\boldsymbol{\tilde{x}}_{i}\right)-m\left(\boldsymbol{z}_{i}^{*}\right)\right|\geq\frac{C_{2}\log n}{n^{r\left/\left(p^{\mathsf{c}\ast}+2r\right)\right.}}\right|\hat{E}\cap\hat{F}\right)\lesssim n^{-M+1}\rightarrow0,
\end{align*}
given that $M>1.$ Since $\Pr\left(\hat{E}\cap\hat{F}\right)\rightarrow1$,
we can remove the conditional event because for any event $A$ such
that $\Pr\left(A\left|\hat{E}\cap\hat{F}\right.\right)\rightarrow0$,
we have 
\begin{align*}
\Pr\left(A\right) & \leq\Pr\left(A\left|\hat{E}\cap\hat{F}\right.\right)\cdot\Pr\left(\hat{E}\cap\hat{F}\right)+\Pr\left(A\left|\left\{ \hat{E}\cap\hat{F}\right\} ^{c}\right.\right)\Pr\left(\left\{ \hat{E}\cap\hat{F}\right\} ^{c}\right)\\
 & \leq\Pr\left(A\left|\hat{E}\cap\hat{F}\right.\right)+\Pr\left(\left\{ \hat{E}\cap\hat{F}\right\} ^{c}\right)\rightarrow0.
\end{align*}
Therefore, we reach 
\[
\Pr\left(\max_{i=1,...,n}\left|\hat{m}\left(\boldsymbol{z}_{i}^{\hat{*}}\right)-m\left(\boldsymbol{z}_{i}^{*}\right)\right|\geq\frac{C_{2}\log n}{n^{r\left/\left(p^{\mathsf{c}\ast}+2r\right)\right.}}\right)\rightarrow0,
\]
which is equivalent to 
\[
\max_{i=1,...,n}\left|\hat{m}\left(\boldsymbol{z}_{i}^{\hat{*}}\right)-m\left(\boldsymbol{z}_{i}^{*}\right)\right|=O_{P}\left(\frac{\log n}{n^{r\left/\left(p^{\mathsf{c}\ast}+2r\right)\right.}}\right),
\]
as desired. 

\hfill{}$\square$

\section{Tables} \label{appendix:tables}
\subsection{Simulation Results of Section \ref{sec:simu_1}} \label{appendix:tables_1}
\begin{table}[H]  
	\centering
	\footnotesize
	\setlength{\tabcolsep}{3.5pt}
	\caption{Screening Results for Design 1}
	\label{tab:design1_top4_screening}
	\begin{tabular}{@{}llrrrrrrrrrrrr@{}}
		\toprule
		 &  
		& \multicolumn{4}{c}{$p = 20$} 
		& \multicolumn{4}{c}{$p = 50$} 
		& \multicolumn{4}{c}{$p = 100$} \\
		\cmidrule(lr){3-6}\cmidrule(lr){7-10}\cmidrule(lr){11-14}
		& Method & $X_{1}^{\mathsf{c}}$ &  $X_{2}^{\mathsf{c}}$&  $X_{1}^{\mathsf{d}}$ & All
		& $X_{1}^{\mathsf{c}}$ & $X_{2}^{\mathsf{c}}$ & $X_{1}^{\mathsf{d}}$ & All
		& $X_{1}^{\mathsf{c}}$ & $X_{2}^{\mathsf{c}}$ & $X_{1}^{\mathsf{d}}$ & All \\
		\midrule
		$n=250$  & $\rho$           & 0.976 & 0.964 & 0.705 & 0.652 & 0.953 & 0.918 & 0.546 & 0.459 & 0.910 & 0.868 & 0.463 & 0.333 \\
		      & $\textrm{Quantile-}\rho$ & 0.987 & 0.959 & 0.956 & 0.904 & 0.959 & 0.915 & 0.896 & 0.778 & 0.924 & 0.880 & 0.866 & 0.688 \\
		      & CDCSIS       & 0.998 & 0.999 & 1.000 & 0.997 & 0.999 & 0.991 & 0.999 & 0.989 & 0.994 & 0.983 & 0.999 & 0.976 \\
		      & FOCI         & 0.468 & 0.705 & 0.720 & 0.196 & 0.316 & 0.559 & 0.484 & 0.058 & 0.226 & 0.484 & 0.357 & 0.020 \\\hline
		$n=500$  & $\rho$           & 1.000 & 1.000 & 0.966 & 0.966 & 0.997 & 0.995 & 0.919 & 0.911 & 0.995 & 0.994 & 0.905 & 0.896 \\
		      & $\textrm{Quantile-}\rho$ & 1.000 & 1.000 & 0.999 & 0.999 & 0.997 & 0.995 & 0.997 & 0.989 & 0.997 & 0.993 & 1.000 & 0.990 \\
		      & CDCSIS       & 1.000 & 1.000 & 1.000 & 1.000 & 1.000 & 1.000 & 1.000 & 1.000 & 1.000 & 1.000 & 1.000 & 1.000 \\
		      & FOCI         & 0.622 & 0.927 & 0.842 & 0.450 & 0.484 & 0.860 & 0.720 & 0.243 & 0.330 & 0.772 & 0.605 & 0.125 \\\hline
		$n=1000$ & $\rho$           & 1.000 & 1.000 & 1.000 & 1.000 & 1.000 & 1.000 & 1.000 & 1.000 & 1.000 & 1.000 & 1.000 & 1.000 \\
		      & $\textrm{Quantile-}\rho$ & 1.000 & 1.000 & 1.000 & 1.000 & 1.000 & 1.000 & 1.000 & 1.000 & 1.000 & 1.000 & 1.000 & 1.000 \\
		      & CDCSIS       & 1.000 & 1.000 & 1.000 & 1.000 & 1.000 & 1.000 & 1.000 & 1.000 & 1.000 & 1.000 & 1.000 & 1.000 \\
		      & FOCI         & 0.805 & 0.998 & 0.953 & 0.764 & 0.661 & 0.988 & 0.930 & 0.597 & 0.586 & 0.972 & 0.856 & 0.460 \\\hline
		$n=2000$ & $\rho$           & 1.000 & 1.000 & 1.000 & 1.000 & 1.000 & 1.000 & 1.000 & 1.000 & 1.000 & 1.000 & 1.000 & 1.000 \\
		      & $\textrm{Quantile-}\rho$ & 1.000 & 1.000 & 1.000 & 1.000 & 1.000 & 1.000 & 1.000 & 1.000 & 1.000 & 1.000 & 1.000 & 1.000 \\
		      & CDCSIS       & / & / & / & / & / & / & / & / & / & / & / & / \\
		      & FOCI         & 0.945 & 1.000 & 1.000 & 0.945 & 0.902 & 1.000 & 0.995 & 0.897 & 0.853 & 1.000 & 0.990 & 0.843 \\
		\bottomrule
	\end{tabular}
\end{table}

\begin{table}[H]  
    \centering
    \footnotesize
    \setlength{\tabcolsep}{3.5pt}
    \caption{Screening Results for Design 2}
    \label{tab:design2_top4_screening}
    \begin{tabular}{@{}llrrrrrrrrrrrr@{}}
		\toprule
		  &  
		& \multicolumn{4}{c}{$p = 20$} 
		& \multicolumn{4}{c}{$p = 50$} 
		& \multicolumn{4}{c}{$p = 100$} \\
		\cmidrule(lr){3-6}\cmidrule(lr){7-10}\cmidrule(lr){11-14}
		& Method & $X_{1}^{\mathsf{c}}$ &  $X_{2}^{\mathsf{c}}$&  $X_{1}^{\mathsf{d}}$ & All
		& $X_{1}^{\mathsf{c}}$ & $X_{2}^{\mathsf{c}}$ & $X_{1}^{\mathsf{d}}$ & All
		& $X_{1}^{\mathsf{c}}$ & $X_{2}^{\mathsf{c}}$ & $X_{1}^{\mathsf{d}}$ & All \\
		\midrule
        $n=250$  & $\rho$           & 1.000 & 0.997 & 0.973 & 0.970 & 1.000 & 0.999 & 0.957 & 0.956 & 0.998 & 0.989 & 0.928 & 0.915 \\
              & $\textrm{Quantile-}\rho$ & 1.000 & 0.997 & 0.999 & 0.996 & 0.999 & 0.996 & 0.995 & 0.990 & 0.997 & 0.990 & 0.989 & 0.976 \\
              & CDCSIS       & 1.000 & 1.000 & 1.000 & 1.000 & 1.000 & 1.000 & 1.000 & 1.000 & 1.000 & 1.000 & 1.000 & 1.000 \\
              & FOCI         & 0.468 & 0.705 & 0.720 & 0.196 & 0.316 & 0.559 & 0.484 & 0.058 & 0.226 & 0.484 & 0.357 & 0.020 \\\hline
        $n=500$  & $\rho$           & 1.000 & 1.000 & 0.966 & 0.966 & 0.997 & 0.995 & 0.919 & 0.911 & 0.995 & 0.994 & 0.905 & 0.896 \\
              & $\textrm{Quantile-}\rho$ & 1.000 & 1.000 & 0.999 & 0.999 & 0.997 & 0.995 & 0.997 & 0.989 & 0.997 & 0.993 & 1.000 & 0.990 \\
              & CDCSIS       & 1.000 & 1.000 & 1.000 & 1.000 & 1.000 & 1.000 & 1.000 & 1.000 & 1.000 & 1.000 & 1.000 & 1.000 \\
              & FOCI         & 0.622 & 0.927 & 0.842 & 0.450 & 0.484 & 0.860 & 0.720 & 0.243 & 0.330 & 0.772 & 0.605 & 0.125 \\\hline
        $n=1000$ & $\rho$           & 1.000 & 1.000 & 1.000 & 1.000 & 1.000 & 1.000 & 1.000 & 1.000 & 1.000 & 1.000 & 1.000 & 1.000 \\
              & $\textrm{Quantile-}\rho$ & 1.000 & 1.000 & 1.000 & 1.000 & 1.000 & 1.000 & 1.000 & 1.000 & 1.000 & 1.000 & 1.000 & 1.000 \\
              & CDCSIS       & 1.000 & 1.000 & 1.000 & 1.000 & 1.000 & 1.000 & 1.000 & 1.000 & 1.000 & 1.000 & 1.000 & 1.000 \\
              & FOCI         & 0.805 & 0.998 & 0.953 & 0.764 & 0.661 & 0.988 & 0.930 & 0.597 & 0.586 & 0.972 & 0.856 & 0.460 \\\hline
        $n=2000$ & $\rho$           &  1.000 &  1.000 &  1.000 &  1.000 &  1.000 & 1.000 &  1.000 &  1.000 &  1.000 &  1.000 &  1.000 &  1.000 \\
              & $\textrm{Quantile-}\rho$ &  1.000 &  1.000 &  1.000 &  1.000 &  1.000 &  1.000 &  1.000 &  1.000 &  1.000 &  1.000 &  1.000 &  1.000 \\
              & CDCSIS       &  / & / & / & / & / & / & / & / & / & / & / & / \\
              & FOCI         &  1.000 &  1.000 &  1.000 &  1.000 &  1.000 &  1.000 &  1.000 &  1.000 &  1.000 &  1.000 &  1.000 &  1.000 \\
        \bottomrule
    \end{tabular}
\end{table}

\begin{table}[H]   
    \centering
    \footnotesize
    \setlength{\tabcolsep}{3.5pt}
    \caption{Screening Results for Design 3}
    \label{tab:design3_top4_screening}
    \begin{tabular}{@{}llrrrrrrrrrrrr@{}}
		\toprule
		  &  
		& \multicolumn{4}{c}{$p = 20$} 
		& \multicolumn{4}{c}{$p = 50$} 
		& \multicolumn{4}{c}{$p = 100$} \\
		\cmidrule(lr){3-6}\cmidrule(lr){7-10}\cmidrule(lr){11-14}
		& Method & $X_{1}^{\mathsf{c}}$ &  $X_{2}^{\mathsf{c}}$&  $X_{1}^{\mathsf{d}}$ & All
		& $X_{1}^{\mathsf{c}}$ & $X_{2}^{\mathsf{c}}$ & $X_{1}^{\mathsf{d}}$ & All
		& $X_{1}^{\mathsf{c}}$ & $X_{2}^{\mathsf{c}}$ & $X_{1}^{\mathsf{d}}$ & All \\
		\midrule
        $n=250$  & $\rho$           & 0.992 & 0.974 & 0.840 & 0.807 & 0.975 & 0.948 & 0.690 & 0.629 & 0.958 & 0.902 & 0.606 & 0.509 \\
              & $\textrm{Quantile-}\rho$ & 0.980 & 0.956 & 0.971 & 0.908 & 0.949 & 0.911 & 0.910 & 0.776 & 0.925 & 0.848 & 0.829 & 0.631 \\
              & CDCSIS       & 0.997 & 0.994 & 1.000 & 0.991 & 0.997 & 0.982 & 1.000 & 0.979 & 0.994 & 0.977 & 1.000 & 0.971 \\
              & FOCI         & 0.443 & 0.678 & 0.629 & 0.151 & 0.285 & 0.508 & 0.401 & 0.040 & 0.206 & 0.448 & 0.298 & 0.015 \\\hline
        $n=500$  & $\rho$           & 1.000 & 1.000 & 0.988 & 0.986 & 0.999 & 0.997 & 0.941 & 0.936 & 0.997 & 0.995 & 0.921 & 0.906 \\
              & $\textrm{Quantile-}\rho$ & 1.000 & 1.000 & 1.000 & 0.998 & 0.997 & 0.994 & 0.997 & 0.989 & 0.995 & 0.991 & 0.995 & 0.984 \\
              & CDCSIS       & 1.000 & 1.000 & 1.000 & 1.000 & 1.000 & 1.000 & 1.000 & 1.000 & 1.000 & 1.000 & 1.000 & 1.000 \\
              & FOCI         & 0.600 & 0.897 & 0.833 & 0.349 & 0.502 & 0.830 & 0.730 & 0.176 & 0.389 & 0.744 & 0.603 & 0.090 \\\hline
        $n=1000$ & $\rho$           & 1.000 & 1.000 & 1.000 & 1.000 & 1.000 & 1.000 & 1.000 & 1.000 & 1.000 & 1.000 & 1.000 & 1.000 \\
              & $\textrm{Quantile-}\rho$ & 1.000 & 1.000 & 1.000 & 1.000 & 1.000 & 1.000 & 1.000 & 1.000 & 1.000 & 1.000 & 1.000 & 1.000 \\
              & CDCSIS       & 1.000 & 1.000 & 1.000 & 1.000 & 1.000 & 1.000 & 1.000 & 1.000 & 1.000 & 1.000 & 1.000 & 1.000 \\
              & FOCI         & 0.748 & 0.991 & 0.950 & 0.654 & 0.633 & 0.987 & 0.927 & 0.494 & 0.554 & 0.968 & 0.847 & 0.341 \\\hline
        $n=2000$ & $\rho$           & 1.000 & 1.000 & 1.000 & 1.000 & 1.000 & 1.000 & 1.000 & 1.000 & 1.000 & 1.000 & 1.000 & 1.000 \\
              & $\textrm{Quantile-}\rho$ & 1.000 & 1.000 & 1.000 & 1.000 & 1.000 & 1.000 & 1.000 & 1.000 & 1.000 & 1.000 & 1.000 & 1.000 \\
              & CDCSIS       & / & /  & /  & /  & /  & /  & /  & /  & /  & /  & /  & /  \\
              & FOCI         & 0.997 & 1.000 & 1.000 & 0.997 & 0.980 & 1.000 & 0.999 & 0.979 & 0.969 & 1.000 & 1.000 & 0.969 \\
        \bottomrule
    \end{tabular}
\end{table}

\begin{table}[htbp] 
    \centering
    \caption{Performance of CV Refining for Designs 1--3}
    \label{tab:prob1_refine}
    \renewcommand{\arraystretch}{1.2}
    \small
    \begin{tabular}{clc cc cc cc}
        \toprule[1.2pt]
        \multicolumn{3}{c}{%
            \multirow{2}{*}{$\Pr\!\bigl(\boldsymbol{X}(\mathcal{M}^{*}) = \boldsymbol{X}(\hat{\mathcal{M}}^{(2)})\bigr)$}
        }
        & \multicolumn{2}{c}{$p = 20$}
        & \multicolumn{2}{c}{$p = 50$}
        & \multicolumn{2}{c}{$p = 100$} \\
        \cmidrule(lr){4-5}\cmidrule(lr){6-7}\cmidrule(lr){8-9}
        & & & CV & Modified-CV & CV & Modified-CV & CV & Modified-CV \\
        \midrule
\multirow{4}{*}{Design 1}
& $n=250$  & & 0.552 & 0.640 & 0.453 & 0.542 & 0.343 & 0.466 \\
& $n=500$  & & 0.748 & 0.915 & 0.741 & 0.897 & 0.719 & 0.894 \\
& $n=1000$ & & 0.869 & 0.990 & 0.865 & 0.993 & 0.824 & 0.985 \\
& $n=2000$ & & /     & 1.000     & /     & 1.000     & /     & 1.000     \\
\midrule
\multirow{4}{*}{Design 2}
& $n=250$  & & 0.641 & 0.740 & 0.589 & 0.723 & 0.539 & 0.699 \\
& $n=500$  & & 0.782 & 0.920 & 0.724 & 0.918 & 0.717 & 0.911 \\
& $n=1000$ & & 0.860 & 0.991 & 0.836 & 0.988 & 0.846 & 0.981 \\
& $n=2000$ & & /     & 0.999     & /     & 1.000     & /    & 1.000       \\
\midrule
\multirow{4}{*}{Design 3}
& $n=250$  & & 0.542 & 0.697 & 0.448 & 0.571 & 0.344 & 0.466 \\
& $n=500$  & & 0.758 & 0.930 & 0.699 & 0.898 & 0.682 & 0.875 \\
& $n=1000$ & & 0.813 & 0.983 & 0.774 & 0.981 & 0.783 & 0.975 \\
& $n=2000$ & & /     & 1.000     & /     & 0.998     & /     & 0.997     \\
        \bottomrule[1.2pt]
    \end{tabular}
\end{table}

\subsection{Simulation Results of Section \ref{sec:simu_2}} \label{appendix:tables_2}
\begin{table}[H]
    \centering
    \footnotesize
    \setlength{\tabcolsep}{3.5pt}
    \caption{Screening Results for Design 4}
    \label{tab:design4_top4_screening}
    \begin{tabular}{@{}llrrrrrrrrrrrr@{}}
		\toprule
		  &  
		& \multicolumn{4}{c}{$p = 20$} 
		& \multicolumn{4}{c}{$p = 50$} 
		& \multicolumn{4}{c}{$p = 100$} \\
		\cmidrule(lr){3-6}\cmidrule(lr){7-10}\cmidrule(lr){11-14}
		& Method & $X_{1}^{\mathsf{c}}$ &  $X_{2}^{\mathsf{c}}$&  $X_{1}^{\mathsf{d}}$ & All
		& $X_{1}^{\mathsf{c}}$ & $X_{2}^{\mathsf{c}}$ & $X_{1}^{\mathsf{d}}$ & All
		& $X_{1}^{\mathsf{c}}$ & $X_{2}^{\mathsf{c}}$ & $X_{1}^{\mathsf{d}}$ & All \\
		\midrule
        $n=250$ & $\rho$ & 0.978 & 0.929 & 0.923 & 0.833 & 0.934 & 0.846 & 0.858 & 0.665 & 0.884 & 0.786 & 0.776 & 0.514 \\
         & CDCSIS & 0.987 & 0.929 & 0.999 & 0.915 & 0.961 & 0.870 & 0.994 & 0.830 & 0.937 & 0.807 & 0.989 & 0.745 \\
         & FOCI & 0.446 & 0.577 & 0.582 & 0.108 & 0.270 & 0.391 & 0.379 & 0.024 & 0.205 & 0.321 & 0.238 & 0.012 \\\hline
        $n=500$ & $\rho$ & 1.000 & 0.999 & 0.997 & 0.996 & 0.996 & 0.990 & 0.998 & 0.984 & 0.997 & 0.983 & 0.997 & 0.977 \\
         & CDCSIS & 1.000 & 0.999 & 1.000 & 0.999 & 0.999 & 0.992 & 1.000 & 0.991 & 1.000 & 0.991 & 1.000 & 0.991 \\
         & FOCI & 0.524 & 0.754 & 0.704 & 0.219 & 0.354 & 0.637 & 0.523 & 0.079 & 0.244 & 0.566 & 0.429 & 0.032 \\\hline
        $n=1000$ & $\rho$ & 1.000 & 1.000 & 1.000 & 1.000 & 1.000 & 1.000 & 1.000 & 1.000 & 1.000 & 1.000 & 1.000 & 1.000 \\
         & CDCSIS & 1.000 & 1.000 & 1.000 & 1.000 & 1.000 & 1.000 & 1.000 & 1.000 & 1.000 & 1.000 & 1.000 & 1.000 \\
         & FOCI & 0.644 & 0.947 & 0.873 & 0.517 & 0.484 & 0.894 & 0.718 & 0.263 & 0.395 & 0.811 & 0.615 & 0.157 \\\hline
         $n=2000$ & $\rho$           & 1.000 & 1.000 & 1.000 & 1.000 & 1.000 & 1.000 & 1.000 & 1.000 & 1.000 & 1.000 & 1.000 & 1.000 \\
          & CDCSIS       & / & / & / & / & / & / & / & / & / & / & / & / \\
          & FOCI         & 0.824 & 0.997 & 0.971 & 0.794 & 0.681 & 0.986 & 0.940 & 0.616 & 0.596 & 0.989 & 0.894 & 0.516 \\
        \bottomrule
    \end{tabular}
\end{table}

\begin{table}[H]
    \centering
    \footnotesize
    \setlength{\tabcolsep}{3.5pt}
    \caption{Screening Results for Design 5}
    \label{tab:design5_top4_screening}
    \begin{tabular}{@{}llrrrrrrrrrrrr@{}}
		\toprule
		  &  
		& \multicolumn{4}{c}{$p = 20$} 
		& \multicolumn{4}{c}{$p = 50$} 
		& \multicolumn{4}{c}{$p = 100$} \\
		\cmidrule(lr){3-6}\cmidrule(lr){7-10}\cmidrule(lr){11-14}
		& Method & $X_{1}^{\mathsf{c}}$ &  $X_{2}^{\mathsf{c}}$&  $X_{1}^{\mathsf{d}}$ & All
		& $X_{1}^{\mathsf{c}}$ & $X_{2}^{\mathsf{c}}$ & $X_{1}^{\mathsf{d}}$ & All
		& $X_{1}^{\mathsf{c}}$ & $X_{2}^{\mathsf{c}}$ & $X_{1}^{\mathsf{d}}$ & All \\
		\midrule
        $n=250$ & $\rho$ & 1.000 & 0.994 & 0.995 & 0.989 & 0.997 & 0.983 & 0.984 & 0.965 & 0.992 & 0.973 & 0.956 & 0.921 \\
         & CDCSIS & 1.000 & 0.999 & 1.000 & 0.999 & 1.000 & 0.996 & 1.000 & 0.995 & 0.999 & 0.991 & 1.000 & 0.991 \\
         & FOCI & 0.560 & 0.793 & 0.765 & 0.275 & 0.436 & 0.688 & 0.592 & 0.132 & 0.335 & 0.612 & 0.482 & 0.053 \\\hline
        $n=500$ & $\rho$ & 1.000 & 1.000 & 0.998 & 0.998 & 0.998 & 0.996 & 0.990 & 0.988 & 0.997 & 0.992 & 0.989 & 0.983 \\
         & CDCSIS & 1.000 & 1.000 & 1.000 & 1.000 & 1.000 & 1.000 & 1.000 & 1.000 & 1.000 & 1.000 & 1.000 & 1.000 \\
         & FOCI & 0.643 & 0.929 & 0.860 & 0.438 & 0.513 & 0.897 & 0.767 & 0.233 & 0.389 & 0.810 & 0.661 & 0.155 \\\hline
        $n=1000$ & $\rho$ & 1.000 & 1.000 & 1.000 & 1.000 & 1.000 & 1.000 & 1.000 & 1.000 & 1.000 & 1.000 & 1.000 & 1.000 \\
         & CDCSIS & 1.000 & 1.000 & 1.000 & 1.000 & 1.000 & 1.000 & 1.000 & 1.000 & 1.000 & 1.000 & 1.000 & 1.000 \\
         & FOCI & 0.799 & 0.996 & 0.950 & 0.765 & 0.650 & 0.987 & 0.928 & 0.593 & 0.595 & 0.973 & 0.852 & 0.465 \\\hline
          $n=2000$ & $\rho$           & 1.000 & 1.000 & 1.000 & 1.000 & 1.000 & 1.000 & 1.000 & 1.000 & 1.000 & 1.000 & 1.000 & 1.000 \\
          & CDCSIS       & / & / & / & / & / & / & / & / & / & / & / & / \\
          & FOCI         & 0.984 & 1.000 & 0.997 & 0.981 & 0.942 & 0.996 & 0.997 & 0.935 & 0.926 & 0.995 & 0.998 & 0.919 \\
        \bottomrule
    \end{tabular}
\end{table}

\begin{table}[H]
    \centering
    \footnotesize
    \setlength{\tabcolsep}{3.5pt}
    \caption{Screening Results for Design 6}
    \label{tab:design6_top4_screening}
    \begin{tabular}{@{}llrrrrrrrrrrrr@{}}
		\toprule
		  &  
		& \multicolumn{4}{c}{$p = 20$} 
		& \multicolumn{4}{c}{$p = 50$} 
		& \multicolumn{4}{c}{$p = 100$} \\
		\cmidrule(lr){3-6}\cmidrule(lr){7-10}\cmidrule(lr){11-14}
		& Method & $X_{1}^{\mathsf{c}}$ &  $X_{2}^{\mathsf{c}}$&  $X_{1}^{\mathsf{d}}$ & All
		& $X_{1}^{\mathsf{c}}$ & $X_{2}^{\mathsf{c}}$ & $X_{1}^{\mathsf{d}}$ & All
		& $X_{1}^{\mathsf{c}}$ & $X_{2}^{\mathsf{c}}$ & $X_{1}^{\mathsf{d}}$ & All \\
		\midrule
        $n=250$ & $\rho$ & 0.947 & 0.901 & 0.892 & 0.749 & 0.894 & 0.799 & 0.782 & 0.527 & 0.844 & 0.705 & 0.654 & 0.350 \\
         & CDCSIS & 0.979 & 0.917 & 0.994 & 0.810 & 0.951 & 0.846 & 0.995 & 0.760 & 0.924 & 0.746 & 0.987 & 0.674 \\
         & FOCI & 0.395 & 0.575 & 0.509 & 0.101 & 0.248 & 0.389 & 0.329 & 0.028 & 0.171 & 0.331 & 0.247 & 0.015 \\\hline
        $n=500$ & $\rho$ & 0.996 & 0.992 & 0.979 & 0.969 & 0.995 & 0.987 & 0.930 & 0.912 & 0.990 & 0.979 & 0.906 & 0.873 \\
         & CDCSIS & 0.999 & 0.995 & 1.000 & 0.995 & 0.997 & 0.989 & 1.000 & 0.989 & 0.996 & 0.988 & 1.000 & 0.987 \\
         & FOCI & 0.491 & 0.800 & 0.694 & 0.248 & 0.401 & 0.715 & 0.611 & 0.127 & 0.311 & 0.646 & 0.512 & 0.076 \\\hline
        $n=1000$ & $\rho$ & 1.000 & 1.000 & 1.000 & 1.000 & 1.000 & 1.000 & 1.000 & 1.000 & 1.000 & 1.000 & 1.000 & 1.000 \\
         & CDCSIS & 1.000 & 1.000 & 1.000 & 1.000 & 1.000 & 1.000 & 1.000 & 1.000 & 1.000 & 1.000 & 1.000 & 1.000 \\
         & FOCI & 0.733 & 0.987 & 0.930 & 0.629 & 0.623 & 0.985 & 0.917 & 0.494 & 0.551 & 0.962 & 0.837 & 0.338 \\\hline
           $n=2000$ & $\rho$           & 1.000 & 1.000 & 1.000 & 1.000 & 1.000 & 1.000 & 1.000 & 1.000 & 1.000 & 1.000 & 1.000 & 1.000 \\
          & CDCSIS       & / & / & / & / & / & / & / & / & / & / & / & / \\
          & FOCI         & 0.879 & 0.987 & 0.989 & 0.856 & 0.779 & 0.969 & 0.968 & 0.728 & 0.701 & 0.945 & 0.927 & 0.599 \\
        \bottomrule
    \end{tabular}
\end{table}

\begin{table}[htbp]
    \centering
    \caption{Performance of CV Refining for Designs 4--6}
    \label{tab:prob1_refine_d456}
    \renewcommand{\arraystretch}{1.2}
    \small
    \begin{tabular}{clc cc cc cc}
        \toprule[1.2pt]
        \multicolumn{3}{c}{\multirow{2}{*}{$\Pr\!\bigl(\boldsymbol{X}(\mathcal{M}^{*}) = \boldsymbol{X}(\hat{\mathcal{M}}^{(2)})\bigr)$}}
        & \multicolumn{2}{c}{$p = 20$}
        & \multicolumn{2}{c}{$p = 50$}
        & \multicolumn{2}{c}{$p = 100$} \\
        \cmidrule(lr){4-5}\cmidrule(lr){6-7}\cmidrule(lr){8-9}
        & & & CV & Modified-CV & CV & Modified-CV & CV & Modified-CV \\
        \midrule
\multirow{4}{*}{Design 4}
& $n=250$  & & 0.461 & 0.456 & 0.335 & 0.344 & 0.242 & 0.263 \\
& $n=500$  & & 0.722 & 0.723 & 0.663 & 0.723 & 0.613 & 0.714 \\
& $n=1000$ & & 0.861 & 0.941 & 0.810 & 0.947 & 0.801 & 0.939 \\
& $n=2000$ & & /     & 1.000     & /     & 0.999     & /     & 0.999     \\
\midrule
\multirow{4}{*}{Design 5}
& $n=250$  & & 0.476 & 0.447 & 0.382 & 0.389 & 0.336 & 0.395 \\
& $n=500$  & & 0.642 & 0.665 & 0.614 & 0.622 & 0.571 & 0.642 \\
& $n=1000$ & & 0.787 & 0.894 & 0.776 & 0.886 & 0.773 & 0.886 \\
& $n=2000$ & & /     & 0.993     & /     & 0.997    & /     & 0.998     \\
\midrule
\multirow{4}{*}{Design 6}
& $n=250$  & & 0.353 & 0.357 & 0.213 & 0.222 & 0.151 & 0.161 \\
& $n=500$  & & 0.623 & 0.636 & 0.540 & 0.607 & 0.439 & 0.529 \\
& $n=1000$ & & 0.746 & 0.894 & 0.753 & 0.863 & 0.689 & 0.866 \\
& $n=2000$ & & /     & 0.992     & /     & 0.991     & /     & 0.998     \\
        \bottomrule[1.2pt]
    \end{tabular}
\end{table}

\begin{table}[H]
    \centering
    \footnotesize
    \setlength{\tabcolsep}{3.5pt}
    \caption{ATE Estimation for Design 4}
    \label{tab:design4_ate}
\begin{tabular}{@{}lccccccccccccc@{}}          
    \toprule
      
    & \multicolumn{4}{c}{$p = 20$} 
    & \multicolumn{4}{c}{$p = 50$} 
    & \multicolumn{4}{c}{$p = 100$} \\
    \cmidrule(lr){3-6}\cmidrule(lr){7-10}\cmidrule(lr){11-14}
    & Method 
    & Bias & RMSE & IL & CP
    & Bias & RMSE & IL & CP
    & Bias & RMSE & IL & CP \\
        \midrule
        $n=500$  & $\hat{\psi}_{1}$ & 0.132 & 0.732 & 2.608 & 0.935 & 0.207 & 0.909 & 3.074 & 0.920 & 0.251 & 0.971 & 3.195 & 0.913 \\
             & $\hat{\psi}_{2}$ & 0.137 & 0.395 & 1.123 & 0.877 & 0.120 & 0.445 & 1.251 & 0.860 & 0.131 & 0.444 & 1.262 & 0.877 \\
             & $\hat{\psi}_{3}$ & 0.073 & 0.313 & 0.973 & 0.904 & 0.045 & 0.326 & 0.986 & 0.894 & 0.054 & 0.347 & 1.012 & 0.892 \\
             & $\hat{\psi}_{4}$ & 0.064 & 0.346 & 1.108 & 0.924 & 0.031 & 0.348 & 1.113 & 0.900 & 0.015 & 0.367 & 1.133 & 0.889 \\\hline
        $n=1000$ & $\hat{\psi}_{1}$ & 0.038 & 0.462 & 1.733 & 0.942 & 0.200 & 0.699 & 2.300 & 0.913 & 0.214 & 0.747 & 2.411 & 0.905 \\
             & $\hat{\psi}_{2}$ & 0.050 & 0.210 & 0.649 & 0.887 & 0.122 & 0.316 & 0.814 & 0.836 & 0.099 & 0.314 & 0.849 & 0.832 \\
             & $\hat{\psi}_{3}$ & 0.007 & 0.178 & 0.626 & 0.939 & 0.018 & 0.178 & 0.620 & 0.939 & 0.007 & 0.191 & 0.633 & 0.919 \\
             & $\hat{\psi}_{4}$ & 0.002 & 0.201 & 0.693 & 0.925 & 0.015 & 0.185 & 0.692 & 0.948 & 0.002 & 0.203 & 0.703 & 0.930 \\\hline
        $n=2000$ & $\hat{\psi}_{1}$ & 0.011 & 0.278 & 1.049 & 0.938 & 0.151 & 0.529 & 1.736 & 0.897 & 0.229 & 0.571 & 1.797 & 0.895 \\
             & $\hat{\psi}_{2}$ & 0.021 & 0.111 & 0.392 & 0.923 & 0.074 & 0.207 & 0.516 & 0.786 & 0.111 & 0.241 & 0.575 & 0.779 \\
             & $\hat{\psi}_{3}$ & 0.002 & 0.158 & 0.592 & 0.949 & 0.003 & 0.168 & 0.593 & 0.944 & 0.003 & 0.183 & 0.595 & 0.931 \\
             & $\hat{\psi}_{4}$ & 0.003 & 0.172 & 0.674 & 0.938 & 0.005 & 0.179 & 0.692 & 0.932 & 0.006 & 0.189 & 0.707 & 0.919 \\\hline
        $n=4000$ & $\hat{\psi}_{1}$ & 0.016 & 0.172 & 0.613 & 0.931 & 0.078 & 0.345 & 1.276 & 0.936 & 0.196 & 0.444 & 1.363 & 0.886 \\
         & $\hat{\psi}_{2}$ & 0.003 & 0.071 & 0.258 & 0.932 & 0.039 & 0.110 & 0.307 & 0.845 & 0.078 & 0.180 & 0.377 & 0.724 \\
         & $\hat{\psi}_{3}$ &\!–0.002 & 0.080 & 0.286 & 0.936 & 0.002 & 0.078 & 0.287 & 0.950 & 0.001 & 0.083 & 0.289 & 0.939 \\
         & $\hat{\psi}_{4}$ & 0.004 & 0.083 & 0.307 & 0.950 & 0.007 & 0.082 & 0.306 & 0.946 & 0.007 & 0.088 & 0.308 & 0.935 \\
        \bottomrule
    \end{tabular}
\end{table}

\begin{table}[H]
   \centering
    \footnotesize
    \setlength{\tabcolsep}{3.5pt}
    \caption{ATE Estimation for Design 5}
    \label{tab:design5_ate}
\begin{tabular}{@{}lccccccccccccc@{}}          
    \toprule
 
    & \multicolumn{4}{c}{$p = 20$} 
    & \multicolumn{4}{c}{$p = 50$} 
    & \multicolumn{4}{c}{$p = 100$} \\
    \cmidrule(lr){3-6}\cmidrule(lr){7-10}\cmidrule(lr){11-14}
    & Method 
    & Bias & RMSE & IL & CP
    & Bias & RMSE & IL & CP
    & Bias & RMSE & IL & CP \\
        \midrule
        $n=500$ & $\hat{\psi}_{1}$ & 0.128 & 0.689 & 2.455 & 0.931 & 0.189 & 0.927 & 3.081 & 0.914 & 0.207 & 0.955 & 3.187 & 0.912 \\
            & $\hat{\psi}_{2}$ & 0.104 & 0.358 & 1.106 & 0.887 & 0.132 & 0.424 & 1.219 & 0.871 & 0.124 & 0.432 & 1.241 & 0.866 \\
            & $\hat{\psi}_{3}$ & 0.040 & 0.299 & 0.977 & 0.912 & 0.050 & 0.290 & 0.978 & 0.923 & 0.059 & 0.314 & 0.984 & 0.895 \\
            & $\hat{\psi}_{4}$ & 0.017 & 0.330 & 1.113 & 0.925 & 0.031 & 0.328 & 1.102 & 0.915 & 0.033 & 0.345 & 1.109 & 0.913 \\\hline
        $n=1000$ & $\hat{\psi}_{1}$ & 0.046 & 0.445 & 1.611 & 0.925 & 0.166 & 0.658 & 2.222 & 0.920 & 0.222 & 0.749 & 2.444 & 0.914 \\
             & $\hat{\psi}_{2}$ & 0.044 & 0.197 & 0.645 & 0.896 & 0.096 & 0.295 & 0.793 & 0.852 & 0.103 & 0.298 & 0.827 & 0.855 \\
             & $\hat{\psi}_{3}$ & 0.009 & 0.176 & 0.635 & 0.943 & 0.011 & 0.186 & 0.631 & 0.918 & 0.014 & 0.183 & 0.636 & 0.935 \\
             & $\hat{\psi}_{4}$ & 0.014 & 0.193 & 0.693 & 0.938 & 0.015 & 0.201 & 0.693 & 0.930 & 0.010 & 0.203 & 0.701 & 0.927 \\\hline
        $n=2000$ & $\hat{\psi}_{1}$ & 0.020 & 0.262 & 0.990 & 0.944 & 0.116 & 0.442 & 1.590 & 0.929 & 0.182 & 0.535 & 1.746 & 0.912 \\
             & $\hat{\psi}_{2}$ & 0.012 & 0.104 & 0.397 & 0.948 & 0.053 & 0.167 & 0.492 & 0.860 & 0.088 & 0.217 & 0.555 & 0.844 \\
             & $\hat{\psi}_{3}$ & 0.004 & 0.165 & 0.622 & 0.953 & 0.005 & 0.175 & 0.623 & 0.946 & 0.009 & 0.184 & 0.629 & 0.928 \\
             & $\hat{\psi}_{4}$ & 0.006 & 0.181 & 0.676 & 0.943 & 0.009 & 0.189 & 0.687 & 0.930 & 0.013 & 0.198 & 0.697 & 0.926 \\\hline
        $n=4000$ & $\hat{\psi}_{1}$ & 0.016 & 0.160 & 0.586 & 0.928 & 0.063 & 0.309 & 1.188 & 0.946 & 0.138 & 0.386 & 1.242 & 0.902 \\
         & $\hat{\psi}_{2}$ & 0.005 & 0.065 & 0.263 & 0.958 & 0.023 & 0.091 & 0.303 & 0.902 & 0.051 & 0.136 & 0.347 & 0.815 \\
         & $\hat{\psi}_{3}$ & 0.001 & 0.077 & 0.290 & 0.943 & 0.001 & 0.080 & 0.291 & 0.938 &\!–0.004 & 0.077 & 0.293 & 0.956 \\
         & $\hat{\psi}_{4}$ & 0.006 & 0.081 & 0.306 & 0.955 & 0.004 & 0.084 & 0.308 & 0.933 & 0.002 & 0.082 & 0.308 & 0.948 \\
        \bottomrule
    \end{tabular}
\end{table}

\begin{table}[H]
   \centering
    \footnotesize
    \setlength{\tabcolsep}{3.5pt}
    \caption{ATE Estimation for Design 6}
    \label{tab:design6_ate}
\begin{tabular}{@{}lccccccccccccc@{}}          
    \toprule
 
    & \multicolumn{4}{c}{$p = 20$} 
    & \multicolumn{4}{c}{$p = 50$} 
    & \multicolumn{4}{c}{$p = 100$} \\
    \cmidrule(lr){3-6}\cmidrule(lr){7-10}\cmidrule(lr){11-14}
    & Method 
    & Bias & RMSE & IL & CP
    & Bias & RMSE & IL & CP
    & Bias & RMSE & IL & CP \\
        \midrule
        $n=500$ & $\hat{\psi}_{1}$ & 0.090 & 0.684 & 2.438 & 0.942 & 0.141 & 0.889 & 3.069 & 0.936 & 0.194 & 0.950 & 3.229 & 0.933 \\
            & $\hat{\psi}_{2}$ & 0.099 & 0.371 & 1.106 & 0.886 & 0.125 & 0.414 & 1.228 & 0.887 & 0.153 & 0.450 & 1.239 & 0.870 \\
            & $\hat{\psi}_{3}$ & 0.065 & 0.312 & 0.978 & 0.919 & 0.064 & 0.325 & 1.013 & 0.890 & 0.093 & 0.343 & 1.035 & 0.893 \\
            & $\hat{\psi}_{4}$ & 0.041 & 0.332 & 1.105 & 0.918 & 0.051 & 0.339 & 1.125 & 0.920 & 0.077 & 0.362 & 1.149 & 0.913 \\\hline
        $n=1000$ & $\hat{\psi}_{1}$ & 0.065 & 0.432 & 1.579 & 0.936 & 0.174 & 0.643 & 2.197 & 0.930 & 0.162 & 0.702 & 2.421 & 0.931 \\
             & $\hat{\psi}_{2}$ & 0.049 & 0.195 & 0.644 & 0.914 & 0.112 & 0.281 & 0.790 & 0.864 & 0.104 & 0.294 & 0.826 & 0.863 \\
             & $\hat{\psi}_{3}$ & 0.029 & 0.183 & 0.623 & 0.923 & 0.022 & 0.184 & 0.629 & 0.934 & 0.019 & 0.185 & 0.631 & 0.921 \\
             & $\hat{\psi}_{4}$ & 0.041 & 0.208 & 0.687 & 0.935 & 0.034 & 0.193 & 0.686 & 0.933 & 0.031 & 0.204 & 0.694 & 0.928 \\\hline
        $n=2000$ & $\hat{\psi}_{1}$ & 0.048 & 0.264 & 0.975 & 0.933 & 0.112 & 0.462 & 1.579 & 0.924 & 0.171 & 0.509 & 1.754 & 0.912 \\
             & $\hat{\psi}_{2}$ & 0.023 & 0.109 & 0.399 & 0.929 & 0.055 & 0.168 & 0.487 & 0.868 & 0.081 & 0.207 & 0.551 & 0.844 \\
             & $\hat{\psi}_{3}$ & 0.013 & 0.162 & 0.622 & 0.946 & 0.013 & 0.169 & 0.615 & 0.941 & 0.021 & 0.175 & 0.622 & 0.928 \\
             & $\hat{\psi}_{4}$ & 0.017 & 0.181 & 0.676 & 0.943 & 0.020 & 0.190 & 0.689 & 0.931 & 0.028 & 0.198 & 0.705 & 0.918 \\\hline
        $n=4000$ & $\hat{\psi}_{1}$ & 0.032 & 0.165 & 0.579 & 0.927 & 0.058 & 0.312 & 1.175 & 0.932 & 0.125 & 0.356 & 1.254 & 0.922 \\
         & $\hat{\psi}_{2}$ & 0.003 & 0.070 & 0.262 & 0.947 & 0.025 & 0.091 & 0.302 & 0.901 & 0.052 & 0.127 & 0.345 & 0.827 \\
         & $\hat{\psi}_{3}$ &\!–0.001 & 0.074 & 0.283 & 0.946 &\!–0.001 & 0.078 & 0.285 & 0.946 & 0.002 & 0.076 & 0.286 & 0.945 \\
         & $\hat{\psi}_{4}$ & 0.007 & 0.078 & 0.301 & 0.958 & 0.008 & 0.085 & 0.302 & 0.942 & 0.013 & 0.083 & 0.303 & 0.935 \\

        \bottomrule
    \end{tabular}
\end{table}

\subsection{Empirical Results of Section \ref{sec:application}} \label{appendix:application_tables}

\begin{table}[htbp]
\centering
\footnotesize
\begin{threeparttable}
\caption{Mean and Standard Deviation (SD) of Outcome Variables and Selected Covariates}
\label{tab:app_summary}
\renewcommand{\arraystretch}{1.15}
\begin{tabular}{l d{3.3} d{3.3} d{3.3} d{3.3} d{3.3} d{3.3} d{3.3}}
\toprule
& \multicolumn{2}{c}{All} & \multicolumn{2}{c}{Eligible} & \multicolumn{2}{c}{Ineligible} & \multicolumn{1}{c}{$p$-value} \\
& \multicolumn{1}{c}{Mean} & \multicolumn{1}{c}{SD} & \multicolumn{1}{c}{Mean} & \multicolumn{1}{c}{SD} & \multicolumn{1}{c}{Mean} & \multicolumn{1}{c}{SD} & \\
\midrule
\textbf{Panel A: Outcome Variables} \\
$\Delta A_2^{\textrm{401(k)}}=$ Change in 401(k) assets in Year 2 & 3.093 & 37.100 & 3.930 & 42.415 & 1.840 & 27.220 & 0.014^{**} \\
$\Delta A_2^{\textrm{IRA}}=$ Change in IRA assets in Year 2 & 3.503 & 42.872 & 4.377 & 47.681 & 2.193 & 34.396 & 0.029^{**} \\
$\Delta A_2^{\textrm{Other}}=$ Change in other assets in Year 2 & 19.261 & 1243.410 & 26.966 & 1595.096 & 0.766 & 224.341 & 0.455 \\
$\Delta A_2^{\textrm{SDebt}}=$ Change in secured debt in Year 2 & 0.725 & 86.635 & 1.736 & 94.787 & -0.791 & 72.719 & 0.217 \\
$\Delta A_2^{\textrm{UDebt}}=$ Change in unsecured debt in Year 2 & -0.034 & 89.861 & -0.926 & 112.688 & 1.302 & 33.849 & 0.238 \\
$\Delta A_2^{\textrm{Car}}=$ Change in car value in Year 2 & 0.812 & 8.558 & 0.910 & 8.766 & 0.665 & 8.237 & 0.244 \\
\midrule
\textbf{Panel B: Selected Covariates} \\
$\textrm{HHIncome}$ = Household income in Year 1 & 65.017 & 44.235 & 70.603 & 47.850 & 56.646 & 36.631 & 0.000^{***} \\
 $\Delta A_0^{\textrm{401(k)}}=$  Change in 401(k) assets in Year 0 & 3.497 & 29.959 & 5.270 & 33.982 & 0.840 & 22.373 & 0.000^{***} \\
 $\Delta A_0^{\textrm{Other}}=$ Change in other assets in Year 0 & -3.588 & 775.758 & -11.136 & 997.896 & 7.723 & 224.392 & 0.234 \\
$\Delta A_0^{\textrm{UDebt}}=$ Change in unsecured debt in Year 0 & 1.250 & 77.510 & 1.977 & 81.803 & 0.162 & 70.592 & 0.332 \\
$\textrm{Female}= \mathbf{1}$(individual is female) & 0.452 & 0.498 & 0.410 & 0.492 & 0.515 & 0.500 & 0.000^{***} \\
$\textrm{PHI}= \mathbf{1}$(individual has private health insurance) & 0.522 & 0.500 & 0.590 & 0.492 & 0.417 & 0.493 & 0.000^{***} \\
 $\textrm{Firm100}= \mathbf{1}$(number of employees $\geq$ 100) & 0.271 & 0.445 & 0.185 & 0.388 & 0.383 & 0.486 & 0.000^{***} \\
$\textrm{Industry4}=\mathbf{1}$(industry code = 4) & 0.139 & 0.346 & 0.261 & 0.378 & 0.084 & 0.284 & 0.000^{***} \\
\midrule
Number of observations & \multicolumn{2}{c}{6741} &  \multicolumn{2}{c}{4043 (59.98\%)} &    \multicolumn{2}{c}{2698 (40.02\%)} & \\
\bottomrule
\end{tabular}
\begin{tablenotes}
\scriptsize
\item Notes: * $p<0.10$; ** $p<0.05$; *** $p<0.01$. $p$-values are based on two-sided $t$-tests for differences in means between eligible and ineligible individuals.
\end{tablenotes}
\end{threeparttable}
\end{table}

\begin{table}[htbp]
  \centering
  \small
  \begin{threeparttable}
  \caption{Doubly Robust Estimates of Average Treatment Effects of 401(k) Eligibility}
    \label{tab:app_estimation}
    \begin{tabular}{l d{3.3} d{3.3} d{3.3} d{3.3} d{3.3} d{3.3}}
    \toprule
          & \multicolumn{1}{c}{$\Delta A_2^{\textrm{401(k)}}$} & \multicolumn{1}{c}{$\Delta A_2^{\textrm{IRA}}$} & \multicolumn{1}{c}{$\Delta A_2^{\textrm{Other}}$} & \multicolumn{1}{c}{$\Delta A_2^{\textrm{SDebt}}$} & \multicolumn{1}{c}{$\Delta A_2^{\textrm{UDebt}}$} & \multicolumn{1}{c}{$\Delta A_2^{\textrm{Car}}$} \\
    \midrule
    \textbf{Panel A: Kernel $\hat{m}$, MLP $(\hat{g}_0, \hat{g}_1)$} &  \multicolumn{1}{c}{}     &    \multicolumn{1}{c}{}   &   \multicolumn{1}{c}{}    &    \multicolumn{1}{c}{}   &   \multicolumn{1}{c}{}    &  \\
    Observations trimmed: 2.58\% & 1.874^{***} & 0.417^{*} & 1.513^{*} & 1.249 & -0.064 & 0.075 \\
          & (0.387) & (0.247) & (0.839) & (0.778) & (0.181) & (0.173) \\
    \midrule
    \textbf{Panel B: MLP $\hat{m}$, MLP $(\hat{g}_0, \hat{g}_1)$} &  \multicolumn{1}{c}{}     &   \multicolumn{1}{c}{}    &   \multicolumn{1}{c}{}    &    \multicolumn{1}{c}{}   &   \multicolumn{1}{c}{}    &  \\
    Observations trimmed: 0.19\% & 1.621^{***} & -0.242 & 0.950 & 1.189 & -0.017 & 0.107 \\
          & (0.418) & (0.331) & (1.004) & (0.868) & (0.187) & (0.178) \\
    \bottomrule
    \end{tabular}
        \begin{tablenotes}
\scriptsize
\item Notes: * $p<0.10$; ** $p<0.05$; *** $p<0.01$.
\end{tablenotes}
\end{threeparttable}
\end{table}

\begin{table}[htbp]
  \centering
  \small
    \begin{threeparttable}
  \caption{$p$-values of Balance Tests for Selected Covariates within Propensity Score Subclasses}
    \label{tab:app_balance}
    \begin{tabular}{c*{8}{d{3.3}}}
    \toprule
    Kernel $\hat{m}$ & \multicolumn{1}{c}{$\textrm{HHIncome}$} & \multicolumn{1}{c}{$\Delta A_0^{\textrm{401(k)}}$}  & \multicolumn{1}{c}{$\Delta A_0^{\textrm{Other}}$} & \multicolumn{1}{c}{$\Delta A_0^{\textrm{UDebt}}$} & \multicolumn{1}{c}{$\textrm{Female}$} & \multicolumn{1}{c}{$\textrm{PHI}$} & \multicolumn{1}{c}{$\textrm{Firm100}$} & \multicolumn{1}{c}{$\textrm{Industry4}$} \\
    \midrule
    $[0.1, 0.2)$ & \multicolumn{1}{c}{$\textrm{\quad NA}$} & \multicolumn{1}{c}{$\textrm{\quad NA}$} & \multicolumn{1}{c}{$\textrm{\quad NA}$} & \multicolumn{1}{c}{$\textrm{\quad NA}$}  & \multicolumn{1}{c}{$\textrm{\quad NA}$} & \multicolumn{1}{c}{$\textrm{\quad NA}$} & \multicolumn{1}{c}{$\textrm{\quad NA}$} & \multicolumn{1}{c}{$\textrm{\quad NA}$} \\
    $[0.2, 0.3)$ & 0.257 & 0.982 & 0.957 & 0.791 & 0.106 & 0.642 & 0.185 & 0.275 \\
    $[0.3, 0.4)$ & 0.006^{***} & 0.321 & 0.722 & 0.181 & 0.066^{*} & 0.275 & 0.034^{**} & 0.029^{**} \\
    $[0.4, 0.5)$ & 0.009^{***} & 0.972 & 0.175 & 0.280 & 0.117 & 0.212 & 0.328 & 0.971 \\
    $[0.5, 0.6)$ & 0.096^{*} & 0.235 & 0.128 & 0.752 & 0.691 & 0.315 & 0.016^{**} & 0.084^{*} \\
    $[0.6, 0.7)$ & 0.240 & 0.586 & 0.221 & 0.321 & 0.188 & 0.122 & 0.748 & 0.136 \\
    $[0.7, 0.8)$ & 0.834 & 0.581 & 0.484 & 0.353 & 0.523 & 0.964 & 0.001^{***} & 0.380 \\
    $[0.8, 0.9)$ & 0.338 & 0.333 & 0.976 & 0.724 & 0.652 & 0.601 & 0.565 & 0.413 \\
    \midrule
        & \multicolumn{1}{c}{} & \multicolumn{1}{c}{} & \multicolumn{1}{c}{} & \multicolumn{1}{c}{} & \multicolumn{1}{c}{} & \multicolumn{1}{c}{} & \multicolumn{1}{c}{} & \multicolumn{1}{c}{}  \\
    \midrule
    MLP $\hat{m}$ & \multicolumn{1}{c}{$\textrm{HHIncome}$} & \multicolumn{1}{c}{$\Delta A_0^{\textrm{401(k)}}$}  & \multicolumn{1}{c}{$\Delta A_0^{\textrm{Other}}$} & \multicolumn{1}{c}{$\Delta A_0^{\textrm{UDebt}}$} & \multicolumn{1}{c}{$\textrm{Female}$} & \multicolumn{1}{c}{$\textrm{PHI}$} & \multicolumn{1}{c}{$\textrm{Firm100}$} & \multicolumn{1}{c}{$\textrm{Industry4}$} \\
    \midrule
    $[0.1, 0.2)$ & \multicolumn{1}{c}{$\textrm{\quad NA}$} & \multicolumn{1}{c}{$\textrm{\quad NA}$} & \multicolumn{1}{c}{$\textrm{\quad NA}$} & \multicolumn{1}{c}{$\textrm{\quad NA}$}  & \multicolumn{1}{c}{$\textrm{\quad NA}$} & \multicolumn{1}{c}{$\textrm{\quad NA}$} & \multicolumn{1}{c}{$\textrm{\quad NA}$} & \multicolumn{1}{c}{$\textrm{\quad NA}$} \\
    $[0.2, 0.3)$ & \multicolumn{1}{c}{$\textrm{\quad NA}$} & \multicolumn{1}{c}{$\textrm{\quad NA}$} & \multicolumn{1}{c}{$\textrm{\quad NA}$} & \multicolumn{1}{c}{$\textrm{\quad NA}$}  & \multicolumn{1}{c}{$\textrm{\quad NA}$} & \multicolumn{1}{c}{$\textrm{\quad NA}$} & \multicolumn{1}{c}{$\textrm{\quad NA}$} & \multicolumn{1}{c}{$\textrm{\quad NA}$} \\
    $[0.3, 0.4)$ & \multicolumn{1}{c}{$\textrm{\quad NA}$} & \multicolumn{1}{c}{$\textrm{\quad NA}$} & \multicolumn{1}{c}{$\textrm{\quad NA}$} & \multicolumn{1}{c}{$\textrm{\quad NA}$}  & \multicolumn{1}{c}{$\textrm{\quad NA}$} & \multicolumn{1}{c}{$\textrm{\quad NA}$} & \multicolumn{1}{c}{$\textrm{\quad NA}$} & \multicolumn{1}{c}{$\textrm{\quad NA}$} \\
    $[0.4, 0.5)$ & 0.389 & 0.094^{*} & 0.090^{*} & 0.665 & 0.000^{***} & 0.183 & 0.000^{***} & 0.000^{***} \\
    $[0.5, 0.6)$ & 0.017^{**} & 0.197 & 0.908 & 0.488 & 0.998 & 0.883 & 0.000^{***} & 0.000^{***} \\
    $[0.6, 0.7)$ & 0.003^{***} & 0.814 & 0.307 & 0.082^{*} & 0.054^{**} & 0.007^{***} & 0.006^{***} & 0.000^{***} \\
    $[0.7, 0.8)$ & 0.000^{***} & 0.001^{***} & 0.545 & 0.321 & 0.201 & 0.437 & 0.045^{**} & 0.004^{***} \\
    $[0.8, 0.9)$ & 0.313 & 0.738 & 0.572 & 0.259 & 0.172 & 0.017^{**} & 0.639 & 0.420 \\
    \bottomrule
    \end{tabular}
    \begin{tablenotes}
\scriptsize
\item Notes: * $p<0.10$; ** $p<0.05$; *** $p<0.01$.
\end{tablenotes}
\end{threeparttable}
\end{table}

\begin{table}[htbp]
  \centering
  \small
  \begin{threeparttable}
  \caption{Doubly Robust (H{\'a}jek) Estimates of Average Treatment Effects of 401(k) Eligibility}
    \label{tab:app_estimation_Hajek}
    \begin{tabular}{l d{3.3} d{3.3} d{3.3} d{3.3} d{3.3} d{3.3}}
    \toprule
          & \multicolumn{1}{c}{$\Delta A_2^{\textrm{401(k)}}$} & \multicolumn{1}{c}{$\Delta A_2^{\textrm{IRA}}$} & \multicolumn{1}{c}{$\Delta A_2^{\textrm{Other}}$} & \multicolumn{1}{c}{$\Delta A_2^{\textrm{SDebt}}$} & \multicolumn{1}{c}{$\Delta A_2^{\textrm{UDebt}}$} & \multicolumn{1}{c}{$\Delta A_2^{\textrm{Car}}$} \\
    \midrule
    \textbf{Panel A: Kernel $\hat{m}$, MLP $(\hat{g}_0, \hat{g}_1)$} &  \multicolumn{1}{c}{}     &    \multicolumn{1}{c}{}   &   \multicolumn{1}{c}{}    &    \multicolumn{1}{c}{}   &   \multicolumn{1}{c}{}    &  \\
    Observations trimmed: 2.58\% & 1.819^{***} & 0.317 & 1.276 & 1.329 & -0.068 & 0.037 \\
          & (0.412) & (0.264) & (0.897) & (0.832) & (0.193) & (0.185) \\
          \midrule
    \textbf{Panel B: MLP $\hat{m}$, MLP $(\hat{g}_0, \hat{g}_1)$} &  \multicolumn{1}{c}{}     &   \multicolumn{1}{c}{}    &   \multicolumn{1}{c}{}    &    \multicolumn{1}{c}{}   &   \multicolumn{1}{c}{}    &  \\
    Observations trimmed: 0.19\% & 1.708^{***} & -0.194 & 1.132 & 1.193 & -0.017 & 0.124 \\
          & (0.420) & (0.331) & (1.005) & (0.869) & (0.188) & (0.178) \\
    \bottomrule
    \end{tabular}
        \begin{tablenotes}
\scriptsize
\item Notes: * $p<0.10$; ** $p<0.05$; *** $p<0.01$.
\end{tablenotes}
\end{threeparttable}
\end{table}

\begin{table}[htbp]
  \centering
  \small
  \begin{threeparttable}
  \caption{DML Estimates of Average Treatment Effects of 401(k) Eligibility}
    \label{tab:app_estimation_split}
    \begin{tabular}{l d{3.3} d{3.3} d{3.3} d{3.3} d{3.3} d{3.3}}
    \toprule
          & \multicolumn{1}{c}{$\Delta A_2^{\textrm{401(k)}}$} & \multicolumn{1}{c}{$\Delta A_2^{\textrm{IRA}}$} & \multicolumn{1}{c}{$\Delta A_2^{\textrm{Other}}$} & \multicolumn{1}{c}{$\Delta A_2^{\textrm{SDebt}}$} & \multicolumn{1}{c}{$\Delta A_2^{\textrm{UDebt}}$} & \multicolumn{1}{c}{$\Delta A_2^{\textrm{Car}}$} \\
    \midrule
    \textbf{Panel A: Kernel $\hat{m}$, MLP $(\hat{g}_0, \hat{g}_1)$} &  \multicolumn{1}{c}{}     &    \multicolumn{1}{c}{}   &   \multicolumn{1}{c}{}    &    \multicolumn{1}{c}{}   &   \multicolumn{1}{c}{}    &  \\
    Observations trimmed: 4.51\% & 1.894^{***} & 0.302 & 1.726^{*} & 1.523^{*} & -0.107 & 0.070 \\
          & (0.422) & (0.282) & (0.910) & (0.861) & (0.195) & (0.187) \\
            \midrule
    \textbf{Panel B: MLP $\hat{m}$, MLP $(\hat{g}_0, \hat{g}_1)$} &  \multicolumn{1}{c}{}     &   \multicolumn{1}{c}{}    &   \multicolumn{1}{c}{}    &    \multicolumn{1}{c}{}   &   \multicolumn{1}{c}{}    &  \\
    Observations trimmed: 0.22\% & 2.140^{***} & 0.378 & 2.269^{***} & 0.903 & 0.002 & 0.179 \\
          & (0.371) & (0.270) & (0.861) & (0.792) & (0.175) & (0.167) \\
    \bottomrule
    \end{tabular}
        \begin{tablenotes}
\scriptsize
\item Notes: * $p<0.10$; ** $p<0.05$; *** $p<0.01$.
\end{tablenotes}
\end{threeparttable}
\end{table}

\end{document}